\documentclass[reprint,pre,aps 
	]{revtex4-2} 

\usepackage[english]{babel}

\usepackage[utf8]{inputenc} 
\usepackage{lmodern} 
\usepackage[T1]{fontenc}

\usepackage{amssymb}
\usepackage{mathtools} 

\usepackage{bbm} 
	\newcommand{\id}{\mathbbm{1}}

\usepackage[dvipsnames]{xcolor} 
	
\usepackage{graphicx}

\usepackage{tikz}
	\usetikzlibrary{decorations.pathreplacing}
	\usetikzlibrary{decorations.pathmorphing} 
	\usetikzlibrary{matrix,arrows} 

\usepackage{booktabs} 

\usepackage{enumitem}

\DeclarePairedDelimiter{\ket}{\lvert}{\rangle}
\DeclarePairedDelimiterX{\ketbra}[2]{\lvert}{\rvert}{#1\rangle \langle#2}
\DeclarePairedDelimiterX{\braket}[2]{\langle}{\rangle}{#1\vert#2}
	
\usepackage{bm} 

\newcommand{\I}{\mathrm{i}}
\newcommand{\E}{\mathrm{e}}

\newcommand\pp{\mathrm{p}}
\newcommand\G{\mathrm{G}}
\newcommand\HH{\mathrm{H}}

\newcommand\LL{\textsc{l}}
\newcommand\RR{\textsc{r}}

\newcommand\F{\mathrm{F}}
\newcommand\EE{\mathrm{E}}
\newcommand\N{\mathrm{N}}

\newcommand\CC{\mathrm{C}}
\newcommand\PP{\mathrm{P}}
\newcommand\TT{\mathrm{T}}
\newcommand\UU{\mathrm{U}}

\newcommand\q{\mathrm{q}}
\newcommand\V{\mathrm{V}}
\newcommand\Ss{\mathrm{S}}
\newcommand\R{\check{\mspace{1mu}\mathrm{R}\mspace{-1mu}}}

\DeclareMathOperator*{\ordprod}{\prod\limits^{\vbox to -.5ex{\kern-0.5ex\hbox{$\leftharpoonup$}\vss}}}
\DeclareMathOperator*{\ordprodopp}{\prod\limits^{\vbox to -.5ex{\kern-0.5ex\hbox{$\rightharpoonup$}\vss}}}
	
\usepackage[
	ocgcolorlinks, 
	linkcolor=BlueViolet, citecolor=OliveGreen, urlcolor=RawSienna, filecolor=Sepia,
	hyperfootnotes=false,
	]{hyperref} 

\begin{document}
		
\title{A solvable non-unitary fermionic long-range model with extended symmetry}
	
	\author{Adel Ben Moussa\textsuperscript{$a$}, Jules Lamers\textsuperscript{$a \!\curvearrowright\! b$}, Didina Serban\textsuperscript{$a$} and Ayman Toufik\textsuperscript{$a$}}
	\affiliation{\textsuperscript{$a$}\,Université Paris--Saclay, CNRS, CEA, Institut de Physique Théorique, 91191 Gif-sur-Yvette, France} 
	\affiliation{\textsuperscript{$b$}\,Deutsches Elektronen-Synchrotron DESY, Notkestraße 85, 22607 Hamburg, Germany}
	
\begin{abstract}
	\noindent 
	We define and study a long-range version of the \textsc{xx} model, arising as the free-fermion point of the \textsc{xxz}-type Haldane--Shastry (HS) chain. It has a description via non-unitary fermions, based on the free-fermion Temperley--Lieb algebra, and may also be viewed as an alternating $\mathfrak{gl}(1|1)$ spin chain. Even and odd length behave very differently; we focus on odd length. The model is integrable, and we explicitly identify two commuting hamiltonians. While non-unitary, their spectrum is real by PT-symmetry. One hamiltonian is chiral and quadratic in fermions, while the other is parity-invariant and quartic. Their one-particle spectra have two linear branches, realising a massless relativistic dispersion on the lattice. The appropriate fermionic modes arise from `quasi-translation' symmetry, which replaces ordinary translation symmetry. The model exhibits exclusion statistics, like the isotropic HS chain, with even more `extended symmetry' and larger degeneracies.
\end{abstract}
	
\hfill DESY-24-051
	
\maketitle

\noindent{\itshape\textbf{Introduction.}} Strongly-interacting quantum many-body systems lie at the core of condensed-matter physics. In 2d such systems exhibit rich collective behaviours, e.g.\ fractional excitations and spin-charge separation.
Several particularly interesting disorder-driven critical phenomena, like the plateau transition in the integer quantum Hall effect, and geometric problems, e.g.\ polymers or percolation, are inherently non-unitary \cite{chalker1988percolation,zirnbauer1994towards,gruzberg1999exact}.
The few tools available to treat such systems analytically are mostly based on super-spin chains, loop models and the Heisenberg \textsc{xxz} chain \cite{koo1994representations,Read_07,Gainutdinov:2011ab}. Despite their integrability, these non-unitary models remain challenging to analyse, as it is not yet well understood how their non-unitary infinite-dimensional symmetries are realised.

Models with \emph{long-range} interactions constitute an important chapter of integrability. Prominent examples are Calogero--Sutherland systems \cite{calogero1975exactly,*sutherland1975exact} and the associated spin chains \cite{Hal_88,*Sha_88,Ino_90}, 
which are deeply related to matrix models, exclusion statistics and 2d CFT \cite{Hal_94,Pol_99}. 
Long-range spin chains arise in AdS/CFT integrability too \cite{rej2012review}.
Rather than a Bethe ansatz, such models are tackled via symmetry-based algebraic methods.
In particular, the trigonometric spin-Calogero--Sutherland system and the associated Haldane--Shastry (HS) chain have \emph{extended} (Yangian) spin symmetry \cite{HH+_92,Bernard:1993va} rendering the spectrum very simple and degenerate \cite{Haldane:1991cxr,Bernard:1993va}.
Yet there are few, if any, examples of \emph{non-unitary} spin chains with extended symmetry to serve as finite discretisations of the non-unitary CFTs with current-algebra symmetry expected for disordered critical systems \cite{bernard1995perturbed}. 
\smallskip

\textit{Main result.} 
We introduce a new integrable system that can be viewed as a long-range \textsc{xx} model, a long-range model of non-unitary fermions, or a long-range alternating $\mathfrak{gl}(1|1)$ super-spin chain. It has 
\begin{itemize}[topsep=.5ex,itemsep=-.75ex]
 	\item[i)] a family of conserved charges, 
 	\item[ii)] extended symmetry,
 	\item[iii)] an extremely degenerate and simple spectrum.
\end{itemize}
\smallskip

This model arises as the free-fermion point of the HS chain. Underlying it is a `parent model' whose symmetries and spectrum are understood in detail: the \textsc{xxz}-type HS chain \cite{Bernard:1993va,Uglov:1995di,*Lamers:2018ypi,Lamers:2020ato}. It generalises the HS chain by breaking the $\mathfrak{su}(2)$ spin symmetry to $\mathfrak{u}(1)$ while preserving its key properties. 
Crucially, the extended spin symmetry persists \cite{Bernard:1993va}, where the Yangian is replaced by quantum-affine $\mathfrak{su}(2)$.
A deformation parameter $\q$ plays the role of the anisotropy $\Delta=(\q+\q^{-1})/2$ of the Heisenberg \textsc{xxz} chain, 
with $\q=1$ the isotropic case.
For generic $\q$, the parent model behaves like the HS chain, yet new features appear at roots of unity. Here we consider the important case $\q=\I$. For the Heisenberg \textsc{xxz} chain this gives the \textsc{xx} model ($\Delta=0$), equivalent to free fermions via the Jordan--Wigner transformation.
The model introduced in this Letter is the point $\q=\I$ of the parent model, described from a new fermionic perspective. We combine general knowledge from the parent model with fermionic techniques capturing the special features at $\q=\I$.

Our model has several striking features. Its properties depend sensitively on the parity of the system size. The parent model's antiferromagnetic (ground)state is non-/doubly degenerate for even/odd length, respectively. At $\q=1$, this yields two different CFT sectors in the scaling limit \cite{Haldane:1991cxr,BPS_94,Bouwknegt:1994sj}. At $\q=\I$, the difference for finite size is even more dramatic, presumably due to the global symmetry discussed below. 
In this Letter we focus on an \emph{odd} number of sites. 
We can give three conserved charges from (i) explicitly. One is a `quasi-translation', replacing the lattice translation since ordinary translation invariance is broken. The second charge is free fermionic, and parity odd (chiral). The third charge has quartic interactions and is parity even, thus naturally assuming the role of hamiltonian.
Like in \cite{Pasquier:1989kd}, for $\q\neq 1$ extended symmetry~(ii) is incompatible with periodicity. Instead, the hamiltonians are `quasi-periodic', i.e.\ they commute with the quasi-translation. Although the model is non-unitary, 
the spectrum is real by $\PP\TT$-invariance, cf.~\cite{bender2007making}. The reward for having complicated interactions is that this spectrum is extremely simple as in (iii): sums of quasiparticle energies with \emph{linear} dispersions,
comprising two branches associated with even and odd mode numbers. Linear dispersions also occur for the (antiperiodic) $1/r$ Hubbard model \cite{gebhard1992exact,*bares1995asymptotic,*gohmann1996yangian} and spin chains in $\mathrm{AdS}_3/\mathrm{CFT}_2$ integrability~\cite{dei2018integrable,*gaberdiel2023beyond,*Frolov:2023pjw}.
The interaction in the quartic hamiltonian implements a statistical selection rule that excludes successive occupied mode numbers,
originating in the parent model and matching the 
`motif' description of the HS chain \cite{Haldane:1991cxr}. This selection rule comes with high degeneracies caused by (ii). The extended symmetry, inherited from the parent model, contains $\mathfrak{gl}(1|1)$.
\smallskip

\textit{Significance.} 
The HS chain captures the salient properties of the Heisenberg \textsc{xxx} chain, notably its description in terms of spinons as quasiparticles with fractional statistics. This is particularly useful in the scaling limit in the anti\-ferromagnetic regime, which is captured by the $SU\mspace{-1mu}(2)_{k=1}$ Wess--Zumino--Witten CFT \cite{HH+_92,BPS_94,Bouwknegt:1994sj}.
The present model, combining exclusion statistics with a fermionic realisation, will play a similar role for non-unitary spin chains and their logarithmic CFT limits, which are notoriously hard to analyse. An explicit lattice regularisation, with a 
new kind of realisation for the symmetry algebra, is a great theoretical asset.
\medskip

\noindent
{\itshape\textbf{Methodology.}} Our results are obtained as follows \cite{Supp_Mat}.

Recall that the Heisenberg \textsc{xxx} chain has a hidden algebraic structure, the $\mathfrak{su}(2)$ Yangian, providing both
commuting charges and their diagonalisation via the algebraic Bethe ansatz. For \textsc{xxz} the Yangian is $\q$-deformed to quantum-affine $\mathfrak{su}(2)$ \cite{Faddeev:1996iy}. At $\q=\I$ the Jordan--Wigner transformation provides a more direct treatment by fermionic methods. If periodicity is replaced by special boundary conditions, the $\mathfrak{su}(2)$ spin symmetry persists in the guise of  $U_{\q}\mspace{2mu}\mathfrak{sl}(2)$, enabling techniques based on the Temperley--Lieb (TL) algebra, both for general $\q$ and at $\q=\I$ \cite{Pasquier:1989kd,Gainutdinov:2011ab}.

For the HS chain, the Yangian plays \emph{another} role: its generators are different, and commute with the hamiltonian \cite{HH+_92}, providing extended spin \emph{symmetry} rather than the full spectrum. Instead, an algebraic formalism\,---\,using so-called Dunkl operators and `freezing'\,---\,allows one to construct the  commuting charges, Yangian, and eigenvectors
\cite{Bernard:1993va,polychronakos1993lattice,polychronakos1994exact,talstra1995integrals}, cf.~\cite{lamers2022fermionic}.
This lifts to the \textsc{xxz}-type level~\cite{Lamers:2020ato}, providing the commuting charges, \emph{extended} (quantum-affine) spin symmetry, \emph{explicit} spectrum and eigenvectors of the parent model.

We leverage this knowledge of the parent model, reviewed in \cite{Supp_Mat}, supplemented by fermionic techniques. We extract the first few explicit commuting charges from the parent model by specialising to $\q=\I$. The TL algebra provides the bridge to Jordan--Wigner-like fermions. We exhibit discrete symmetries, and describe the extended symmetry and exact spectrum inherited from the parent model. Exploiting the quasi-translation operator, we define fermions that can be Fourier transformed to bring the hamiltonians to a simple form and describe part of the eigenstates in the corresponding Fock basis.
\medskip

\noindent{\itshape\textbf{The model.}}
Consider fermions hopping on a 1d lattice with an \emph{odd} number $N$ sites. The simplest formulation of our model uses \emph{non-unitary} fermionic operators with anticommutation relations \cite{Gainutdinov:2011ab}
\begin{equation} \label{fermtrans}
	\{f_i,f_j^+\}=(-1)^i\,\delta_{ij} \, , \quad 
	\{f_i,f_j\} = \{f^+_i,f^+_j\} = 0 \, .
\end{equation}
They are related to canonical Jordan--Wigner fermions as $f_j = (-\I)^{\mspace{2mu}j} \, c_j$, $f^+_j = (-\I)^{\mspace{2mu}j} \, c^\dag_j$. The $f$s will avoid a proliferation of factors of~$\I$ and make the symmetries more transparent. From the two-site fermionic operators
\begin{equation} \label{gfer}
	g_i \equiv f_i + f_{i+1} \, , \quad g_i^+ = f_i^+ + f_{i+1}^+ \, , \qquad 1\leqslant i < N \, ,
\end{equation}
we construct the quadratic combinations
\begin{equation} \label{TLdef}
	e_i \equiv g_i^+ g_i \, , \qquad 1\leqslant i < N \, ,
\end{equation}
which obey the free-fermion TL algebra relations
\begin{equation} \label{eq:TL_ff}
	 e_i^2=0 \, , \quad  e_i \, e_{i \pm 1} \, e_i = e_i \, , \quad [e_i,e_j]=0 \ \mathrm{if} \ |i-j|>1 \, .
\end{equation}
Further define the nested TL commutators \cite[\textsection\ref{sec_sm:fermionic_rep}]{Supp_Mat}
\begin{equation} \label{TLgen}
	\begin{aligned} 
	e_{[i,j]} & \equiv [[\cdots[e_i,e_{i+1}],\cdots],e_{j}] \\
	& = s_{ij} \, \bigl(g_{j}^+ \, g_i + (-1)^{i-j} g_i^+ \, g_{j} \bigr) \, , \qquad i\neq j \, ,
	\end{aligned}
\end{equation}
where $s_{ij}\equiv(-1)^{(i-j)(i+j-1)/2}$, and we set $e_{[i,i]} \equiv e_i$. Note that \eqref{TLgen} is bilinear in the fermions~\eqref{fermtrans}. Finally set
\begin{equation} \label{def t_k t_kl}
	t_k \equiv \tan \! \tfrac{\pi k}{N}\, , \quad t_{k,l} \equiv \textstyle \prod_{i=k}^{l-1} t_{i} \ \, (k<l)\, , \quad t_{k,k} \equiv 1 \, . \!
\end{equation}
Then the chiral hamiltonian reads
\begin{equation} \label{Ham_L}
	\begin{aligned} 
	\HH^\LL = {} & \frac{\I}{2}
	\sum_{1 \leqslant i \leqslant j < N}\!\!\!\!\! h^\LL_{ij} \; e_{[i,j]} \, , \\
	& h^\LL_{ij} \equiv \! \sum_{n=j+1}^{N} \!\! t_{n-j,n-i} \, \bigl( 1-(-1)^i \, t^2_{n-i,n} \bigr) \, .
	\end{aligned}
\end{equation}
We do not imply any periodicity of the sites; \eqref{Ham_L} is not translation invariant. Instead, it commutes with the \emph{quasi-translation} 
operator 
\begin{equation} \label{Gop}
	\G = (1+t_{N-1} \,e_{N-1}) \cdots (1+t_{1} \,e_{1}) \, , \quad \G^N=1 \, ,
\end{equation}
which takes over the role of the usual lattice translation.

The next charge is a linear combination of anti\-commutators of the nested commutators \eqref{TLgen},
\begin{gather} \label{Htildedef}
	\HH = -\frac{1}{4N} \! \sum_{i\leqslant j<k\leqslant l}^{N-1} \! \bigl(h^\LL_{ij;kl} + h^\RR_{ij;kl} \bigr) \, \bigl\{ e_{[i,j]},e_{[k,l]} \bigr\} \, , \\
	h^\LL_{ij;kl} \equiv (-1)^{k-j} \! \sum_{n(>l)}^{N} \! t_{n-l,n-j} \, t_{n-k,n-i} \, \bigl(1-(-1)^i \, t^2_{n-i,n} \bigr) \, , \nonumber \\
	h^\RR_{ij;kl} \equiv (-1)^ {l-j+k-i} \, h^\LL_{N-l,N-k;N-j,N-i} \, . \nonumber
\end{gather}
Stemming from the parent model, these quantities, and higher charges that we do not give here, commute \cite{Supp_Mat},
\begin{equation} \label{GHop}
	\bigl[ \G,\HH^\LL \bigr] = \bigl[ \G,\HH \bigr] = \bigl[\HH^\LL,\HH \bigr] = 0 \, .
\end{equation}

\noindent{\itshape\textbf{Symmetries.}} The commuting charges \eqref{Ham_L}--\eqref{Htildedef} have several transformation properties and symmetries.
\smallskip

\textit{Parity.}  Parity reverses the lattice sites, $\PP(f_i) = f_{N+1-i}$. This preserves \eqref{fermtrans} since $N$ is odd. The TL generators transform as $\PP(e_i)=e_{N-i}$. We have
\begin{equation} \label{parity}
	\PP(\HH^\LL) = -\HH^\LL\, , \quad  \PP(\HH) = \HH \, , \quad \PP(\G) = \G^{-1} \, .
\end{equation}
The first of these relations is highly nontrivial \cite{Supp_Mat}.
\pagebreak[3]

\textit{Time reversal.} We define time reversal as complex conjugation of coefficients with respect to the Fock basis 
$f_{i_1}^+ \cdots f_{i_M}^+ \ket{\varnothing}$. 
By using the $f$s, $\TT(e_i)=e_i$ and \eqref{Ham_L}--\eqref{Htildedef} have real matrix elements, except for the prefactor in \eqref{Ham_L}.
Thus 
\begin{equation} \label{charges under T}
	\TT(\HH^\LL) = -\HH^\LL \, , \quad \TT(\HH) = \HH \, , \quad \TT(\G) = \G\, .
\end{equation}
Since the hamiltonians (and their eigenstates) are $\PP\TT$-invariant, their spectrum is real \cite{Korff_2007,Korff_2008,korff2007ptinvariance,morin2016reality}. The same is true for the `quasi-momentum' $\pp = -\I\log \G$.
\smallskip

\textit{Charge conjugation.} 
The particle-hole transformation
\begin{equation} \label{particle hole}
	\CC'(f_i) = f_i^+ \,, \quad 
	\CC'(f^+_i) = f_i \, ,
\end{equation}
gives $\CC'(e_i)= - e_i$, preserving~\eqref{eq:TL_ff}.
Including a suitable antilinear transformation $\UU$, see \cite{Supp_Mat}, gives charge conjugation $\CC = \CC' \, \UU$. It acts on the conserved charges by
\begin{equation} \label{charges under charge transf}
	\CC(\HH^\LL) = -\HH^\LL \, , \quad \CC(\HH) = \HH \, , \quad \CC(\G) = \G \, .
\end{equation} 

\textit{Global symmetry.} The model can be seen as a long-range spin chain with alternating $\mathfrak{gl}(1|1)$-representations. Recall that $\mathfrak{gl}(1|1)$ has two bosonic and two fermionic generators, which we denote by $\N,\EE$ and $\F_1,\F_1^+$ respectively. The nontrivial (anti)commutation relations are
\begin{equation} \label{gl11comm}
	\!\!
	\bigl[ \N ,\F_1 \bigr] = -\F_1 \,, \quad
	\bigl[ \N, \F_1^+\bigr] = \F_1^+ \,, \quad 
	\bigl\{ \F_1 , \F_1^+ \bigr\} = \EE \,, 
\end{equation}
and $\EE$ is central. This is just a fermionic version of the usual spin algebra. Each site~$i$ carries a $\mathfrak{gl}(1|1)$-representation generated by $f_i$, $f^+_i$, the number operator $(-1)^i \, f_i^+ f_i$ and central charge $(-1)^i$. From this perspective, our model is a long-range $\mathfrak{gl}(1|1)$ super-spin chain. 
It has a global $\mathfrak{gl}(1|1)$-symmetry generated by
\begin{equation} \label{global_gl11}
	\begin{gathered}
	\F_1 = \sum_{i=1}^N f_i \, , \quad \F_1^+ = \sum_{i=1}^N f^+_i \,, \\
	\N = \sum_{i=1}^N \, (-1)^i \, f^+_i \mspace{1mu} f_i \, , \quad \EE =
	\sum_{i=1}^N \, (-1)^i = -1 \, ,
	\end{gathered}
\end{equation}
Indeed, these operators commute with all $e_i$, and thus with the conserved charges \eqref{Ham_L}--\eqref{Htildedef}.

Since $\F_1^2 = (\F_1^+)^2 = 0$, $\mathfrak{gl}(1|1)$ produces fewer descendants than $\mathfrak{su}(2)$ does at $\q=1$. This is compensated by additional bosonic generators
\begin{equation} \label{global_additional}
	\F_2 = \sum_{i<j}^N f_i \, f_j \, , \quad
	\F_2^+ = \sum_{i<j}^N f_i^+ \mspace{1mu} f_j^+ \, ,
\end{equation}
which commute with the $e_i$, whence with \eqref{Ham_L}--\eqref{Htildedef}. Together, 
\eqref{global_gl11}--\eqref{global_additional} generate the full global-symmetry algebra \cite{Read_07,Note1}.
\footnotetext[1]{To be precise, $\mathfrak{gl}(1|1)$ is the `little quantum group' and $\mathcal{A}_{1|1}$ the `Lusztig quantum group' of $U_{\q}\mspace{2mu}\mathfrak{sl}(2)$ at $\q=\I$.}
It is the $U_{\q}\mspace{2mu}\mathfrak{sl}(2)|_{\q=\I}$ symmetry from the parent model in fermionic language, cf.~\cite[\textsection2.3]{Gainutdinov:2011ab}.
\smallskip

\textit{Extended symmetry.} The parent model has quantum-affine $\mathfrak{sl}(2)$ symmetry, which underpins its large degeneracies.
Its specialisation to $\q=\I$ is tricky and seems absent in the mathematical literature. 
A detailed study will be performed elsewhere. 
The extended symmetry is visible in the highly degenerate spectrum. 
\medskip

\begin{figure}[h]
	\centering
	\begin{tikzpicture}[xscale=1.3,yscale=.7]
		\draw[->] (0,-1.1) -- (0,1.2) node[above]{$\varepsilon^\LL$};
		\draw (0,-1) -- (-.1/1.3,-1) node[left] {$-N$};
		\draw (0,0) -- (-.1/1.3,0) node[left] {$0$};
		\draw (0,1) -- (-.1/1.3,1) node[left] {$N$};
		\draw[->] (0,0) -- (2+.2/1.3,0) node[right] {$n$};
		\draw (0,0) -- (0,-.1/.7); 
		\draw (2,0) -- (2,-.1/.7) node[below] {$N$};
		\foreach \x in {1,...,16} {\draw[very thin] (\x*2/17,0) -- (\x*2/17,-.05/.7);};
		\draw[very thin,gray!40] (-.15,-.075) -- (0,0);
		\draw[very thin,gray!40] (0,0) -- (2,1);
		\draw[very thin,gray!40] (0,-1) -- (2,0);
		\draw[very thin,gray!40] (2,0) -- (2.15,0.075);
		\foreach \n in {0,2,...,16} \fill[black] (\n*2/17,\n/17) ellipse (1.25/1.3 pt and 1.25/.7 pt);
		\foreach \n in {1,3,...,16} \fill[black] (\n*2/17,\n/17-1) ellipse (1.25/1.3 pt and 1.25/.7 pt);
		\draw (2,0) ellipse (1.25/1.3 pt and 1.25/.7 pt);
	\end{tikzpicture}
	\!\!\!\!
	\begin{tikzpicture}[xscale=1.3,yscale=.7]
		\draw[->] (0,0) -- (0,1.2) node[above]{$\varepsilon$};
		\node at (-.1/1.3,-1) [left] {\phantom{$-N$}};
		\draw (0,0) -- (-.1/1.3,0) node[left] {$0$};
		\draw (0,1) -- (-.1/1.3,1) node[left] {$N$};
		\draw[->] (0,0) -- (2+.2/1.3,0) node[right] {$n$};
		\draw (0,0) -- (0,-.1) node[below] {$0$};
		\draw (2,0) -- (2,-.1) node[below] {$N$};
		\foreach \x in {1,...,16} {\draw[very thin] (\x*2/17,0) -- (\x*2/17,-.05/.7);};
		\draw[very thin,gray!40] (0,0) -- (2,1);
		\draw[very thin,gray!40] (0,1) -- (2,0);
		\foreach \n in {0,2,...,16} \fill[black] (\n*2/17,\n/17) ellipse (1.25/1.3 pt and 1.25/.7 pt);
		\foreach \n in {1,3,...,16} \fill[black] (\n*2/17,1-\n/17) ellipse (1.25/1.3 pt and 1.25/.7 pt);
		\draw (2,0) ellipse (1.25 / 1.3 pt and 1.25/.7 pt);
	\end{tikzpicture}
	\caption{The dispersion relations~\eqref{eLodd} alternate between two linear branches, realising chiral and `full' (up to a shift) massless relativistic dispersions on the lattice.}
	\label{fg:dispersions}
\end{figure}
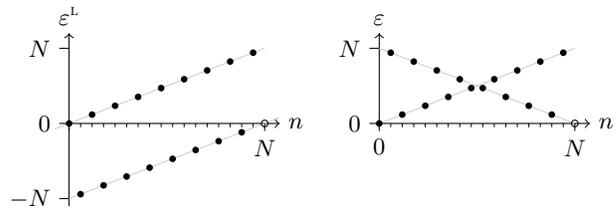

\noindent{\itshape\textbf{Exact spectrum.}} The spectrum and degeneracies of the parent model are known explicitly \cite{Bernard:1993va, Uglov:1995di,*Lamers:2018ypi, Lamers:2020ato}. Like for the HS chain, the quantum numbers are `motifs' \cite{Haldane:1991cxr} $\{\mu_m\}$, consisting
of integers $1\leqslant \mu_m < N$ increasing as
\begin{equation} \label{repulsion}
	\mu_{m+1} > \mu_m + 1 \, , \qquad 1\leqslant m <M \, .
\end{equation}
Such a motif labels an $M$-fermion state with quasi-momentum $p = \frac{2\pi}{N} \sum_m \mu_m \, \mathrm{mod} \, 2\pi$ setting the eigenvalue $\E^{\I \mspace{2mu} p}$ of $\G$. Its energy is additive:
\begin{equation} \label{E^L_and_E}
	E^\LL_{\{\mu_m\}} = \sum_{m=1}^M \! \varepsilon^\LL_{\mu_m} \, , \qquad
	E_{\{\mu_m\}} = \sum_{m=1}^M \! \varepsilon_{\mu_m} \, ,
\end{equation}
with dispersions having two linear branches (Fig.~\ref{fg:dispersions}):
\begin{equation} \label{eLodd}
	\varepsilon^{\LL}_n =
	\begin{cases*} 
		n\,, & $n$ even\,,\\ 
		n-N\,, \ & $n$ odd\,,
	\end{cases*} 
	\qquad \ \varepsilon_n = |\varepsilon_n^{\LL}| \, .
\end{equation}
This state has (often many) descendants due to the extended symmetry. Its multiplicity is \cite{HH+_92,Lamers:2020ato} $N+1$ for the empty motif (at $M=0$), and otherwise
\begin{equation} \label{degeneracy}
	\mu_1 \, (N-\mu_M) \! \prod_{m=1}^{M-1} \! (\mu_{m+1} - \mu_m - 1) \, .
\end{equation}
Given the extremely simple dispersion, further (`accidental') degeneracies between different motifs occur much more often than even for the HS chain.
\medskip

\noindent{\itshape\textbf{Fermionic approach.}}
Lacking periodicity, we cannot simply Fourier transform the $f$s as usual.
The key to defining a good basis of fermions is to start at one end of the lattice and use the quasi-translation operator:
\begin{equation} \label{eq:Phi_def}
	\Phi_i \equiv \G^{1-i} \, f_1 \, \G^{i-1} \, , \qquad 
	\Phi_i^+ \equiv \G^{1-i} \, f_1^+ \, \G^{i-1} \, .
\end{equation}
These dressed fermions, reminiscent of \cite{fendley2019free}, \emph{are} periodic,
\begin{equation} \label{eq:Phi_periodic}
	\Phi_{i+N} = \Phi_i \, , \qquad  \Phi^+_{i+N}=\Phi^+_i \, ,
\end{equation}
and obey non-local anticommutation relations \cite{Supp_Mat}
\begin{equation} \label{psicomm}
	\bigl\{ \Phi_i , \Phi_j^+ \bigr\} = -(1+t_{j-i}) \,, \quad
	\{\Phi_i,\Phi_j\} = \{\Phi^+_i,\Phi^+_j\} = 0 \, .
\end{equation}
The nontrivial relation only depends on the distance. 
\pagebreak[2]
Due to \eqref{eq:Phi_periodic} we may Fourier transform \eqref{eq:Phi_def} as usual.
Set $a_0 \equiv \I$ and $a_n \equiv \I^{n+1/2}$ else. The \emph{rescaled} Fourier modes
\begin{equation} \label{Psi_tilde} 
	\tilde{\Psi}_n \equiv \frac{a_n}{N} \sum_{\smash{j=1}}^{N}  \E^{-2\I\pi\mspace{1mu}n\mspace{1mu}j/N} \, \Phi_{j} \, , \ \ 
	\tilde{\Psi}_n^+ \equiv \frac{a_n}{N} \sum_{\smash{j=1}}^{N}  \E^{2\I\pi\mspace{1mu}n\mspace{1mu}j/N} \, \Phi_{j}^+ ,
\end{equation}
obey canonical anticommutation relations
\begin{equation} \label{psitildecomm}
	\!\! \bigl\{ \tilde{\Psi}_n , \tilde{\Psi}_m^+ \bigr\} = \delta_{nm} \, , \quad 
	\{\tilde{\Psi}_n,\tilde{\Psi}_m \} = \{\tilde{\Psi}_n^+,\tilde{\Psi}_m^+\} = 0 \, .
\end{equation}
They are covariant under quasi-translations in the sense
\begin{equation} \label{psitildeG}
	\G \, \tilde{\Psi}_n \, \G^{-1} = \E^{-2\I\pi\mspace{1mu}n/N} \, \tilde{\Psi}_n \, , \quad 
	\G \, \tilde{\Psi}_n^+ \, \G^{-1} = \E^{2\I\pi\mspace{1mu}n/N}  \, \tilde{\Psi}_n^+ \, .
\end{equation}
The relation to the original fermions is strikingly simple.
The zero-modes commute with the hamiltonians: they are just the fermionic $\mathfrak{gl}(1|1)$ generators from \eqref{global_gl11} \cite{Supp_Mat},
\begin{equation} \label{zero_modes}
	\frac{1}{a_0} \, \tilde{\Psi}_0 = \sum_{i=1}^N f_i = \F_1 \,, \quad 
	\frac{1}{a_0} \, \tilde{\Psi}^+_0 = \sum_{i=1}^N f^+_i = \F_1^+ \, .
\end{equation}
The other modes are explicit \emph{linear} combinations of the two-site fermions~\eqref{gfer}, with coefficients given in \cite[\textsection{}C.2]{Supp_Mat}. In these terms, $\HH^\LL$ is diagonal:
\begin{equation} \label{H^L_ff}
	\HH^\LL = \sum_{n=1}^{N-1} \varepsilon^\LL_n \, \tilde{\Psi}_n^+ \, \tilde{\Psi}_n \, . 
\end{equation}
Numerics for low~$N$ confirms the equality with \eqref{Ham_L}.
If $\ket{\varnothing}$ is the fermionic vacuum, then by \eqref{psitildecomm} the Fock states
\begin{equation} \label{freespectrum}
 	\ket{n_1,\dots,n_M} \equiv \tilde{\Psi}^+_{n_1} \dots \tilde{\Psi}^+_{n_M}\, \ket{\varnothing} \,,
\end{equation}
form an $\HH^\LL$-eigenbasis labelled by all $2^N$ fermionic mode numbers $0 \leqslant n_1 < \dots < n_M <N$. The quasi-momentum is $p = \frac{2\pi}{N} \sum_m n_m \, \mathrm{mod} \, 2\pi$, and the chiral energy $E^\LL_{\{n_m\}} = \sum_m \varepsilon^\LL_{n_m}$ matches \eqref{E^L_and_E}--\eqref{eLodd} when $\{n_m\}$ is a motif. 

Next, \eqref{Htildedef} takes the quartic form \cite{Supp_Mat}
\begin{equation} \label{ham}
	\HH = \sum_{n=1}^{N-1} \varepsilon_n \, \tilde{\Psi}_n^+ \, \tilde{\Psi}_n + \ \, \sum_{\mathclap{\substack{1\leqslant m<n<N \\ 1\leqslant r<s<N}}}
	\tilde{V}_{mn;rs} \, \tilde{\Psi}_m^+ \, \tilde{\Psi}_n^+ \, \tilde{\Psi}_r \, \tilde{\Psi}_s \, .
\end{equation}
The commutation~\eqref{GHop} only allows $\tilde{V}_{mn;rs} \neq 0$ if \cite{Supp_Mat} the quasi-momentum and chiral energy are conserved:
\begin{equation} \label{eq:mtm chiral energy conserved}
	m+n = r +s \ \, \mathrm{mod} \ N\,, \qquad 
 	\varepsilon^\LL_m + \varepsilon^\LL_n = \varepsilon^\LL_r + \varepsilon^\LL_s \, .
\end{equation}
Numerics for odd $N \leqslant 9$ suggest the stronger selection rule that $m+n=r+s$ be odd, with nonzero values $\tilde{V} = \pm4$ determined by $\tilde{V}_{mn;rs} = \tilde{V}_{rs;mn}$ and
\begin{equation} \label{pot}
 	\tilde{V}_{mn;m+k,n-k} = (-1)^{k+1} 4 \, \delta_{m \,\mathrm{odd}} \, , \ \ 0 \leqslant 2k < n-m \, .
\end{equation}
For one-particle states~$\ket{n}$ only the quadratic part of \eqref{ham} contributes, reproducing the non-chiral dispersion~\eqref{eLodd}. The quartic part implements the statistical repulsion rule~\eqref{repulsion}: $\HH$ is genuinely interacting. Thus, \eqref{freespectrum} are generally \emph{not} eigenstates of $\HH$. 
In \cite[\textsection{}D]{Supp_Mat} we illustrate this for the two-fermion spectrum, showing how $\tilde{V}$ `squeezes' adjacent modes to extended-symmetry descendants.
E.g., for $N=5$ we know from the parent model that $\ket{1,2}$ is an extended-symmetry descendant of the motif $\{3\}$. This matches $\varepsilon^\LL_n + \varepsilon^\LL_{n+1} = \varepsilon^\LL_{2\mspace{1mu}n+1\,\mathrm{mod}\,N}$ (cf.~Fig.~\ref{fg:dispersions}), yet for $\HH$ relies on the contribution of $\tilde{V}_{12;12} = -4$ to $\varepsilon_1 + \varepsilon_2 + \tilde{V}_{12;12} = \varepsilon_3$.
A fermionic description of the full spectrum requires a deeper understanding of the extended symmetry.
\medskip

\noindent{\itshape\textbf{Outlook.}}
We obtained and analysed a long-range fermi\-onic model with extended symmetry from the \textsc{xxz}-type HS chain (`parent model') at $\q=\I$. It can be also viewed as a long-range alternating super-spin chain.
Periodicity, being incompatible with the symmetry, is modified to `quasi-translation' invariance. As for Heisenberg \textsc{xxz}, we get a free-fermion hamiltonian, $\HH^\LL$. Albeit quadratic in fermions from the start, its explicit diagonalisation relies on quasi-translated fermions akin to \cite{fendley2019free}. 
Another conserved charge, $\HH$, has a four-fermion interaction realising the statistical repulsion known from the isotropic HS chain. Both hamiltonians have a Weyl-like spectrum on the lattice. Together, these features render the model rather simple, although a
full understanding requires an explicit fermionic realisation of the extended symmetry. This, and a systematic construction of the eigenvectors, which are known for the parent model, is left for future work. 

In spin chains, even/odd length may lead to different properties due to the excitations' topological nature. The distinction is more pronounced for alternating $\mathfrak{gl}(1|1)$ spin chains, cf.~\cite{Gainutdinov:2011ab}. Presently, it becomes extreme due to a pole in the parent model's long-range potential. For even $N$, the parent hamiltonian \emph{diverges} as $\q\to\I$, and regularisation sets all energies to \emph{zero}. Yet, the wave functions remain non-trivial, and numerics indicates Jordan blocks up to size $N\!/2 \, + 1$. While indecomposable representations are expected at central charge $\EE = 0$, their dimensions signal that these are not just zigzag modules for systems with merely global symmetry \cite{Schomerus_2006}. We will soon report on 
even $N$ \cite{BMLS_24u}. A spin chain with a similar spectrum arose in AdS/CFT integrability \cite{ahn2021integrable}.

Another important direction is the continuum limit, where we expect conformal invariance. Explicit identification of the CFT requires determining the extended symmetry. The isotropic HS chain suggests Kac--Moody symmetry in the CFT limit, perhaps level-1 $\mathfrak{gl}(1|1)$ \cite{creutzig2013w}. It will be interesting to find the continuum counterpart of $\HH^\LL$, reminding the Virasoro generator $L_0$, and see what happens with the dispersion relation's staggering. The relativistic-like dispersions for odd $N$ seems well adapted to the continuum limit, albeit at odds with the vanishing spectrum for even $N$. In the scaling limit, the $f$s will yield symplectic fermions, but the even/odd-length difference we see goes beyond that of the Ramond/Neveu--Schwarz sectors, requiring further insight.

For Heisenberg \textsc{xxz}, other root-of-unity cases, notably $\q^3 = 1$ \cite{Razumov_2001}, are special too. Their analogues for the parent model will also exhibit extended symmetry, quasi-translation invariance, non-unitarity and statistical level repulsion. More generally, it would be exciting to study root-of-unity points of the \textsc{xxz}-type Inozemtsev chain \cite{klabbers2023deformed}, which interpolates between a quasiperiodic Heisenberg \textsc{xxz} chain and our parent model.
\medskip

\noindent{\itshape\textbf{Acknowledgements.}}
We thank P.~Di Francesco, F.~Essler, P.~Fendley, H.~Frahm, A.~Gainutdinov, F.~G\"ohmann, J.~Jacobsen, V.~Pasquier, H.~Saleur, V.~Schomerus, R.~Weston and J.-B.~Zuber for inspiring discussions. We further thank the referees for useful feedback. JL was funded by LabEx Math\'ematique Hadamard (LMH), and in the final stage by ERC-2021-CoG\,--\,BrokenSymmetries 101044226. ABM and DS thank \textsc{cern}, where part of this work was done, for hospitality.

\let\oldaddcontentsline\addcontentsline
\renewcommand{\addcontentsline}[3]{}
\bibliography{MyBib}
\let\addcontentsline\oldaddcontentsline

\onecolumngrid
\clearpage

\let\oldaddcontentsline\addcontentsline
\renewcommand{\addcontentsline}[3]{}
\section*{Supplemental Material}
\setcounter{subsection}{0}
\let\addcontentsline\oldaddcontentsline

\numberwithin{equation}{subsection}

\renewcommand{\thefootnote}{\fnsymbol{footnote}}

\tableofcontents

\subsection{The parent model}

\noindent
The model introduced in this Letter is the $\q=\I$ limit of
an integrable long-range version of the Heisenberg \textsc{xxz} spin chain: the \textsc{xxz}-type Haldane--Shastry (HS) chain with spin 1/2. This `parent model' was proposed in \cite{Bernard:1993va} and explicitly described and studied further in \cite{Uglov:1995di,*Lamers:2018ypi,Lamers:2020ato}. Here we review the basics. 

\subsubsection{The Haldane--Shastry chain}
\noindent To set up our notation we start with the isotropic case. For $N$ spin-1/2 sites, the HS chain \cite{Hal_88,*Sha_88} has pairwise spin exchange interactions with an inverse square potential,
\begin{equation} \label{eq:HS}
	\HH^\textsc{hs} = \sum_{i<j}^N \V\mspace{-1mu}(i-j) \, (1- \PP_{\! ij}) \, , \qquad \V\mspace{-1mu}(k) = \frac{1}{4\mspace{1mu} \sin(\pi k/N)^2} \, .
\end{equation}
This spin chain has many remarkable properties. For us the most important ones are that it possesses
\begin{itemize}[topsep=1ex,itemsep=-.5ex]
	\item[i)] a family of conserved charges: a hierarchy of commuting operators including \eqref{eq:HS} 
	\cite{Ino_90, HH+_92, talstra1995integrals,ferrando2023bethe}, 
	\item[ii)]  extended (Yangian) $\mathfrak{su}(2)$ spin symmetry \cite{HH+_92,Bernard:1993va},
	\item[iii)]  a very simple and highly degenerate spectrum.
\end{itemize}
Let us elaborate on the last point. The eigenspaces are labelled by `motifs' $\{\mu_m\}_m$, which are
\begin{equation} \label{rrepulsion}
	\text{integers} \quad 1\leqslant \mu_1 < \dots < \mu_M < N \quad \text{obeying the statistical repulsion rule} \quad \mu_{m+1} > \mu_m + 1 \, .
\end{equation}
The momentum is 
\begin{equation} \label{momentum} 
	p = \frac{2\pi}{N} \sum_m \mu_m \ \mathrm{mod}\,2\pi \, , 
\end{equation}
which means that the lattice translation 
\begin{equation} \label{eq:transl}
	\G^\textsc{hs} = \PP_{\mspace{-1mu}N-1,N} \cdots \PP_{\mspace{-1mu}12} \, , 
\end{equation}
which is one of the conserved charges, has eigenvalue $\E^{\I \mspace{2mu} p}$. The energy is (strictly) additive with a quadratic dispersion: 
\begin{equation} \label{HS_disp}
	E^\textsc{hs}_{\{\mu_m\}} = \sum_{m=1}^M \! \varepsilon^\textsc{hs}_{\mu_m} \, , \qquad 
	\varepsilon^\textsc{hs}_n = \frac{1}{2} \, n \, (N-n) \, .
\end{equation}
Therefore all energies are half integers, and the motifs control which energy levels occur. 
For example, the empty motif corresponds to the ferromagnetic vacuum~$\ket{\uparrow\cdots\uparrow}$ with zero momentum and energy, while if $N$ is even the fully packed motif $\{1,3,\dots,N-3,N-1\}$ corresponds to the antiferromagnetic vacuum with maximal energy.
In general, the motif~\eqref{rrepulsion} has a (Yangian) highest-weight state at $S^z = N/2 - M$, with an explicit wave function containing a Jack polynomial~\cite{Haldane:1991cxr,Bernard:1993va}.
It typically has many (ordinary as well as `affine') descendants thanks to the extended spin symmetry. The degeneracy equals \cite{HH+_92,Bernard:1993va}
\begin{equation} \label{sm:degeneracy}
	N+1 \quad \text{for the empty motif} \, , \qquad 
	\mu_1 \, (N-\mu_M) \, \prod_{m=1}^{M-1} (\mu_{m+1} - \mu_m - 1) \quad \text{else} \, ,
\end{equation}
which is \eqref{degeneracy} from the main text. Additional `accidental' degeneracies between different motifs occur.

\subsubsection{The \textsc{xxz}-type Haldane--Shastry chain}

\noindent 
The HS chain has an \textsc{xxz}-type generalisation where the $\mathfrak{su}(2)$ spin symmetry is broken to $\mathfrak{u}(1)$, or more precisely: deformed. It still possesses
\begin{itemize}[topsep=1ex,itemsep=-.5ex]
	\item[i)] a family of conserved charges 	\cite{Bernard:1993va,Uglov:1995di,*Lamers:2018ypi,Lamers:2020ato}, 
	\item[ii)] extended spin symmetry, now deformed to quantum-affine $\mathfrak{sl}(2)$ \cite{Bernard:1993va},
	\item[iii)] a very simple and highly degenerate spectrum \cite{Uglov:1995di,*Lamers:2018ypi,Lamers:2020ato}.
\end{itemize}
The price to pay for keeping (i)--(ii) in the presence of anisotropy is that the long-range spin exchange $1-\mathrm{P}_{\!ij}$ from \eqref{eq:HS} becomes more complicated. This goes as follows. 

We start from the two-site operators
\begin{equation} \label{eq:TL_spin}
	e_i(\q) \equiv \id^{\otimes(i-1)} \otimes \, 
	\begin{pmatrix} 
		\, 0 & \color{gray!80}{0} & \color{gray!80}{0} & \color{gray!80}{0} \, \\
		\, \color{gray!80}{0} & \q^{\mspace{-1mu}-1} & -1\hphantom{-} & \color{gray!80}{0} \, \\
		\, \color{gray!80}{0} & -1\hphantom{-} & \q^{\mspace{-1mu}-1} & \color{gray!80}{0} \, \\
		\, \color{gray!80}{0} & \color{gray!80}{0} & \! \color{gray!80}{0} & 0 \, \\
	\end{pmatrix}
	\otimes \id^{\otimes(N-i-1)} \, , \qquad 1\leqslant i<N \, ,
\end{equation}
where $\q \in \mathbb{C}$ is the anisotropy (deformation) parameter. These local operators obey the commutation relations
\begin{equation} \label{TLrel}
	\begin{aligned}
		e_i(\q)^2 & = \bigl(\q + \q^{\mspace{-1mu}-1}\bigr) \, e_i(\q) \, , \\
		e_i(\q) \, e_j(\q) \, e_i(\q) & = e_i(\q) \, , \qquad && \text{if} \ |i-j|=1 \, , \\ 
		\bigl[ e_i(\q), e_j(\q) \bigr] & = 0 \, ,\qquad && \text{if} \ |i-j|>1 \, .
	\end{aligned}
\end{equation}
That is, \eqref{eq:TL_spin} furnish a representation of the Temperley--Lieb (TL) algebra with `loop fugacity' $\q + \q^{\mspace{-1mu}-1}$ on the spin chain. This TL representation also appears for the Heisenberg \textsc{xxz} chain with $\Delta = (\q + \q^{\mspace{-1mu}-1})/2$, cf.\ e.g.\ \cite{Pasquier:1989kd}. Since $e_i(1) = 1 - \mathrm{P}_{\!i,i+1}$ the TL generators~\eqref{eq:TL_spin} are deformed nearest-neighbour spin antisymmetrisers.

Next we define the trigonometric (\textsc{xxz}) \textit{R}-matrix
\begin{equation} \label{Rcheck}
	\R_{i,i+1}(u;\q) = \mathrm{P}_{\!i,i+1} \, \mathrm{R}_{i,i+1}(u;\q) \equiv 1- f(u;\q) \, e_i(\q) \, , \qquad f(u;\q) \equiv \frac{u-1}{\q \,u-\q^{-1}}\, .
\end{equation}
It obeys the Yang--Baxter equation in the form $\R_{12}(u/v;\q) \, \R_{23}(u;\q) \, \R_{12}(v;\q) = \R_{23}(v;\q) \, \R_{12}(u;\q) \, \R_{23}(u/v;\q)$, as well as the unitarity relation $\R(u;\q) \, \R(1/u;\q) = \mathbbm{1} \otimes \mathbbm{1}$, and `initial' condition $\R(0;\q) = \mathbbm{1} \otimes \mathbbm{1}$. These relations, together with $\R(u;1) = \mathrm{P}$, means that we can think of \eqref{Rcheck} as a fancy version of the spin permutation operator.  

The \textsc{xxz}-type long-range spin exchange operator turns out to come in two `chiral' versions:
\begin{equation} \label{SqHSL}
	\begin{aligned} 
		\Ss^\LL_{[i,j]}(\q) \equiv
		\, \Biggl(\mspace{1mu} \ordprod_{j>k>i} \!\!\! \R_{k,k+1}\bigl({\omega^{j-k}};\q\bigr) \Biggr) 
		\ e_i(\q) \
		\Biggl(\mspace{1mu} \ordprodopp_{i<k<j} \!\!\! \R_{k,k+1}\bigl({\omega^{k-j}};\q\bigr) \Biggr)
		\, & = \!
		\tikz[baseline={([yshift=-.5*11pt*0.13+5pt]current bounding box.center)},xscale=0.4,yscale=0.2,font=\footnotesize]{
			\draw[->] (10.5,0) node[below]{$\ \omega^N$} -- (10.5,10); 
			\draw[->] (9,0) -- (9,10);
			\draw[rounded corners=2pt,->] (8,0) node[below]{$\,\omega^j$} -- (8,1.5) -- (5,4.5) -- (5,5.5) -- (8,8.5) -- (8,10); 
			\draw[rounded corners=2pt,->] (7,0) -- (7,1.5) -- (8,2.5) -- (8,7.5) -- (7,8.5) -- (7,10); 
			\draw[rounded corners=2pt,->] (6,0) -- (6,2.5) -- (7,3.5) -- (7,6.5) -- (6,7.5) -- (6,10);
			\draw[rounded corners=2pt,->] (5,0) -- (5,3.5) -- (6,4.5) -- (6,5.5) -- (5,6.5) -- (5,10); 
			\draw[->] (4,0) node[below]{$\,\omega^i$} -- (4,10);
			\draw[->] (3,0) -- (3,10);
			\draw[->] (1.5,0) node[below]{$\,\omega^1$} -- (1.5,10);
			\draw[style={decorate, decoration={zigzag,amplitude=.5mm,segment length=1mm}}] (4,5) -- (5,5);
			\foreach \x in {-1,...,1} \draw (2.25+.2*\x,5) node{$\cdot\mathstrut$};
			\foreach \x in {-1,...,1} \draw (9.75+.2*\x,5) node{$\cdot\mathstrut$};	
		}
		, \\[.5ex]
		\Ss^\RR_{[i,j]}(\q) \equiv
		\Biggl(\mspace{1mu} \ordprodopp_{i<k<j} \!\!\! \R_{k-1,k}\bigl(\omega^{k-i};\q\bigr) \Biggr) 
		\, e_{j-1}(\q) \,
		\Biggl(\mspace{1mu} \ordprod_{j>k>i}  \!\!\! \R_{k-1,k}\bigl(\omega^{i-k};\q\bigr) \Biggr)
		& = \!
		\tikz[baseline={([yshift=-.5*11pt*0.13+5pt]current bounding box.center)},xscale=-0.4,yscale=0.2,font=\footnotesize]{
			\draw[->] (10.5,0) node[below]{$\,\omega^1$} -- (10.5,10);
			\draw[->] (9,0) -- (9,10);
			\draw[rounded corners=2pt,->] (8,0) node[below]{$\,\omega^i$} -- (8,1.5) -- (5,4.5) -- (5,5.5) -- (8,8.5) -- (8,10);
			\draw[rounded corners=2pt,->] (7,0) -- (7,1.5) -- (8,2.5) -- (8,7.5) -- (7,8.5) -- (7,10);
			\draw[rounded corners=2pt,->] (6,0) -- (6,2.5) -- (7,3.5) -- (7,6.5) -- (6,7.5) -- (6,10);
			\draw[rounded corners=2pt,->] (5,0) -- (5,3.5) -- (6,4.5) -- (6,5.5) -- (5,6.5) -- (5,10);
			\draw[->] (4,0) node[below]{$\,\omega^j$} -- (4,10);
			\draw[->] (3,0) -- (3,10);
			\draw[->] (1.5,0) node[below]{$\ \omega^N$} -- (1.5,10);
			\draw[style={decorate, decoration={zigzag,amplitude=.5mm,segment length=1mm}}] (4,5) -- (5,5);
			\foreach \x in {-1,...,1} \draw (2.25+.2*\x,5) node{$\cdot\mathstrut$};
			\foreach \x in {-1,...,1} \draw (9.75+.2*\x,5) node{$\cdot\mathstrut$};	
		} ,
	\end{aligned}
	\qquad\ \ i<j \, .
\end{equation}
Here the arrows on the products indicate the direction in which the subscripts of the factors, which do not commute, increase. The diagrams show how we can think of these interactions: one of the two interacting spins is transported to the site next to the other interacting spin, where it interacts with is neighbour, after which it is transported back. The transport $\tikz[baseline={([yshift=-2*11pt*.15]current bounding box.center)},xscale=.3,yscale=.15]{
\draw[rounded corners=2pt,->] (1,-.1) -- (1,.5) -- (0,1.5) -- (0,2.3);
\draw[rounded corners=2pt,->] (0,-.1) -- (0,.5) -- (1,1.5) -- (1,2.3);
}$ is taken care of by the deformed spin permutation~\eqref{Rcheck}, and the nearest-neighbour exchange
$\tikz[baseline={([yshift=-2*11pt*0.15]current bounding box.center)},xscale=0.3,yscale=0.15]{
\draw[->] (1,0) -- (1,2.4);
\draw[->] (2,0) -- (2,2.4);
\draw[style={decorate, decoration={zigzag,amplitude=.5mm,segment length=1mm}}] (1,1) -- (2,1);}$ 
by the antisymmetriser $e_i(\q)$. More precisely, in \eqref{SqHSL} the diagrams are equivalent to the formulas via the rules
\begin{equation}
	\R(u/v;\q) = 
	\tikz[baseline={([yshift=-2*11pt*.15]current bounding box.center)},xscale=.4,yscale=.2]{
		\draw[rounded corners=2pt,->] (1,-.1) node[below]{$v$} -- (1,.5) -- (0,1.5) -- (0,2.3)  node[above]{\textcolor{gray!50}{$v$}};
		\draw[rounded corners=2pt,->] (0,-.1) node[below]{$u$} -- (0,.5) -- (1,1.5) -- (1,2.3)   node[above]{\textcolor{gray!50}{$u$}};
	} \, , \qquad
	e(\q) = 
	\tikz[baseline={([yshift=-2*11pt*.15]current bounding box.center)},xscale=.4,yscale=.2]{
		\draw[->] (1,0) node[below]{$u$} -- (1,2.4) node[above]{\textcolor{gray!50}{$u$}};
		\draw[->] (2,0) node[below]{$v$} -- (2,2.4) node[above]{\textcolor{gray!50}{$v$}};
		\draw[style={decorate, decoration={zigzag,amplitude=.5mm,segment length=1mm}}] (1,1) -- (2,1);
	} \, .
\end{equation}
At the isotropic point we recover $\Ss^\LL_{[i,j]}(1) = \Ss^\RR_{[i,j]}(1) = 1 - \mathrm{P}_{\!ij}$. In general the chiral operators~\eqref{SqHSL} differ from each other and involve multi-spin interactions, as the intermediate spins do feel the transport.

The final ingredients that we need are the `quantum integers' 
\begin{equation} \label{N_q}
	[N]_\q \equiv \frac{\q^N-\q^{-N}}{\q-\q^{-1}} = \frac{\sin(N \eta)}{\sin(\eta)} \, ,
\end{equation}
and the appropriate modification of the pair potential~\eqref{eq:HS}, 
\begin{equation} \label{VqHS}
	\V(k;\q) \equiv \frac{1}{(\q \, \omega^k - \q^{-1})(\q \, \omega^{-k} - \q^{-1})} = \frac{1}{4\,\sin(\pi k/N + \eta) \, \sin(\pi k/N - \eta)} \, , \qquad 
	\omega \equiv \E^{2\pi\mspace{1mu}\I/N} \, ,
\end{equation}
where on the right-hand sides we wrote $\q = \E^{\I \eta}$. 
Then the \textsc{xxz}-type HS chain has chiral hamiltonians
\begin{equation} \label{qHSL}
	\HH^\LL(\q)=\frac{[N]_\q}{N} \sum_{i<j}^N \V(i-j;\q) \, \Ss^\LL_{[i,j]}(\q)\, , \qquad
	\HH^\RR(\q)=\frac{[N]_\q}{N}\sum_{i<j}^N \V(i-j;\q) \, \Ss^\RR_{[i,j]}(\q)\, ,
\end{equation}
where by `$\sum_{i<j}^N$' we always mean the sum over all pairs of spins, i.e.\ over $1\leqslant i < j\leqslant N$. 

Comparing the nearest-neighbour bulk terms $\Ss^\LL_{[i,i+1]}(\q) = e_i(\q) = \Ss^\RR_{[i,i+1]}(\q)$ with the highly non-local boundary terms $\Ss^\LL_{[1,N]}(\q) \neq \Ss^\RR_{[1,N]}(\q)$ shows that \eqref{qHSL} are not invariant under the usual lattice translation~\eqref{eq:transl}. Its role is taken over by the deformed shift operator, which is again built from the deformed permutations~\eqref{Rcheck} \cite{Lamers:2018ypi}:
\begin{equation} \label{eq:qshift}
	\G(\q) \equiv 
	\ordprod_{N > k \geqslant 1} \!\!\!\! \R_{k,k+1}\bigl(\omega^{-k};\q\bigr) = 
	\R_{N-1,N}\bigl(\omega^{1-N};\q\bigr) \cdots \R_{12}\bigl(\omega^{-1};\q\bigr) 
	= \!
	\tikz[baseline={([yshift=-.5*11pt*0.13+5pt]current bounding box.center)},xscale=0.4,yscale=0.2,font=\footnotesize]{
		\draw[rounded corners=2pt,->] (4,3) node[below]{$\,\omega^1$} -- (4,4.5) -- (8,8.5) -- (8,10); 
		\draw[rounded corners=2pt,->] (8,3) node[below]{$\ \omega^N$}  -- (8,7.5) -- (7,8.5) -- (7,10); 
		\draw[rounded corners=2pt,->] (7,3) -- (7,6.5) -- (6,7.5) -- (6,10);
		\draw[rounded corners=2pt,->] (6,3) -- (6,5.5) -- (5,6.5) -- (5,10); 
		\draw[rounded corners=2pt,->] (5,3) -- (5,4.5) -- (4,5.5) -- (4,10); 
	}
	, \qquad \G(\q)^N = 1 \, .
\end{equation}
We will call this operator the `quasi-translation', and denote its eigenvalues by $\E^{\I \, p(\q)}$ where $p(\q)$ is the `quasi-momentum'. At $\q=1$, \eqref{eq:qshift} reduces to~\eqref{eq:transl} and $p(\q)$ becomes the usual lattice momentum.
\bigskip

\noindent
The deformed hamiltonians and quasi-translation are constructed such that the special properties of the HS chain are preserved for $\q\neq 1$: the \textsc{xxz}-type HS chain is more complicated in position space (hamiltonians) precisely so that it remains extremely simple in momentum space (spectrum). More precisely,
\begin{itemize}[topsep=1ex,itemsep=-.5ex]
	\item[i)] The chiral hamiltonians~\eqref{SqHSL} and quasi-translation operator~\eqref{eq:qshift} all commute, 
	\begin{equation} \label{qHSLR}
		\bigl[ \G(\q) , \HH^\LL(\q) \bigr] = \bigl[ \G(\q) , \HH^\RR(\q) \bigr] = \bigl[ \HH^\LL(\q) , \HH^\RR(\q) \bigr] = 0 \, ,
	\end{equation}
	belonging to the conserved charges of the parent model.
	In particular, the parity-invariant hamiltonian
	\begin{equation} \label{qHS}
		\HH^\mathrm{full}(\q) \equiv \frac{\HH^\LL(\q) + \HH^\RR(\q)}{2}
	\end{equation}
	is also a conserved charge. At $\q=1$, all three deformed hamiltonians reduce to the HS hamiltonian~\eqref{eq:HS}.
	\item[ii)] The extended spin symmetry persists, again suitably deformed. 
	We refer to \cite{Lamers:2020ato} for details. 
	\item[iii)] The spectrum is still given in terms of the motifs~\eqref{rrepulsion}, with degeneracies~\eqref{sm:degeneracy}. The (highest-weight) eigenstate of each motif is known in closed form~\cite{Lamers:2020ato}. Its quasi-momentum has the same value~\eqref{momentum} as for the HS chain: the \emph{meaning} of $p(\q)$ depends on $\q$ through the definition of $\G(\q)$, but its \emph{values} are constant. The chiral and full energies remain additive as in \eqref{HS_disp}, with dispersion relations
	\begin{equation} \label{dispRL}
		\varepsilon^{\LL}_n (\q) = \frac{1}{\q-\q^{-1}} \biggl(\q^{N-n} \, [n]_\q - \frac{[N]_\q}{N} \, n \biggr) \, , \qquad
		\varepsilon^\mathrm{full}_n (\q) = \frac{1}{2} \, [n]_\q \, [N-n]_\q \, , \qquad
		\varepsilon^{\RR}_n (\q) = \varepsilon^{\LL}_n (\q^{\mspace{-1mu}-1}) \, .
	\end{equation}
	The `full' dispersion follows from the chiral ones by \eqref{qHS}; note that it is real for $\q$ real or unimodular ($|\q|=1$). When $\q\to1$ all three dispersions reduce to \eqref{HS_disp}. For generic $\q$, the `accidental' degeneracies of the HS chain are lifted, while for $\q$ a root of unity there are more accidental degeneracies between different motifs.
\end{itemize} 

For general $\q$ we work in the spin representation~\eqref{eq:TL_spin} of the TL algebra. However, it is known that this representation is faithful for any $\q \in \mathbb{C}$ (\cite{morin2016reality} Section 2.B),
which means that many
results carry over to any other representation, or even to the `abstract' setting directly inside the TL algebra itself. 
To be precise, any equality established in the representation carries over to the TL algebra as long as it is algebraic, i.e.\ involving addition and multiplication only. This is relevant for us because of the limit~\eqref{HL H defs}, which does not make sense in the TL algebra properly speaking, but will still give results that are valid in the TL algebra.
In particular, the commutativity \eqref{qHSLR} of the conserved charges in the \textsc{xxz} representation \cite{Lamers:2020ato} persists at the level of the TL algebra. This will be important in the following.

\subsection{The model at q = i} \label{sec: model at i}
	
\subsubsection{Specialising the parent model}
\noindent
Now consider the special point $\q=\I$. The TL generators are regular at $\q=\I$ but 
\begin{equation} \label{eq:TL_q=i}
	e_j \equiv e_j(\I) \, , \qquad 1\leqslant j <N \, ,
\end{equation}
are nilpotent due to \eqref{TLrel}:
\begin{equation} \label{TLrel_freefermion}
	\begin{aligned}
	e_i^2 & = 0 \, , \\
	e_i \, e_j \, e_i & = e_i \, , \qquad && \text{if} \ |i-j|=1 \, , \\ 
	\bigl[ e_i, e_j\bigr] & = 0 \, ,\qquad && \text{if} \ |i-j|>1 \, .
	\end{aligned}
\end{equation}
We will call this the free-fermion TL algebra.
Since the function~\eqref{Rcheck} becomes
\begin{equation} \label{f at i}
	f(u;\I) = \I \, \frac{1-u}{1+u} \, , \quad 
	f(\E^{\I x};\I) = \tan x \, ,
\end{equation}
we shall have ample opportunity to use the short-hand notation 
\begin{equation} \label{def tk}
	t_k \equiv f(\omega^k;\I) = \tan \! \tfrac{\pi\,k}{N} \, .
\end{equation}
Thus the quasi-translation \eqref{eq:qshift} becomes \eqref{Gop} from the main text, 
\begin{equation} \label{qshift at i}
	\G \equiv \G(\I) = (1+t_{N-1} \, e_{N-1}) \cdots (1+t_{1}\,e_{1}) \, .
\end{equation}

Note, however, that $t_k$ has a simple pole at $k=N/2$, which appears in \eqref{qshift at i} when $N$ is even. This is the first hint that the specialisation of the parent model to $\q=\I$ behaves very differently for even vs odd length. This is more pronounced still for the hamiltonians. On the one hand, the quantum integers~\eqref{N_q} simplify to
\begin{equation} \label{N_q=i}
	[2k]_{\I} =0 \, , \qquad [2k+1]_{\I} = (-1)^k \, ,
\end{equation}
so the prefactor in \eqref{qHSL} contributes a simple zero when $N$ is even. On the other hand, the potential~\eqref{VqHS} becomes
\begin{equation} \label{eq:V_q=i}
	\V\mspace{-1mu}(k;\I) = -\frac{1}{4\cos(\pi\,k/N)^2} = -\frac{1}{4} \, (1+t_k^2) \, ,
\end{equation}
which has a second-order pole at antipodal points, $k=N/2$. The sign in \eqref{eq:V_q=i} is irrelevant and will be absorbed in a rescaling below. The point is that we see that some of the matrix elements of the hamiltonians also become infinite when $N$ is even. One can renormalise the quasi-translation and hamiltonians in order to remove these poles (by taking residues), but this sets their spectra to zero. Numerical investigations show that there are Jordan blocks of size up to $(M+1) \times (M+1)$ in the $M$-magnon sector, $M\leqslant N/2$. We will come back to even length, and investigate these intriguing features, in a separate publication. 
\bigskip

\noindent
In everything that follows, we will exclusively consider the case where \emph{$N$ is odd}. None of the preceding singularities appear, and all matrix elements remain finite. It will require quite some effort to work out what happens with the long-range interactions at $\q=\I$. Let us first see what happens to the energies. The chiral dispersions are purely imaginary and become linear with the mode number~$n$:
\begin{equation} \label{epLodd}
	\varepsilon^{\LL}_n(\I) = -\varepsilon^{\RR}_n(\I) = \frac{(-1)^{(N+1)/2}}{2\,\I\,N} \times 
	\begin{cases} 
		n \, , & n \ \text{even} , \\
		n-N \, , & n \ \text{odd} \, ,
	\end{cases}
\end{equation}
This is the origin of the chiral dispersion~\eqref{eLodd} from the main text. Next, the full dispersion~\eqref{dispRL} vanishes identically at $\q=\I$ when $N$ is odd, because either $[n]_{\q=\I}$ or $[N-n]_{\q=\I}$ is zero. To extract a nonzero result we divide by $\q+\q^{-1}$ before we specialise,
\begin{equation} \label{eptLodd}
	\lim_{\q\to\I} \frac{\varepsilon^\mathrm{full}_n (\q)}{\q+\q^{-1}} = 	\frac{(-1)^{(N+1)/2}}{4} \times 
	\begin{cases} 
		n \,, & n \ \text{even} , \\
		N-n \,, & n \ \text{odd} .
	\end{cases}
\end{equation}
This is where the second part of~\eqref{eLodd} from the main text comes from. More precisely, removing the prefactors from \eqref{epLodd} and \eqref{eptLodd} we obtain the dispersions~\eqref{eLodd} from the main text. 

Combining these dispersions with the motifs \eqref{rrepulsion} yields energy levels that are equispaced with steps of $2$. $E^\LL$ is bounded by $\pm (N^2 - 1)/4$ and reaches these extremes at motifs $\{1,3,\dots,N-2\}$ and $\{2,4,\dots,N-1\}$, cf.~Fig.~\ref{fg:dispersions}. $\HH$ has eigenvalues $\geqslant0$, with $E=0$ for the empty motif, and maximal energy $E=2\ell(3\ell+1)$ or $E=2(\ell+1)(3\ell+1)$ depending on whether $N=4\ell+3$ or $N=4\ell+1$, respectively. This maximum corresponds to the one or two motifs $\{1,3,\dots,N-4,N-2\}$ switching halfway between the branches of $\varepsilon_n$ in Fig.~\ref{fg:dispersions}.

It is a highly nontrivial fact that the vanishing of $\varepsilon^\mathrm{full}$ at $\q=\I$ already occurs at the level of the hamiltonian:
\begin{equation} \label{Hfull=0 at q=i}
	\HH^\mathrm{full}(\I) = \frac{\HH^\LL(\I) + \HH^\RR(\I)}{2} = 0 \, ,
\end{equation}
as will be established in \textsection\ref{proof of vanishing hij}. We are thus led to define
\begin{equation} \label{HL H defs}
	\HH^{\LL} \equiv 
	2\,\I\,N (-1)^{(N+1)/2} \; \HH^{\LL}(\I) \,, \qquad
	\HH \equiv 2 \, (-1)^{(N+1)/2} \, \lim_{\q\to \I} \frac{ \HH^\mathrm{full}(\q) }{(\q+\q^{-1})/2} \,, \qquad
	\HH^{\RR} \equiv 2\,\I\,N (-1)^{(N+1)/2} \; \HH^{\RR}(\I) \, ,
\end{equation}
where we note that $2\,\HH^\mathrm{full}(\q) /(\q+\q^{-1})=(\HH^\LL(\q)+\HH^\RR(\q))/[2]_\q$ also $q$-deforms $\HH^\textsc{hs}$.
We have $\HH^{\LL} = -\HH^{\RR}$ according to \eqref{Hfull=0 at q=i}.
The commutativity~\eqref{qHSLR} of the quasi-translation and chiral hamiltonians survives at $\q=\I$. 
The vanishing \eqref{Hfull=0 at q=i} guarantees that those operators furthermore commute with $\HH$, as one sees by expanding the commutators like~\eqref{qHSLR} in $\q+\q^{-1}$. Therefore we have \eqref{GHop} from the main text, i.e.\
\begin{equation} \label{eq:comm_charges}
	\bigl[ \G , \HH^\LL \bigr] = \bigl[ \G , \HH \bigr] = \bigl[ \HH^\LL , \HH \bigr] =  0 \, .
\end{equation} 
See Table~\ref{tb:table} for an overview of the commuting charges and symmetries obtained from the parent model.

\begin{table}[h]
	\begin{tabular}[c]{lcccccccl} 
		\toprule
		& \rlap{\ \ conserved charges} & & & & & \rlap{\ \ symmetry} & & \\
		\midrule
		HS chain & \ \ & $\G^\textsc{hs}$ & \textcolor{gray!50}{$\HH^\textsc{hs}$} & $\HH^\textsc{hs}$ & \textcolor{gray!50}{$\HH^\textsc{hs}$} & \ \ & $\mathfrak{su}(2)$ & $\mathfrak{su}(2)$ Yangian \\[1.5ex]
		parent model & & $\G(\q)$ & $\HH^\textsc{l}(\q)$ & $\displaystyle \frac{\HH^\textsc{l}(\q)+\HH^\textsc{r}(\q)}{\q+\q^{-1}}$ & $\HH^\textsc{r}(\q)$ & & $U_\q\mspace{2mu}\mathfrak{sl}(2)$ & quantum-affine $\mathfrak{gl}(2)$ \\[3ex]
		this paper & & $\G$ & $\HH^\textsc{l}$ & $\HH$ & $-\HH^\textsc{l}$ & & $\mathcal{A}_{1|1}$ & extended \textcolor{gray!50}{($\mathfrak{gl}(1|1)$ Yangian?)} \\
		\bottomrule\hline
	\end{tabular}
	\caption{Correspondence of the conserved charges and symmetries of the HS chain, parent model, and the model studied here.}
	\label{tb:table}
\end{table}

More precisely, the limit in \eqref{HL H defs} is taken in the spin representation~\eqref{eq:TL_spin} of the TL generators, where the dependence on $\q$ is meromorphic. It turns out that the result can again be expressed in terms of the TL generators~\eqref{eq:TL_q=i}. It follows from the comments just preceding \textsection{\ref{sec: model at i}} that \eqref{qshift at i} and \eqref{HL H defs} correspond to pairwise commutative elements of the TL algebra at the free-fermion point. Next we will provide explicit formulas for the hamiltonians~\eqref{HL H defs}.

\subsubsection{Explicit formulas for the hamiltonians}

\noindent 
Let us expand the chiral spin interactions around $\q=\I$ as
\begin{equation} \label{spinlin}
	\Ss^{\LL,\RR}_{[i,j]}(\q) = \Ss^{\LL,\RR}_{[i,j]} + \frac{\q+\q^{\mspace{-2mu}-1}}{2} \, \tilde{\Ss}^{\LL,\RR}_{[i,j]} + \mathcal{O}\bigl((\q+\q^{\mspace{-2mu}-1})^2\bigr) \, , \qquad
	\Ss^{\LL,\RR}_{[i,j]} \equiv \Ss^{\LL,\RR}_{[i,j]}(\I) \, , \quad
	\tilde{\Ss}^{\LL,\RR}_{[i,j]} \equiv \partial_\q\big|_{\q = \I} \, \Ss^{\LL,\RR}_{[i,j]}(\q) \, .
\end{equation} 
Since \eqref{eq:V_q=i} holds up to quadratic corrections we have 
\begin{equation} \label{HamLLi}
	\begin{aligned}
	\HH^{\LL,\RR} & = \frac{\I}{2}\sum_{i=1}^{N-1}\sum_{k=1}^{N-i} \bigr(1+t_k^2\bigl) \, \Ss^{\LL,\RR}_{[i,i+k]} \, , \\
  	\HH & = \frac{1}{4N} \sum_{i=1}^{N-1} \sum_{k=1}^{N-i} \bigr(1+t_k^2\bigl) \, \bigl(\tilde{\Ss}^{\LL}_{[i,i+k]}+\tilde{\Ss}^\RR_{[i,i+k]}\bigr) \, ,
	\end{aligned}
\end{equation}
Remarkably, the complicated spin operators can be written explicitly in terms of nested commutators of \emph{adjacent} TL generators,
\begin{equation} \label{eq:nested_TL}
	e_{[l,m]} \equiv [e_l,[e_{l+1},\ldots[e_{m-1},e_m]\ldots]] = 
	[[\ldots [e_l,e_{l+1}],\ldots e_{m-1}],e_m] \, , \qquad i\leqslant l<m<j \, .
\end{equation}
Note that $e_{[k,k]} = e_k$ is just a single TL generator. The next examples are $e_{[k,k+1]} = [e_k,e_{k+1}]$ and $e_{[k,k+2]} = [e_k,[e_{k+1},e_{k+2}]] = [[e_k,e_{k+1}],e_{k+2}]$. More generally, the commutators in \eqref{eq:nested_TL} can be nested from left to right or from right to left, as can be proven by induction using the Jacobi identity. Note that it suffices to consider commutators of strings of successive TL generators, since non-successive generators commute, which implies that the nested commutators vanish if any $e_k$ with $l\leqslant k\leqslant m$ is missing from the string. 
\medskip

\paragraph{Chiral hamiltonians.}
In \textsection\ref{sec:interaction_recursion} we will use the recursive structure of the spin interactions to show that the 
\begin{equation} \label{SijLR}
	\begin{aligned}
	\Ss^{\LL}_{[i,i+k]} & = \! \sum_{0\leqslant l \leqslant m < k} \!\!\!\!\!
	(-1)^{l} \, t_{k-m,k-l} \, t^2_{k-l,k} \, e_{[i+l,i+m]} \, , \\
	\Ss^{\RR}_{[i,i+k]} & = \! \sum_{0\leqslant l \leqslant m < k} \!\!\!\!\! 
	(-1)^{k-l-1} \, t_{l+1,m+1} \, t^2_{m+1,k} \, e_{[i+l,i+m]} \, ,
	\end{aligned}
\end{equation}
where denote products of the tangents~\eqref{def tk} as
\begin{equation} \label{def t_kl SM}
	t_{k, l} \equiv  \prod_{i=k}^{l-1} t_{i} \quad (k<l)\, , \qquad t_{k,k} \equiv 1 \, .
\end{equation}
For our purposes it will be more convenient to rewrite \eqref{SijLR} in the more symmetric form
\begin{equation} \label{def slmk}
	\Ss^{\LL, \RR}_{[i,i+k]} = \! \sum_{0\leqslant l \leqslant m < k} \!\!\!\!\!\! 
	\varsigma^{\LL,\RR}_{l,m,k-1} \, e_{[i+l,i+m]} \, , \qquad
	\varsigma_{l,m,k-1}^{\LL} \equiv (-1)^{l} \, t_{k- l,k} \, t_{k - m, k} \, , \qquad 
	\varsigma_{l, m, k}^{\RR} \equiv \varsigma^{\LL}_{k - l, k - m, k} \, .
\end{equation}
Note that \eqref{def t_kl SM} makes sense for any $k\leqslant l$, and the coefficients in \eqref{def slmk} for arbitrary $k$ and nonnegative $l$, $m$.

Plugging \eqref{def slmk} into the expression~\eqref{HamLLi} for the chiral hamiltonians one obtains
\begin{equation} \label{Hamioddter}
	\HH^{\LL,\RR} = \frac{\I}{2} \! \sum_{1\leqslant i\leqslant j < N} \!\!\!\!\!\! h^{\LL,\RR}_{ij} \, e_{[i,j]} \,,
\end{equation}
with the nested commutators have coefficients
\begin{equation} \label{notation hij}
	h^{\LL,\RR}_{ij} \equiv \sum_{l=1}^{i} \sum_{m=j}^{N-1} \bigl(1+t^{2}_{m-l+1}\bigr) \, \varsigma^{\LL,\RR}_{i-l,j-l,m-l} \, , \quad 
	h^{\LL, \RR}_{ji} \equiv (-1)^{j-i} \, h^{\LL, \RR}_{ij} \, , 
	\qquad i\leqslant j \, .
\end{equation}
Now, using the $\LL$-$\RR$-symmetry $\varsigma_{l, m, k}^{\RR} = \varsigma_{k - l, k - m, k}^{\LL}$ and changing indices, it can be checked that the coefficients of the chiral hamiltonians~\eqref{Hamioddter} are related by
\begin{equation} \label{hij LR symm}
	h^{\RR}_{ij} = (-1)^{j-i} \, h^{\LL}_{N-j, N-i} \, .
\end{equation}
Furthermore, $h^{\LL}_{ij}$ can be simplified using $t^{2}_{m-l+1} \, \varsigma^{\LL}_{i-l,j-l,m-l} = -\varsigma^{\LL}_{i-(l-1),j-(l-1),m-(l-1)}$, which follows from the definitions. As such, coefficients in $h_{ij}^{\LL}$ telescope over the $l$ variable to yield 
\begin{equation} \label{telescoped h^L}
	h^{\LL}_{i j} = \sum_{m=j}^{N-1} \bigl(\varsigma^{\LL}_{0,\, j-i, \, m-i} - \varsigma^{\LL}_{i, j, m} \bigr) \, .
\end{equation}
This can be more explicitly factorised as 
\begin{equation}
	h^{\LL}_{i j} = \! \sum_{n=j+1}^N \!\! t_{n-j,n-i} \, \bigl(1-(-1)^{i} \, t_{n-i,n}^{2}\bigr) \, ,
\end{equation}
which is useful for numerical computations. This proves \eqref{Ham_L} from the main text once we show~\eqref{SijLR}. In \textsection\ref{proof of vanishing hij} we will furthermore establish, using a rather technical analytic proof, that for any $i,j$
\begin{equation} \label{hL=-hR}
    h^\RR_{ij} = -h^\LL_{ij} \, ,
\end{equation}
which means that \eqref{Hfull=0 at q=i} holds coefficient by coefficient in terms of the nested TL commutators. This is the origin of the first equality in \eqref{parity}.
\medskip 

\paragraph{Full hamiltonian.}
To find an explicit expression for $\HH$ we need the next term in the expansion~\eqref{spinlin} of the spin operators. As we will outline in \textsection\ref{sec:interaction_recursion}, they can be written as anticommutators of the nested commutators~\eqref{eq:nested_TL}:
\begin{equation} \label{tildeSLR}
	\begin{aligned}
	\tilde{\Ss}^{\LL}_{[i,i+k]} & = \!\!\sum_{0\leqslant j\leqslant l<m\leqslant n<k} \mspace{-34mu} (-1)^{n-1} \, t_{k-n,k-l} \, t_{k-m,k-j} \, t^2_{k-j,k} \, \bigl\{e_{[i+n,i+m]},e_{[i+l,i+j]}\bigl\} \, , \\
	\!\! \tilde{\Ss}^{\RR}_{[i,i+k]} & = \!\! \sum_{0\leqslant j\leqslant l<m\leqslant n<k} \mspace{-34mu} (-1)^{k-j} \; t_{j+1,m+1} \, t_{l+1,n+1} \, t^2_{n+1,k} \, \bigl\{e_{[i+j,i+l]},e_{[i+m,i+n]}\bigl\}\,.
	\end{aligned}
\end{equation} 
After combining all the factors and telescoping the sums like before we obtain
\begin{equation} \label{Hamioddfull}
	\HH = -\frac{1}{4N} \! \sum_{1\leqslant i\leqslant j<k\leqslant l <N} \mspace{-33mu} \bigl(h^\LL_{ij;kl} + h^\RR_{ij;kl} \bigr) \, \bigl\{ e_{[i,j]},e_{[k,l]} \bigr\} \, ,
\end{equation}
with
\begin{equation} \label{eq:hLRijkl}
	\begin{aligned} 
	h^\LL_{ij;kl} & = (-1)^{k-j} \!\! \sum_{n=l+1}^{N} \!\! t_{n-l,n-j} \, t_{n-k,n-i} \, \bigl(1-(-1)^i \, t^2_{n-i,n} \bigr) \, , \\
	h^\RR_{ij;kl} & = (-1)^{l-i} \sum_{n=0}^{i-1} t_{i-n,k-n} \, t_{j-n,l-n} \, \bigl(1-(-1)^{N-l} \, t^2_{l-n,N-n} \bigr) \\
	& = (-1)^{l-j+k-i} \, h^\LL_{N-l,N-k;N-j,N-i} \, .
	\end{aligned}
\end{equation}
This is \eqref{Htildedef} from the main text.
\medskip

\paragraph{Example.} For instance, at $N = 3$, we have
\begin{equation} \label{H N3}
	\begin{aligned}
    \HH^\LL & = \frac{\I}{2} \, \Bigl( t_1 \bigl(1+t_2^2 \bigr) \, [e_1,e_2] + \bigl(2+t_1^2 + t_2^2\bigr) \, e_1 + \bigl(1 - t_1^2 \, t_2^2\bigr) \, e_2 \Bigr) = 2\,\I\, \bigl( \sqrt{3} \, [e_1,e_2] + 2 \, (e_1 - e_2) \bigr) \, , \\
    \HH & = \frac{1}{6} \, \bigl(1+t_2^2\bigr) \, t_1^2 \, \{ e_1, e_2 \} = 2 \, \{ e_1,e_2 \} \, .
    \end{aligned}
\end{equation}

\subsubsection{Sketch of the derivation of the spin interaction} \label{sec:interaction_recursion}

\noindent
In order to prove \eqref{SijLR} it is useful to introduce the slight generalisation of the spin interaction,
\begin{equation} \label{SqHSLR}
	\begin{aligned}
	\Ss^{\LL}_{ [i,j];n}(\q) & \equiv \Biggl(\mspace{1mu} \ordprod_{j>k>i} \!\! \R_{k,k+1}\bigl(\omega^{n-k};\q\bigr) \Biggr) \, e_i(\q) \, \Biggl( \ordprodopp_{i<k<j} \!\! \R_{k,k+1}\bigl(\omega^{k-n};\q\bigr) \Biggr) \, , \\
	\Ss^\RR_{[i,j];n}(\q) & \equiv \Biggl(\mspace{1mu} \ordprodopp_{j>k>i} \!\! \R_{k-1,k}(\omega^{k-n};\q) \Biggr) \, e_{j-1}(\q) \, \Biggl( \ordprod_{i<k<j} \!\! \R_{k-1,k}(\omega^{n-k};\q) \Biggr) \,,
	\end{aligned} 
	\qquad i<j \, ,
\end{equation}
so that 
\begin{equation} \label{SqHSLRn}
	\Ss^{\LL}_{[i,j]}(\q) = \Ss^{\LL}_{[i,j];j}(\q) \, , \qquad \Ss^{\RR}_{[i,j]}(\q) = \Ss^{\RR}_{[i,j];i}(\q) \, .
\end{equation}
Now set $\q=\I$ and write
\begin{equation}
	\Ss^{\LL}_{[i,j];n} \equiv \Ss^{\LL}_{ [i,j];n}(\I) \, , \qquad
	\Ss^\RR_{[i,j];n} \equiv \Ss^\RR_{[i,j];n}(\I) \, .
\end{equation}
These operators obey the recursion relations
\begin{equation} \label{SqHSLRN}
	\begin{aligned}
	\Ss^{\LL}_{[i,j+1 ];n} & = \bigl( 1-t_{n-j} \,e_j \bigr) \, \Ss^{\LL}_{ [i,j];n}\, \bigl(1+t_{n-j} \,e_j \bigr) \,, \\
	\Ss^{\RR}_{ [i,j];n}& = \bigl(1-t_{i-n}\, e_i \bigr) \, \Ss^{\RR}_{[i+1,j ];n} \, \bigl(1+t_{i-n}\, e_i \bigr) \,.
	\end{aligned}
\end{equation}
Let us show, using this structure, that $\Ss^{\LL}_{ [i,j];n}$ and $\Ss^{\RR}_{ [i,j];n}$ are linear combinations of nested commutators of TL generators. We will use induction on the distance $j-i$. The base case is
\begin{equation} \label{SqHSLR1}
	\Ss^{\LL}_{[i,i+1 ];n} = \Ss^{\RR}_{[i,i+1 ];n} = e_i \, .
\end{equation}
The recursion relations \eqref{SqHSLRn} can be rewritten as
\begin{equation} \label{SqHSLRrec}
	\begin{aligned}
	\Ss^{\LL}_{[i,j+1 ];n} & = \Ss^{\LL}_{[i,j];n} + t_{n-j} \, \ \bigl[\Ss^{\LL}_{ [i,j];n},e_j\bigr] \ \, - t^2_{n-j} \ \: e_j \ \ \Ss^{\LL}_{ [i,j];n} \ e_j \, , \\
	\Ss^{\RR}_{[i-1,j ];n} & = \Ss^{\RR}_{ [i,j];n} - t_{i-n} \, \bigl[ e_{i-1},\Ss^{\RR}_{ [i,j];n}\bigr] - t^2_{i-n} \, e_{i-1} \, \Ss^{\RR}_{[i,j];n} \, e_{i-1}\, .
	\end{aligned}
\end{equation}
The first terms on the right-hand sides of \eqref{SqHSLRrec} reproduce the structure of \eqref{SqHSLRn}. The second terms increase the length of the nested commutators by one to the right or left, respectively. The third terms give a result proportional to $e_j$ or $e_{i-1}$, respectively, and thus move a TL generator to the right or left, respectively. At each level of nesting, we get a factor $\pm t$, and every time the index of a solitary TL generator is shifted by one to the right or left we get a factor of $-t^2$. In this way we obtain linear combinations of nested TL commutators, as we claimed.

For example, 
\begin{equation} \label{Sii2L}
	\begin{aligned}
	\Ss^{\LL}_{[i,i+2]} & = e_{[i,i]} -t_1^2\, e_{[i+1,i+1]} + t_1 \, e_{[i,i+1]}\,,\\ 
	\Ss^{\LL}_{[i,i+3]} & = e_{[i,i]}-t_2^2 \, e_{[i+1,i+1]} + t_2^2 \, t_1^2\, e_{[i+2,i+2]} + t_2 \, e_{[i,i+1]} - t_2^2 \, t_1 \, e_{[i+1,i+2]} + t_2  \, t_1 \, e_{[i,i+2]} \, ,
  	\end{aligned}
\end{equation}
and
\begin{equation} \label{Sii2R}
	\begin{aligned}
	\Ss^{\RR}_{[j-2,j]} & = e_{[j-1,j-1]} - t_1^2\, e_{[j-2,j-2]} - t_1 \, e_{[j-2,j-1]}\,,\\
	\Ss^{\RR}_{[j-3,j]} & = e_{[j-1,j-1]} - t_2^2 \, e_{[j-2,j-2]} + t_2^2 \, t_1^2\, e_{[j-3,j-3]} - t_2 \, e_{[j-2,j-1]} + t_2^2 \, t_1\, e_{[j-3,j-2]} + t_2 \, t_1 \, e_{[j-3,j-1]} \, .
	\end{aligned}
\end{equation}
This structure generalises for any separation between the spins to the result \eqref{SijLR} for the chiral hamiltonians.

In order to compute $\HH$ one has to expand the spin interactions $\Ss^{\LL,\RR}_{[i,j]}(\q)$ to linear order in $\q+\q^{-1}$ as in \eqref{spinlin}. It can be proved that, thanks to \eqref{hL=-hR}, we may keep the TL generators as they are, and expand
\begin{equation}
	\begin{aligned}
	f(u;\q) & = f(u) - \frac{\q+\q^{-1}}{2} \, f(u)^2 + \dots \,, & \qquad f(u) \equiv f(u;\I) \, , \\
	f(u^{-1};\q) & = f(u^{-1}) - \frac{\q+\q^{-1}}{2} \, f(u^{-1})^2 \\
	& = -f(u) - \frac{\q+\q^{-1}}{2} \, f(u)^2 + \dots \, .
	\end{aligned}
\end{equation}
Following the recursion relations \eqref{SqHSLRn}, the typical structure of the terms linear in $\q+\q^{-1}$ in $\Ss^{\LL}_{[i,j]}(\q)$ is
\begin{equation}
	f(u)^2 \, e_{i+1} \, N_i \, \bigl(1+f(u) \, e_{i+1}\bigr) + \bigl(1-f(u)\,e_{i+1}\bigr)\,N_i \,f(u)^2 \, e_{i+1} = f(u)^2 \, \{e_{i+1},N_i\} \, ,
\end{equation} 
where $N_i$ contains nested commutators, and $u$ will be specified to a particular value. The next term is of the form
\begin{equation}
	\begin{aligned}
	f^2(u) \, \bigl( 1-f(u/\omega) \, e_{i+2} \bigr) \{e_{i+1},N_i\} \, \bigl(1-f(u/\omega) \, e_{i+2} \bigr) = {} & f(u)^2 \, \{e_{i+1},N_i\} - f(u)^2 \, f(u/\omega)^2 \, \bigl\{e_{i+2},N_i\bigr\} \\
	& - f(u)^2 \, f(u/\omega) \, \bigl[e_{i+2},\{e_{i+1},N_i\}\bigr] \, .
	\end{aligned}
\end{equation} 
In conclusion, the typical structure of each term consists of nested commutators and one anticommutator. As is clear from the preceding formula, the index of the TL generator in the anticommutator can be transported away from those in the commutators. Using the properties of the TL operators at the free-fermion point we conclude that
\begin{align}
	[e_k,\ldots,\{e_l,[e_m,\ldots,[e_{n-1},e_n]\ldots\}\ldots] = \bigl\{ e_{[k,l]}, e_{[m,n]} \bigl\} = (-1)^{k-l+m-n} \bigl\{ e_{[n,m]}, e_{[l,k]} \bigl\} \, , \qquad  k\leqslant l<m\leqslant n \, .
\end{align} 
Then, considering again carefully the recursion relations \eqref{SqHSLRn}, we are able to identify the coefficients in front of these structures to obtain \eqref{tildeSLR}.

\subsection{Fermionic representation} \label{sec_sm:fermionic_rep}

\noindent
So far we have mostly considered the spin representation~\eqref{eq:TL_spin} of the TL generators. In this setting we do not only know the spectrum in closed form, but also the corresponding exact (highest-weight) eigenstates. In addition, to make sense of the limit in \eqref{HL H defs}, we strictly speaking worked abstractly, i.e.\ inside the TL algebra, cf.\ the comments just preceding \textsection\ref{sec: model at i}. Now we will be interested in a fermionic realisation of our model. 

\subsubsection{Fermionic realisation of the Temperley-Lieb algebra}
Let $\sigma^\pm_j \equiv (\sigma^x_j \pm \I\, \sigma^y_j )/2$ and $\sigma^z_j$ denote the Pauli matrices acting at site $j$ of the spin chain. The translation to the fermionic setting goes via the Jordan--Wigner transformation 
\begin{equation}  \label{JW transf}
	c_j \equiv \biggl(\, \prod_{i=1}^{j-1} (-\sigma_i^z) \biggr) \, \sigma_j^- \, , \quad
	c_j^\dag \equiv \biggl(\, \prod_{i=1}^{j-1} (-\sigma_i^z) \biggr) \, \sigma_j^+ \, , \qquad 
	\{ c_i , c_j^\dag \} = \delta_{ij} \, , \quad \{ c_i , c_j \} = \{ c_i^\dag , c_j^\dag \} = 0 \, ,
\end{equation}
providing canonical fermionic creation and annihilation operators that act on the usual fermionic Fock space. For our purposes it will be convenient to follow~\cite{Gainutdinov:2011ab} and rescale by defining 
\begin{equation}  \label{eq: def fk}
    f_j = (-\I)^{\mspace{2mu}j} \, c_j \, , \qquad f^+_j = (-\I)^{\mspace{2mu}j} \, c^\dag_j \, .
\end{equation}
Here and below we use superscript `${}^+$' (rather than `${}^\dag$') since we do not mean the hermitian conjugate or adjoint for some scalar product. 
We thus start from non-unitary fermionic creation and annihilation operators with anticommutation relations
\begin{equation} \label{fermtranssm}
	\{f_i,f_j^+\} = (-1)^i \, \delta_{ij} \, , \quad 
	\{f_i,f_j\} = \{f^+_i,f^+_j\} = 0 \, .
\end{equation}
Then the two-site fermions
\begin{equation} \label{gfer_supp}
	g_j \equiv f_j+f_{j+1} \, , \quad g_j^+ \equiv f_j^+ + f_{j+1}^+ \, 
\end{equation}
have nontrivial anticommutation relations 
\begin{equation} \label{gfercomm}
	\{g_j,g_j^+\} = 0 \,, \quad 
	\{g_j,g_{j+1}^+\} = (-1)^{j+1} \, , \quad 
	\{g_j,g_{j-1}^+\} = (-1)^{j} \, .
\end{equation}
It is straightforward to check that
\begin{equation} \label{gTL}
	e_j \equiv g_j^+\,g_j = -g_j\,g_j^+ \, , \qquad 1\leqslant j < N \, ,
\end{equation}
provides a representation of the free-fermion TL algebra~\eqref{TLrel_freefermion} on the fermionic Fock space. It turns out that this representation is isomorphic to the \textsc{xxz} representation~\eqref{eq:TL_spin}. This fact, mentioned in \cite{Gainutdinov:2011ab}, is proved in \textsection\ref{sec: XXZ vs ferm}.
Therefore, everything that we already know about the spectrum and eigenstates of our model transports from the spin-chain setting into the fermionic set-up.

We will need some explicit (anti)commutation relations for the operators we have thus defined. Using $[AB,C]= A\,\{B,C\} - \{A,C\}\,B$ one obtains the following nontrivial commutation relations between TL generators and two-site fermions:
\begin{equation}  \label{gwithTL}
    \begin{gathered}
    [e_j,g_j^+]=[e_j,g_j]=0 \, ,\\
    \begin{aligned} 
    [e_j,g_{j+1}] & =(-1)^j \, g_j \,, \quad 
    && [e_j,g_{j-1}] = (-1)^{j-1} \, g_j \, ,\\
    [e_j,g^+_{j+1}] & = (-1)^{j+1} \, g_j^+ \, , \quad
    && [e_j,g^+_{j-1}] = (-1)^{j} \, g_j^+ \, .
    \end{aligned}
    \end{gathered}
\end{equation}
Note that these relations hold as long as the subscripts of $e$ and $g$ lie between $1$ and~$N-1$. The commutation relations between the TL generators and the one-site fermions will be given in \eqref{fpmwithTL}.

A special feature of the point $\q=\I$ is that nested commutators of TL generators are quadratic in fermions: using $[AB, CD] = A \, \{B,C\} \, D - \{A,C\}\,B\,D + C\,A\,\{B,D\} -C\,\{A,D\}\,B$ successively, the previous expressions yield
\begin{equation}  \label{TLTL}
    \begin{aligned}
    [e_j,e_{j+1}] & = (-1)^j \, \bigl(g_{j+1}^+ \, g_j - g_j^+ g_{j+1} \bigr)\,,\\
    [[e_j,e_{j+1}],e_{j+2}] & = - \bigl(g_{j+2}^+\,g_j + g_j^+g_{j+2}\bigr)\,,\\
    [[[e_j,e_{j+1}],e_{j+2}],e_{j+3}] & =(-1)^{j+1} \bigl(g_{j+3}^+\,g_j - g_j^+g_{j+3}\bigr)\,.
    \end{aligned}
\end{equation}
Continuing in this we we induce that in general
\begin{equation} \label{TLTLgen}
	e_{[j,j+m]}\equiv[\ldots[e_j,e_{j+1}]\ldots,e_{j+m}]=s_{j,j+m}\,\bigl( g_{j+m}^+\,g_j + (-1)^m g_j^+g_{j+m} \bigr)\,, \qquad m> 0 \,,
\end{equation}
where we define the signs
\begin{equation} \label{sij sign}
    s_{ij} \equiv (-1)^{i + \dots +j-1} = (-1)^{(i-j)(i + j - 1)/2}  \ .
\end{equation}
Armed with these relations we turn to the relation between the fermionic operators and our conserved charges.

\subsubsection{Fermions and quasi-translations}
\label{sec:fermions}
\noindent
First we consider the quasi-translation operator~$\G$ defined in~\eqref{qshift at i}.

\paragraph{Quasi-translated fermions.}
In order to find the fermionic degrees of freedom in which the model becomes as simple as possible, we seek a basis with good transformation properties under quasi-translations. It is shown by formula \eqref{Ginv on the fk} that conjugating by $\G$ preserves the space of one-particle fermionic operator. Thus we are led to define the alternative fermionic basis~\eqref{eq:Phi_def}
\begin{align} \label{defPsi}
	\Phi_i\equiv \G^{1-i} f_1 \, \G^{i-1} \,, \qquad \Phi_i^+\equiv \G^{1-i} f_1^+ \, \G^{i-1} \, .
\end{align}
That is, $\Phi_{1} = f_1$ and $\Phi_{i+1} = \G^{-1} \, \Phi_{i} \, \G$, and likewise for $\Phi_j^+$.
Since $\G^N=1$ these quasi-translated fermions are (formally) periodic by definition, 
\begin{equation} \label{periodicity of phi}
	 \Phi_{i+N}=\Phi_i \,, \qquad \Phi^+_{i+N}= \Phi^+_i \,.
\end{equation}
As expected for some kind of translated fermions, this new basis is triangular in the $f_i$ in the sense that $\Phi_{i}$ is a linear combination of $f_{1},\dots,f_i$, and $\Phi_{i}^+$ of $f_{1}^{+},\dots,f_{i-1}^+$, see \eqref{fourier fermions leading term}

In \textsection\ref{sec:techn_G_fermions} we obtain an explicit formula for the action of $\G^{-1}$ by conjugation on the two-site fermions, see \eqref{Ginv on gk - triangular}. That formula can be expressed via $(N-1)\times (N-1)$ matrices $\gamma=(\gamma_{ij})_{1\leqslant i,j\leqslant N-1}$ and $\bar \gamma=(\bar \gamma_{ij})_{1\leqslant i,j\leqslant N-1}$ defined by the relations
\begin{equation} \label{linearity of G on the g}
    \G^{-1} \, g_{i} \, \G = \sum_{j=1}^{N-1} \gamma_{ij} \, g_j \, , \qquad
    \G^{-1} \, g_{i}^+ \, \G = \sum_{j=1}^{N-1} \bar{\gamma}_{ij} \, g^+_j \, .
\end{equation}
Explicitly, the matrix elements read 
\begin{equation} \label{gamma matrix}
	(-1)^{i-j} \, \gamma_{ij} = \bar{\gamma}_{ij} = 
	\begin{cases}
		0 \, , & i < j-1 \, , \\
		(-1)^i \, t_{i+1} \, , & i = j-1 \, , \\
		s_{ij} \,  (1+t_i \, t_{i+1}) \, t_{j,i} \, & i \geqslant j \, .
	\end{cases}
\end{equation}
In addition, since $\G^N=1$, we have $\gamma^N = \bar{\gamma}^N=1$. 
It is now possible to express the quasi-translated fermions in those terms, since it is shown in \eqref{explicit Phi1} that 
\begin{equation}
    \G^{-1} \, f_{1} \, \G = f_{1} - t_1 \, g_1 \, , \qquad \G^{-1} \, f_{1}^+ \, \G = f_{1}^+  + t_1 \, g^+_1 \, ,
\end{equation}
which immediately leads to the relation
\begin{equation} \label{qorbit recursion}
    \Phi_{i+1} = \Phi_{i} - t_1 \sum_{k=1}^{N-1} (\gamma^{i-1})_{1k} \; g_k \ , \qquad
    \Phi_{i+1}^+ = \Phi_{i}^+ + t_1 \sum_{k=1}^{N-1} (\bar{\gamma}^{i-1})_{1k} \; g_k^+ \, .
\end{equation}
By induction, we then obtain
\begin{equation} \label{qorbit in the gs}
    \begin{gathered}
        \Phi_{i+1}= f_1 - t_1 \! \sum_{k=1}^{N-1} \bigl(\gamma^0 + \dots + \gamma^{i-1} \bigr)_{1k} \; g_k \,, \qquad \Phi_{i+1}^+= f_1^+ + t_1 \! \sum_{k=1}^{N-1} \bigl(\bar{\gamma}^0 + \dots + \bar{\gamma}^{i-1} \bigr)_{1k} \; g_k^+ \, .
    \end{gathered}
\end{equation}
It is remarkable that the transformed fermionic operators can be expressed in terms of the $g^+_k$ and $f_1^+$ in such a way. 

Using definition \eqref{defPsi} it is easy to see that anticommutation relations between the quasi-translated fermions are determined by their distances only, i.e.\ 
\begin{equation}
    \{\Phi_i^+ , \Phi_j\} = \alpha(i-j)  \, ,
\end{equation}
where $\alpha(k) \equiv \{\Phi_{k+1}^+, f_1\}$ satisfies
\begin{equation}
    \alpha(k)=\alpha(N+k) \, .
\end{equation}
A more explicit expression for this quantity follows from \eqref{qorbit in the gs}:
\begin{equation}
	\alpha(k) = -1-t_1 \, \bigl(\gamma^0 + \dots + \gamma^{k-1} \bigr)_{11} \, , \qquad 
	\alpha(-k) = -1 + t_1 \, \bigl(\bar\gamma^0 + \dots + \bar\gamma^{k-1} \bigr)_{11} \, , \qquad 
	k\geqslant 0 \, , 
\end{equation} 
Numerical evidence suggests that the result is exceedingly simple:
\begin{equation} \label{explicit alpha}
    \alpha(k) = -(1+t_k) \ .
\end{equation}
This gives the anticommutation-relations~\eqref{psicomm} in the main text. However, note that we do not yet have a complete proof of this fact, see \eqref{tilde alpha}.
\medskip

\paragraph{Fourier-transformed fermions.}
To build an eigenbasis of the quasi-translation in the one-particle sector, we can use the periodicity \eqref{periodicity of phi} to introduce the Fourier transforms 
\begin{equation} \label{Phi_modes}
	\tilde{\Phi}_n = \frac{1}{N} \sum_{j=1}^{N} \omega^{-nj} \, \Phi_{j} \, , \qquad 
	\tilde{\Phi}_n^+ = \frac{1}{N} \sum_{j=1}^{N} \omega^{nj} \, \Phi_{j}^+ \, ,
\end{equation}
where we recall the notation $\omega \equiv \E^{2\pi\mspace{1mu}\I/N}$ from \eqref{VqHS}.
Since $\G \, \Phi_j \, \G^{-1} = \Phi_{j-1}$ by construction, under quasi-translation we indeed have
\begin{equation}
	\G \, \tilde{\Phi}_n \, \G^{-1} = \omega^{-n} \, \tilde{\Phi}_n \, , \qquad 
	\G \, \tilde{\Phi}_n^+ \, \G^{-1} = \omega^{n} \, \tilde{\Phi}_n^+ \, .
\end{equation}
It is shown in \eqref{matrix extension} that 
\begin{equation} \label{eq:zero_modes}
	\tilde{\Phi}_0 = \sum_{i=1}^{N} f_i \, , \qquad
	\tilde{\Phi}_0^+ = \sum_{i=1}^{N} f_i^+ \, ,
\end{equation}
yielding \eqref{zero_modes} from the main text up to the rescaling \eqref{sm:Psi_tilde} discussed below. 
It follows that \eqref{eq:zero_modes} commute with every $e_j$. These zero modes thus do not only commute with $\G$, but also with the hamiltonians. 
\bigskip

By Fourier-transforming both sides of \eqref{qorbit recursion} one can see that the remaining fermionic modes $\Phi_n$ and $\tilde{\Phi}_n$  ($n > 0$), which have positive quasi-momentum, are linear combinations of the $N-1$ \emph{two}-site fermions:
\begin{equation} \label{Psig}
	\tilde{\Phi}_n = \sum_{j=1}^{N-1} M_{nj} \, g_j \, ,
	\qquad \tilde{\Phi}_n^+ = \sum_{j=1}^{N-1} \bar{M}_{nj} \, g^+_j \, . 
\end{equation}
with coefficients expressed in terms of $\gamma, \bar{\gamma}$ by 
\begin{equation}
	M_{nj} \equiv \frac{\omega^{-2n} \, t_1}{N\, (1-\omega^{-n})} \sum_{k=0}^{N-1} \bigl[(\omega^{-n} \, \gamma)^k \bigr]_{1j} \, , \qquad
	\bar{M}_{nj} \equiv \, -\frac{\omega^{2n} \, t_1}{N \, (1-\omega^{n})} \sum_{k=0}^{N-1} \bigl[(\omega^{n} \, \bar{\gamma})^k \bigr]_{1j} \,.
\end{equation}
A numerically more tractable description of these matrix elements, which allows us to compute them up to $N\leqslant19$, will be given in \eqref{equation M Mbar} below.
Observe that the matrices $\gamma, \bar{\gamma}$ are real. Together with the first equality in \eqref{gamma matrix}, which persists to $\gamma^k, \bar{\gamma}^k$, it follows that
\begin{equation}
\label{M Mbar complex}
	\bar M_{nj} =(-1)^j M_{nj}^*\, ,\qquad M_{N-n,j}=M_{n,j}^*\, , \qquad \bar M_{N-n,j}=\bar M_{n,j}^*\, ,
\end{equation}
where the `${}^*$' denotes complex conjugation.
If we take the numerical conjecture \eqref{explicit alpha} as granted, we can obtain the anticommutation relations between the Fourier modes by Fourier transforming the periodic sequence $\alpha(k)$ as follows:
\begin{equation} \label{psitildecommm}
	\begin{aligned}
	\{ \tilde{\Phi}_n^+, \tilde{\Phi}_m \} & = \frac{1}{N^2} \sum_{i,j=1}^{N} \omega^{ni-mj} \, \{ \Phi_i^+ , \Phi_j \} \\
	& = \frac{1}{N^2} \sum_{i,j=1}^{N} \omega^{ni-mj} \, \alpha(i-j) 
	= \delta_{nm} \, \tilde{\alpha}(m)\, ,
	\end{aligned}
\end{equation}
with
\begin{equation} \label{tilde alpha}
	\tilde{\alpha}(n) \equiv \frac{1}{N} \sum_{p=1}^{N} \omega^{np} \, \alpha(p) =
	\begin{cases*}
		\ -1 \, , & $n=0 \, ,$ \\
		(-1)^n\,\I \, ,\quad & $1\leqslant n < N \, .$
	\end{cases*}
\end{equation} 
Note that \eqref{tilde alpha} is equivalent as a statement to the numerical conjecture \eqref{explicit alpha} it is based on. While this result was only numerically observed at the moment, it will be proven \emph{up to a sign} in \eqref{phitilde anticomm unsigned}. We do not have a way to derive this sign as a function of $n$ yet.

We see that, even though we have a long-range interacting model, we can still Fourier transform the suitably translated fermions as usual. Note the orthogonality $\{ \tilde{\Phi}_n^+,  \tilde{\Phi}_m \} \propto \delta_{nm}$. The \emph{nearly} canonical anticommutation relations are the reason for replacing the modes~\eqref{Phi_modes} by the rescaled fermionic modes 
\begin{equation} \label{sm:Psi_tilde} 
	\tilde{\Psi}_n \equiv a_n \, \tilde{\Phi}_n \, , \qquad 
	\tilde{\Psi}_n^+ \equiv a_n \, \tilde{\Phi}_n^+ \, , \qquad
	a_0 \equiv \I \, , \quad 
	a_n \equiv \I^{n+1/2} \quad (n\neq 0) \, ,
\end{equation}
as in \eqref{Psi_tilde} to absorb the signs and factor of $\I$ in~\eqref{tilde alpha} and obtain the canonical anticommutation relations~\eqref{psitildecomm} from the main text. 
We then have
\begin{equation} \label{lin_transf}
	\frac{1}{a_n} \, \tilde{\Psi}_n = \sum_{i=1}^{N-1} \! M_{ni} \; g_i \, , \quad 
	\frac{1}{a_n} \, \tilde{\Psi}_n^+ = \sum_{i=1}^{N-1} \! \bar{M}_{ni} \; g_i^+ \, .
\end{equation}

\bigskip
\subsubsection{Discrete symmetries on Fock basis}

\noindent
Note that in this section, every occurrence of the symbol `$\propto$' refers to an exact formula given in part \textsection\ref{sec:descr_symm}, where it is established in each case that the corresponding coefficient of proportionality is just a phase, i.e.\ of absolute value $1$.

\paragraph{Parity and time reversal.} 
On the basis $\{f_i\}_i$ let us define parity transformation as usual as
\begin{equation} \label{eq: def P}
    \PP(f_i) \equiv f_{N-i+1} \ , \qquad \PP(f_i^+) \equiv f_{N-i+1}^{+} 
\end{equation}
Time reversal acts by the identity on the generators $f^\pm_i$
\begin{equation} \label{eq: def T}
   \TT(f_i) \equiv f_{i} \, , \qquad 
    \TT(f_i^+) \equiv f_{i}^{+} \, ,
\end{equation}
but is $\mathbb{C}$-\emph{anti}linear.

As expected, time-reversal and parity not only square to the identity as $\PP^2=\TT^2= 1$ but also commute with each other, $\PP\,\TT=\TT\,\PP$. Thus, in particular, $(\PP\,\TT)^2=1$ too.
Moreover, it is shown in \eqref{PT phi explicit} that in the basis of rescaled fermions \eqref{sm:Psi_tilde}, as expected both of these transformations reverse the momentum of fermions up to a phase,
\begin{equation} \label{T sym on phitilde}
	\TT\bigl(\tilde{\Psi}_n\bigr) \propto\PP\bigl(\tilde{\Psi}_n\bigr)\propto\tilde{\Psi}_{N-n} \, , \qquad
	\TT\bigl(\tilde{\Psi}_n^+\bigr) \propto \PP\bigl(\tilde{\Psi}_n^+\bigr)\propto \tilde{\Psi}_{N-n}^+ \, .
\end{equation}

\paragraph{Charge conjugation.} This is a little more subtle. We first define the particle-hole transformation $\CC'$ as
\begin{equation} \label{sm: def parthole}
    \CC'(f_i) \equiv f_{i}^+ \ , \qquad 
    \CC'(f_i^+) \equiv f_{i} \, .
\end{equation}
While we do have ${\CC' \,}^2=1$, and $\CC'$ commutes with $\PP$ and $\TT$, this transformation by itself does not play well with our model, and in particular with our Fock basis \eqref{freespectrum}. 
To fix this we introduce an operator $\UU$ implementing an alternating phase. Namely, we define an $\mathbb{C}$-\emph{anti}linear operator characterised by
\begin{equation} \label{def U}
    \begin{aligned}
        \UU \bigl(\tilde\Phi_0\bigr) & \equiv \tilde\Phi_0 \, , \qquad 
        && \; \UU\bigl(g_j\bigr) \equiv (-1)^{j} \,\I\,  g_j \, , \\
        \UU\bigl(\tilde\Phi_0^+\bigr)  & \equiv \tilde\Phi_0^+ \, , \qquad &&\UU \bigl(g_j^+\bigr) \equiv (-1)^{j} \,\I\,  g_j^+ \, ,
    \end{aligned}
    \qquad 1 \leqslant j \leqslant N-1 \, .
\end{equation}
In \textsection\ref{sec: C transf} we establish that 
\begin{equation} \label{eq:C=C'U}
    \CC \equiv \CC' \, \UU = \UU \, \CC' \, 
\end{equation}
plays the role of a $\mathbb{C}$-\emph{anti}linear charge conjugation in our Fock basis in the sense that 
\begin{equation}
    \CC \bigl( \tilde\Psi_n \bigr)\propto \tilde\Psi_n^+ \, .
\end{equation}
We moreover have $\CC^2 = \UU^2 =1$.

\paragraph{Fock basis.}
The canonical fermionic modes \eqref{sm:Psi_tilde} allow us to define a suitable fermionic Fock basis. Starting from the Fock vacuum $\ket{\varnothing}$, characterised by $\tilde{\Psi}_n \, \ket{\varnothing} = 0$ for all $n$, we define
\begin{equation} \label{freespectrum_app}
	\ket{n_1,\dots,n_M} \equiv \tilde{\Psi}^+_{n_1} \cdots \tilde{\Psi}^+_{n_M}\, \ket{\varnothing} \,, 
	\qquad 0\leqslant n_1 < \cdots < n_M <N \, ,
\end{equation}
as in \eqref{freespectrum}. 
The discrete transformations that we just introduced act on this Fock basis as follows. We set
\begin{equation}
    \PP\,\ket{\varnothing} = \TT\,\ket{\varnothing} = \ket{\varnothing} \, , \qquad \CC\,\ket{\varnothing} = \ket{\bar{\varnothing}} \equiv \ket{0,\dots,N-1} \, .
\end{equation}
Then the above expressions for $\PP$, $\TT$ and $\CC$ in the basis of the canonical fermionic modes \eqref{sm:Psi_tilde} readily induce a transformation of the Fock space as
\begin{equation}
    \PP\bigl(F \, \ket{\varnothing}\bigr) \equiv \PP(F)\, \ket{\varnothing} \, , \qquad
    \TT\bigl(F \, \ket{\varnothing}\bigr) \equiv \TT(F) \, \ket{\varnothing} \, , \qquad
    \CC\bigl(F \, \ket{\varnothing}\bigr) \equiv \CC(F) \, \ket{\bar{\varnothing}} \, .
\end{equation}
As such, we obtain
\begin{equation}
	\begin{aligned}
	\PP \, \ket{n_1,\dots,n_M} &  \propto \TT \, \ket{n_1,\dots,n_M} \propto \ket{N-n_1,\dots,N-n_M} \, , \\
	\CC \, \ket{n_1,\dots,n_M} & \propto \ket{\bar{n}_1,\dots,\bar{n}_{N-M}} \, , \qquad \{\bar{n}_1,\dots,\bar{n}_{N-M}\} \equiv \{0,1,\dots,N-1\} \setminus \{n_1,\dots,n_M\} \, .
	\end{aligned}
\end{equation}
In the first line we identify mode numbers $N \equiv 0$. We emphasise once more that `$\propto$' entails equality up to a phase. When quantum numbers in a ket are not properly ordered, it is understood that the state is defined from the corresponding vector basis \eqref{freespectrum} with the appropriate sign. 

In particular, the Fock states are all eigenvectors for $\PP\,\TT$. This can be made explicit using \eqref{PT on psi} as
\begin{equation}
    \PP \, \TT \, \ket{n_1,\dots,n_M} =         (-1)^{\delta_{n_1,0}} \, \omega^{n_1 + \dots + n_M}\ket{n_1,\dots,n_M} , \qquad 
	\qquad 0\leqslant n_1 < \cdots < n_M <N \, ,
\end{equation}
where $\delta$ stands for the Kronecker symbol. Note that this result is compatible with the fact that $\PP \TT$ can only have unimodular eigenvalues as it is antilinear and satisfies $(\PP\, \TT)^2=1$.

\subsubsection{The hamiltonians in terms of fermionic modes}

\noindent
It is now relatively straightforward to write the hamiltonians in terms of the eigenmodes of the quasi-translations. First, we can use \eqref{telescoped h^L} and \eqref{gTL} to express $\HH^\LL$ in terms of the two site fermions as 
\begin{equation} \label{Ham_Lg}
	\begin{aligned} 
	\HH^\LL = {} &
	\frac{\I}{2} \ \ \, \sum_{\mathllap{1\leqslant}i\leqslant j\mathrlap{<N}} s_{ij} \, h^\LL_{ij} \, \bigl(g_{j}^+ \, g_i + (-1)^{i-j} g_i^+ \, g_{j} \bigr) \, 
	= \frac{\I}{2} \sum_{i,j=1}^{N-1} \! s_{ij} \, h^\LL_{ij} \, g_{j}^+ \, g_i \, , 
	\end{aligned}
\end{equation}
where we used the symmetry property \eqref{notation hij} of $h_{ij}^\LL$ in the second equality. Passing to the new fermionic modes yields the quadratic expression 
\begin{equation} \label{psiquad}
	\HH^\LL = \sum_{n=1}^{N-1} \varepsilon^{\LL}_{n} \, \tilde{\Psi}_n^+ \, \tilde{\Psi}_n \,, \qquad 
	\varepsilon^{\LL}_n = \frac{(-1)^{n+1}}{2}  \sum_{i,j=1}^{N-1} \! s_{ij} \, h^\LL_{ij} \; M^{-1}_{in} \, \bar{M}^{-1}_{jn} \, .
\end{equation}
The final equality, between the coefficients coming from~\eqref{Psig} and the chiral dispersion from~\eqref{epLodd}, is not trivial, but follows since it reproduces the one-magnon spectrum known from the parent model. While we do not have a \emph{direct} proof of this equality, we have checked it numerically for low $N$. We thus obtain the quadratic (free-fermionic) expression \eqref{H^L_ff} from the main text.

The next step is rewrite $\HH$. This can be done by rearranging the creation and annihilation operators in the anticommutators of nested commutators in \eqref{Hamioddfull}. In the process one gets quadratic terms (when $k=j+1$) as well as quartic terms. Let us split the hamiltonian accordingly,
 \begin{equation} \label{eq:H=H2+H4}
 	 \HH = \HH_2+ \HH_4 \,.
\end{equation}
Let us combine the coefficients $h^{\LL,\RR}_{ij;kl}$ from \eqref{eq:hLRijkl} into
\begin{equation}
	h_{ij;kl} \equiv  h^\LL_{ij;kl} + h^\RR_{ij;kl} \, .
\end{equation}
Then the quadratic terms read 
\begin{equation}
	\begin{gathered}
	\HH_2 = -\frac{1}{4N} \sum_{i\leqslant j}^{N-1} s_{ij}\, h_{ij} \, \bigl(g_j^+ g_i + (-1)^{j-i+1} g_i^+ g_j \bigr) \,
	= -\frac{1}{4N} \sum_{i, j=1}^{N-1} \! s_{ij}\, h_{ij} \; g_j^+ g_i \,, \\
	h_{ij} \equiv \sum_{k=i}^{j-1} h_{ik;k+1,j} = (-1)^{i-j+1} \, {h}_{ji} \, ,
	\end{gathered}
\end{equation}
and
\begin{equation} \label{psi2quad}
	\HH_2=\sum_{n=1}^{N-1} \varepsilon_{n} \, \tilde{\Psi}_n^+ \, \tilde{\Psi}_n \, ,\qquad 
	\varepsilon_n = \frac{(-1)^n \, \I}{4N} \sum_{i,j=1}^{N-1} \! s_{ij} \, h_{ij} \, M^{-1}_{in} \bar{M}^{-1}_{jn} \,.
\end{equation}
This gives the first part of \eqref{ham} in the main text. As before, we do not have an independent proof of the final equality, but it follows from the parent model and has been checked numerically. Regarding the quartic hamiltonian, after reordering the different contributions we get
\begin{equation}
	\HH_4 = \frac{1}{2N} \! \sum_{i, j<k, l}^{N-1} \!\! s_{ij} \, s_{kl} \, h_{ij;kl} \; g_j^+ g_l^+ g_i \, g_k \, , \quad
	h_{ij;kl} = (-1)^{l-k} \, h_{ij;lk} = (-1)^{i-j} \, h_{ji;kl} = (-1)^{i-j+l-k} \, h_{ji;lk} \, ,
\end{equation}
where the sum is over all $1\leqslant i,j,k,l <N$ such that $\max(i,j) < \min(k,l)$. After the transformation  \eqref{Psig}--\eqref{sm:Psi_tilde} this becomes
\begin{equation} \label{quartic_part}
	\HH_4 = \!\!\!\!\! \sum_{m<n,r<s}^{N-1} \!\!\!\!\! \tilde{V}_{mn;rs} \, \tilde{\Psi}_m^+ \, \tilde{\Psi}_n^+ \, \tilde{\Psi}_r \, \tilde{\Psi}_s \,, \quad 
	\tilde{V}_{mn;rs} \equiv \frac{1}{a_m \, a_n \, a_r \, a_s} \, \frac{2}{N} \! \sum_{i,j<k,l}^{N-1} \!\! s_{ij} \, s_{kl} \, h_{ij;kl} \, \bar{M}^{-1}_{j[m|} \, \bar{M}^{-1}_{l|n]} \, M^{-1}_{i[r|} \, M^{-1}_{k|s]} \, .
\end{equation}	
where the square brackets signify antisymmetrisation in the indices $n,m$ and $r,s$ (but not $l$ or $k$). Again, we have not been able to compute~\eqref{quartic_part} explicitly, and this time we do not know how to obtain a simplification from the parent model. However, numerical evaluation suggests that the coefficients simplify drastically: nonzero values occur when $m+n$ is odd and equals $r+s$, and are given by
\begin{equation} \label{eq:pot}
 	\tilde{V}_{mn;m+k,n-k} = (-1)^{k+1} \, 4 \,\delta_{m,\, \mathrm{odd}} \,, \qquad 0\leqslant k< \frac{n-m}{2} \, ,
\end{equation}
together with values for $k<0$ following from the symmetry $\tilde{V}_{mn;rs} = \tilde{V}_{rs;mn}$.

Note that these selection rules are \emph{stronger} than the condition~\eqref{eq:mtm chiral energy conserved} for the conservation of quasi-momentum and chiral energy, as required by the commutativity~\eqref{eq:comm_charges}. For example, when $N=9$ those charges take the same value for $(m,n)=(1,2)$ and $(r,s)=(5,7)$, but $m+n$ equals $r+s$ only $\mathrm{mod}\,N$, and indeed we observe $\tilde{V}_{12;57} = 0$. There are more and more such examples for larger $N$. The potential~\eqref{eq:pot} is corroborated by the two-particle spectrum, which is our next topic.

\subsection{Examples of the two-fermion spectrum for small systems} \label{sec: examples}

\noindent
To understand the hamiltonian~\eqref{eq:H=H2+H4} better let us describe the few-particle spectrum in the fermionic language, using the Fock basis~\eqref{freespectrum},
\begin{equation}
	\ket{n_1,\dots,n_M} = \tilde{\Psi}^+_{n_1} \dots \tilde{\Psi}^+_{n_M} \, \ket{\varnothing} \, .
\end{equation}
These are eigenvectors for the quasi-translation~$\G$ and $\HH^\LL$, but not necessarily for $\HH$ because of its quartic part. Let us show for $M\leqslant 2$ how it fits with the description of the spectrum via motifs coming from the parent model.
Observe that, cf.~Fig.~\ref{fg:dispersions},
\begin{equation} \label{motifdesc}
	\varepsilon^\LL_n + \varepsilon^\LL_{n+1} = \varepsilon^\LL_{2\mspace{1mu}n+1\,\mathrm{mod}\,N}\, .
\end{equation}

The Fock vacuum $\ket{\varnothing}$ is an eigenvector for all conserved charges, with vanishing quasi-momentum and energies. It is labelled by the empty motif, which has degeneracy $N+1$. All its descendants come from the global symmetry only.

The one-particle spectrum consists of the global descendant $\ket{0} = \F_1^+ \, \ket{\varnothing}$ along with $N-1$ Fock states $\ket{n}$ that correspond to the motifs $\{n\}$. They have quasi-momentum $p=2\pi\,n/N$ and energies given by the dispersions~\eqref{eLodd}.
For the $\binom{N}{2}$-dimensional two-particle sector we see the quartic part~$\HH_4$ in action, and some Fock states cease to be eigenstates for $\HH$. 
The two-fermion spectrum is admits the following explicit description.
\begin{itemize}[leftmargin=\parindent,topsep=1ex,itemsep=-.5ex]
	\item By global symmetry, the descendant $\ket{0,n} \propto \F_1^+ \, \ket{n}$, cf.~\eqref{zero_modes}, is an $\HH$-eigenstate belonging to the motif $\{n\}$.
	\item Any $\ket{m,n}$ with $0<m<n<N$ and $n-m$ even is protected ($\tilde{V}=0$) by the selection rules. It is an $\HH$-eigenstate with motif $\{m,n\}$. 
	\item Any $\ket{1,2\,n'}$ is mixed with $\ket{1+k,2\,n'-k}$ by \eqref{pot}:
	\begin{itemize}[leftmargin=\parindent,topsep=-.5ex,itemsep=0ex]
		\item For $\ket{1,2}$ only $k=0$ contributes, so it is again an $\HH$-eigenstate, with energy $\varepsilon_1 + \varepsilon_2 + \tilde{V}_{12;12} = \varepsilon_3$. It is degenerate with $\ket{3}$ for all charges, cf.\ \eqref{motifdesc}, and belongs to the motif $\{3\}$: $\ket{1,2} \propto \widehat{\F}_1^+ \, \ket{3}$ for some extended-symmetry generator $\widehat{\F}_1^+$.
		\item All $\ket{1,2\,n'}$ with $n'>1$ mix with $\ket{1+k,2\,n'-k}$, $k>0$. Diagonalising this $n'\times n'$ block of $\HH$ gives eigenstates with `squeezed' motifs $\{1+k,2\,n'-k\}$, $0\leqslant k \leqslant n'-2$, plus a state that is proportional to $\widehat{\F}_1^+ \, \ket{2\,n'+1}$ or, if $2\,n'=N-1$, to $\F_2^+ \, \ket{\varnothing}$.
	\end{itemize} 
	\item Likewise for $\ket{N-2\,n'+1,N-1} \propto \PP \, \ket{1,2\,n'}$ by parity.
\end{itemize}
Besides actually diagonalising the blocks, and matching the result with the parent-model eigenstates at $\q=\I$, this gives the full two-particle spectrum. Note the statistical repulsion in action, `squeezing' adjacent modes to extended-symmetry descendants, cf.~\cite[\textsection4.1.6]{KL_22}.

In the following we work out the preceding explicitly for small systems, and link it to the description coming from the parent model in terms of the motifs~\eqref{sm:degeneracy}.
\medskip

\textit{$N=3$.} Here the explicit hamiltonians are~\eqref{H N3}. The 3 two-particle Fock states lie beyond the `equator' $M=N/2$. They are descendants for the global symmetry: $\ket{0,1} \propto \F_1^+ \, \ket{1}$, its parity-conjugate $\ket{0,2} \propto \F_1^+ \, \ket{2}$, and $\ket{1,2} \propto \F_2^+ \, \ket{\varnothing}$. In this case the Fock basis happens to be an eigenbasis for the conserved charges, including $\HH$, thus yielding a fermionic description of the full Hilbert space. This is illustrated in Fig.~\ref{fg:N=3}. 
\medskip

\DeclareRobustCommand\paritylink{\tikz[baseline={([yshift=-2*11pt*0.15]current bounding box.center)},xscale=0.2,yscale=.5]{ \draw[lightgray] (1.75,2.8) -- (1.75,2.9) -- (3.25,2.9) -- (3.25,2.8);}}
\DeclareRobustCommand\verticalline{\tikz[baseline={([yshift=-2*11pt*0.15]current bounding box.center)},scale=0.15]{ \draw[gray] (0,1.5) -- (0,-.5);}}
\DeclareRobustCommand\diagonalline{\tikz[baseline={([yshift=-2*11pt*0.15]current bounding box.center)},scale=.25]{ \draw[gray] (0,1.5) -- (.5,.5);}}
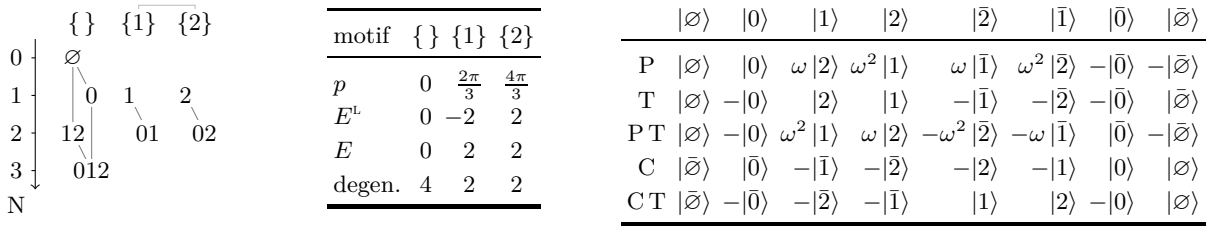
\begin{figure}[h]
	\centering
	\begin{tikzpicture}[scale=0.5,baseline={([yshift=-.5*11pt*0.13]current bounding box.center)}]
		\draw[<-] (-1,-2) -- (-1,2);
		\foreach \y in {-1.5,-.5,...,1.5} \draw (-1,\y) -- (-1.1,\y);
		\node at (-1.5,-2.4) {$\N$};
		\foreach \M in {0,...,3} \node at (-1.5,1.5-\M) {$\M$};
		\draw[gray] (0,1.5) node[black,inner sep=.5pt,fill=white]{$\varnothing$} 
			-- (0,-.5) node[black,inner sep=1.5pt,fill=white]{$12$}
			-- (.5,-1.5) node[black,inner sep=1.5pt,fill=white]{$012$} 
			-- (.5,.5) node[black,inner sep=1.5pt,fill=white]{$0$} 
			-- cycle;
		\draw[gray] (1.5,.5) node[black,inner sep=1.5pt,fill=white]{$1$} 
			-- (2,-.5) node[black,inner sep=2pt,fill=white]{$01$};
		\draw[gray] (3,.5) node[black,inner sep=1.5pt,fill=white]{$2$} 
			-- (3.5,-.5) node[black,inner sep=2pt,fill=white]{$02$};
		\node at (.25,2.4) {$\{\,\}$};
		\node at (1.75,2.4) {$\{1\}$};
		\node at (3.25,2.4) {$\{2\}$};
		\draw[lightgray] (1.75,2.8) -- (1.75,2.9) -- (3.25,2.9) -- (3.25,2.8);
	\end{tikzpicture}
	\qquad\qquad
	\begin{tabular}[c]{lccc} 
		\toprule
		motif & $\{\,\}$ & $\{1\}$ & $\{2\}$ \\
		\midrule
		$p$ & $0$ & $\frac{2\pi}{3}$ & $\frac{4\pi}{3}$ \\
		$E^\LL$ & $0$ & $\mathllap{-}2$ & $2$ \\
		$E$ & $0$ & $2$ & $2$ \\
		degen. & $4$ & $2$ & $2$ \\
		\bottomrule\hline
	\end{tabular}
	\qquad\quad
	\begin{tabular}[c]{ccccccccc} 
		\toprule
		& $\ket{\varnothing}$ & $\hphantom{-}\ket{0}$ & $\hphantom{\omega^2\,}\ket{1}$ & $\hphantom{\omega^2\,}\ket{2}$ & 
		$\hphantom{-\omega^2\,}\ket{\bar{2}}$ & $\hphantom{-\omega\,}\ket{\bar{1}}$ & $\hphantom{-}\ket{\bar{0}}$ & $\hphantom{-}\ket{\bar{\varnothing}}$
		\\
		\midrule
		$\PP$ & $\ket{\varnothing}$ & $\hphantom{-}\ket{0}$ & $\hphantom{\omega^2\,}\mathllap{\omega \,}\ket{2}$ & $\omega^2\,\ket{1}$ & 
		$\hphantom{-\omega^2\,}\mathllap{\omega \,} \ket{\bar{1}}$ & $\hphantom{-\omega\,}\mathllap{\omega^2\,}\ket{\bar{2}}$ & $-\ket{\bar{0}}$ & $-\ket{\bar{\varnothing}}$ \\
		$\TT$ & $\ket{\varnothing}$ & $-\ket{0}$ & $\hphantom{\omega^2\,}\ket{2}$ & $\hphantom{\omega^2\,}\ket{1}$ &
		$\hphantom{-\omega^2\,}\mathllap{-}\ket{\bar{1}}$ & $\hphantom{-\omega\,}\mathllap{-}\ket{\bar{2}}$ & $-\ket{\bar{0}}$ & $\hphantom{-}\ket{\bar{\varnothing}}$ \\
		$\PP\,\TT$ & $\ket{\varnothing}$ & $-\ket{0}$ & $\omega^2\,\ket{1}$ & $\hphantom{\omega^2\,}\mathllap{\omega\,}\ket{2}$ &
		$-\omega^2\,\ket{\bar{2}}$ & $-\omega\,\ket{\bar{1}}$ & $\hphantom{-}\ket{\bar{0}}$ & $-\ket{\bar{\varnothing}}$ \\
		$\CC$ & 
		$\ket{\bar{\varnothing}}$ & $\hphantom{-}\ket{\bar{0}}$ & $\hphantom{\omega^2\,}\mathllap{-}\ket{\bar{1}}$ & $\hphantom{\omega^2\,}\mathllap{-}\ket{\bar{2}}$ & $\hphantom{-\omega^2\,}\mathllap{-}\ket{2}$ & $\hphantom{-\omega\,}\mathllap{-}\ket{1}$ & $\hphantom{-}\ket{0}$ & $\hphantom{-}\ket{\varnothing}$ \\
		$\CC\,\TT$ & 
		$\ket{\bar{\varnothing}}$ & $-\ket{\bar{0}}$ & $\hphantom{\omega^2\,}\mathllap{-}\ket{\bar{2}}$ & $\hphantom{\omega^2\,}\mathllap{-}\ket{\bar{1}}$ & $\hphantom{-\omega^2\,}\ket{1}$ & $\hphantom{-\omega\,}\ket{2}$ & $-\ket{0}$ & $\hphantom{-}\ket{\varnothing}$ 
		\\
		\bottomrule\hline
	\end{tabular}
	\caption{\textit{Left.} The structure of the Hilbert space for $N=3$. Each label represents an eigenstate, which in this case are just Fock states. The vertical axis records the fermion number. The eigenspaces are labelled by motifs, with the parity-conjugate pair linked by a `\,\paritylink\,'. $\CC\,\TT$ acts by reflection in the `equator' $\N = N/2$ up to a possible sign. The lines `\,\diagonalline\,' and `\,\verticalline\,' indicate the action of the global-symmetry generators $\F_1^\pm$ and $\F_2^\pm$, respectively. \textit{Middle.} The corresponding spectrum: quasi-momenta (note that $\frac{4\pi}{3} = -\frac{2\pi}{3} \, \mathrm{mod}\,2\pi$), energies, and degeneracies. \textit{Right.} The action of the discrete symmetries on the Fock basis. The bar denotes the complement, e.g.\ $\ket{\bar{0}} = \ket{12}$, $\ket{\bar{\varnothing}}=\ket{012}$. For parity note that reordering may give a sign, e.g.\ $\PP\,\ket{\bar{\varnothing}} = \ket{210} = -\ket{\bar{\varnothing}}$.}
	\label{fg:N=3}
\end{figure}

\textit{$N=5$.} 
We seek 10 two-particle eigenstates. Their fermionic description is as follows.
\begin{itemize}[topsep=1ex,itemsep=-.5ex]
	\item The 4 Fock states $\ket{0,n} \propto \F_1^+ \, \ket{n}$ with $1\leqslant n \leqslant 4$ are global descendants from the one-particle sector, belonging to the motifs $\{n\}$.
	\item The 2 parity-conjugate Fock states $\ket{1,3}$ and $\ket{2,4}$ are protected by the selection rules. Indeed, `squeezing' produces coinciding mode numbers, which is not allowed for fermions, and $\tilde{V}_{13;13} = \tilde{V}_{24;24} = 0$. Both are eigenvectors for all conserved charges, corresponding to the motifs $\{1,3\}$ and $\{2,4\}$.
	\item The remaining two-particle eigenstates feel the quartic interaction:
	\begin{itemize}[topsep=-.5ex]
		\item The Fock state $\ket{1,2}$ is an eigenstate for $\HH_4$ and therefore $\HH$. From the parent model we know that it is an extended-symmetry descendants of the motif $\{3\}$: $\ket{1,2} \propto \widehat{\F}_1^+ \, \ket{3}$ for some generator $\widehat{\F}_1^+$ of the extended-symmetry algebra. This is consistent with the chiral dispersion thanks to \eqref{motifdesc}, while for $\HH$ it relies on $\tilde{V}_{12;12} = -4$ contributing to $\varepsilon_1 + \varepsilon_2 + \tilde{V}_{12;12} = 4 + 2 - 4 = 2 = \varepsilon_3$.
		\item The Fock states $\ket{1,4}$ and $\ket{2,3}$ are mixed by $\HH_4$. The parent model tells us that diagonalising this $2\times2$ block of $\HH$ produces two eigenvectors that we denote by $\ket{1,4}'$, with motif $\{1,4\}$, and $\ket{2,3}' \propto \F_2^+ \, \ket{\varnothing}$ a global descendant of the empty motif.
		\item The Fock state $\ket{3,4}$ is the parity-conjugate  of $\ket{1,2}$, and is an extended-symmetry descendant of the motif $\{2\}$; note that $2=3+4 \, \mathrm{mod}\,5$
	\end{itemize} 
\end{itemize}
Note that the one-particle motifs $\{2\}$ and $\{3\}$ both have two descendants in the two-particle sector, one for the global symmetry and one for the extended symmetry, while the empty motif, $\{1\}$ and $\{4\}$ each only have a global descendant in this sector. This matches the description known from the parent model. In this case the spectrum with $\leqslant 2$ particles determines the whole Hilbert space by particle-hole symmetry, so we again have complete fermionic description, shown in Fig.~\ref{fg:N=5}.
\medskip

\DeclareRobustCommand\dottedline{\tikz[baseline={([yshift=-2*11pt*0.15]current bounding box.center)},scale=.15]{ \draw[densely dotted,gray] (7.5,1.5) -- (9,.5);}}
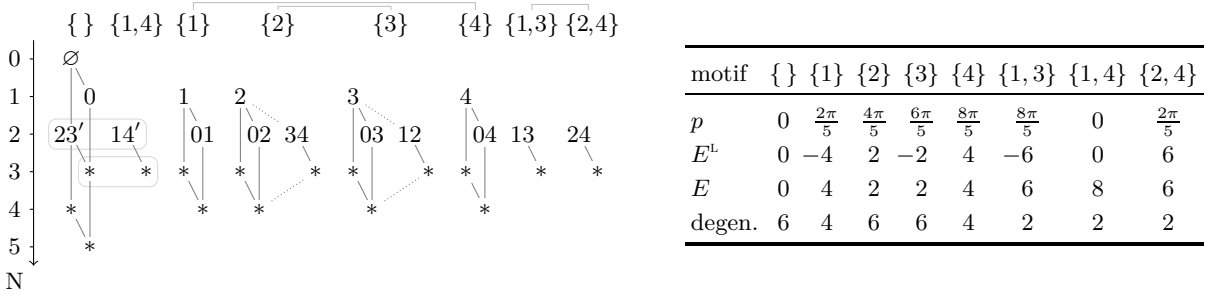
\begin{figure}[h]
	\centering
	\begin{tikzpicture}[scale=0.5,baseline={([yshift=-.5*11pt*0.13]current bounding box.center)}]
		\draw[<-] (-1,-3) -- (-1,3);
		\foreach \y in {-2.5,-1.5,...,2.5} \draw (-1,\y) -- (-1.1,\y);
		\node at (-1.5,-3.4) {$\N$};
		\foreach \M in {0,...,5} \node at (-1.5,2.5-\M) {$\M$};
		\draw[rounded corners=3pt,gray!30] (0-.6,.5+.4) rectangle (1.5+.5,.5-.4);
		\draw[rounded corners=3pt,gray!30] (.5-.3,-.5+.4) rectangle (2+.3,-.5-.4);
		\draw[gray] (0,2.5) -- (0,-1.5) (.5,-2.5) -- (.5,1.5)
			(0,2.5) node[black,inner sep=.5pt,fill=white]{$\varnothing$} 
			-- (.5,1.5) node[black,inner sep=1.5pt,fill=white]{$0$} 
			(0,.5) node[black,inner sep=1pt,fill=white]{$\mspace{-2mu}23\smash{'}$} 
			-- (.5,-.5) node[black,inner sep=1.5pt,fill=white]{$*$}
			(0,-1.5) node[black,inner sep=1.5pt,fill=white]{$*$} 
			-- (.5,-2.5) node[black,inner sep=1.5pt,fill=white]{$*$};
		\draw[gray] (1.5,.5) node[black,inner sep=1pt,fill=white]{$\mspace{-2mu}14\smash{'}$} 
			-- (2,-.5) node[black,inner sep=1.5pt,fill=white]{$*$};
		\draw[gray] (3,1.5) node[black,inner sep=1.5pt,fill=white]{$1$} 
			-- (3,-.5) node[black,inner sep=1.5pt,fill=white]{$*$}
			-- (3.5,-1.5) node[black,inner sep=1.5pt,fill=white]{$*$}
			-- (3.5,.5) node[black,inner sep=2pt,fill=white]{$01$}
			-- cycle;
		\draw[densely dotted,gray] (4.5,1.5) -- (6,.5);
		\draw[densely dotted,gray] (5,-1.5) -- (6.5,-.5);
		\draw[gray] (4.5,1.5) node[black,inner sep=1.5pt,fill=white]{$2$} 
			-- (4.5,-.5) node[black,inner sep=1.5pt,fill=white]{$*$}
			-- (5,-1.5) node[black,inner sep=1.5pt,fill=white]{$*$}
			-- (5,.5) node[black,inner sep=2pt,fill=white]{$02$}
			-- cycle;
		\draw[gray] (6,.5) node[black,inner sep=1.5pt,fill=white]{$34$} 
			-- (6.5,-.5) node[black,inner sep=2pt,fill=white]{$*$};
		\draw[densely dotted,gray] (7.5,1.5) -- (9,.5);
		\draw[densely dotted,gray] (8,-1.5) -- (9.5,-.5);
		\draw[gray] (7.5,1.5) node[black,inner sep=1.5pt,fill=white]{$3$} 
			-- (7.5,-.5) node[black,inner sep=1.5pt,fill=white]{$*$}
			-- (8,-1.5) node[black,inner sep=1.5pt,fill=white]{$*$}
			-- (8,.5) node[black,inner sep=2pt,fill=white]{$03$}
			-- cycle;
		\draw[gray] (9,.5) node[black,inner sep=1pt,fill=white]{$12$} 
			-- (9.5,-.5) node[black,inner sep=1pt,fill=white]{$*$};
		\draw[gray] (10.5,1.5) node[black,inner sep=1.5pt,fill=white]{$4$} 
			-- (10.5,-.5) node[black,inner sep=2pt,fill=white]{$*$}
			-- (11,-1.5) node[black,inner sep=2pt,fill=white]{$*$}
			-- (11,.5) node[black,inner sep=2pt,fill=white]{$04$}
			-- cycle;
		\draw[gray] (12,.5) node[black,inner sep=1.5pt,fill=white]{$13$} 
			-- (12.5,-.5) node[black,inner sep=2pt,fill=white]{$*$};
		\draw[gray] (13.5,.5) node[black,inner sep=1.5pt,fill=white]{$24$} 
			-- (14,-.5) node[black,inner sep=2pt,fill=white]{$*$};
		\node at (.25,3.4) {$\{\,\}$};
		\node at (1.75,3.4) {$\{1\mspace{-1mu},\mspace{-2mu}4\}$};
		\node at (3.25,3.4) {$\{1\}$};
		\node at (5.5,3.4) {$\{2\}$};
		\node at (8.5,3.4) {$\{3\}$};
		\node at (10.75,3.4) {$\{4\}$};
		\node at (12.25,3.4) {$\{1\mspace{-1mu},\mspace{-2mu}3\}$};
		\node at (13.75,3.4) {$\ \{2\mspace{-1mu},\mspace{-2mu}4\}$};
		\draw[lightgray] (3.25,3.8) -- (3.25,4) -- (10.75,4) -- (10.75,3.8);
		\draw[lightgray] (5.5,3.8) -- (5.5,3.9) -- (8.5,3.9) -- (8.5,3.8);
		\draw[lightgray] (12.25,3.8) -- (12.25,3.95) -- (13.75,3.95) -- (13.75,3.8);
	\end{tikzpicture}
	\qquad
	\begin{tabular}[c]{lcccccccc} 
		\toprule
		motif & $\{\,\}$ & $\{1\}$ & $\{2\}$ & $\{3\}$ & $\{4\}$ & $\{1,3\}$ & $\{1,4\}$ & $\{2,4\}$ \\
		\midrule
		$p$ & $0$ & $\frac{2\pi}{5}$ & $\frac{4\pi}{5}$ & $\frac{6\pi}{5}$ & $\frac{8\pi}{5}$ & $\frac{8\pi}{5}$ & $0$ &  $\frac{2\pi}{5}$ \\
		$E^\LL$ & $0$ & $\mathllap{-}4$ & $2$ & $\mathllap{-}2$ & $4$ & $\mathllap{-}6$ & $0$ & $6$ \\
		$E$ & $0$ & $4$ & $2$ & $2$ & $4$ & $6$ & $8$ & $6$\\
		degen. & $6$ & $4$ & $6$ & $6$ & $4$ & $2$ & $2$ & $2$ \\
		\bottomrule\hline
	\end{tabular}
	\caption{\textit{Left.} The structure of the Hilbert space for $N=5$. The states with $\leqslant2$ particles are labelled by their fermionic mode numbers, with a prime for the two eigenstates resulting from diagonalising a $2\times 2$ block of $\HH$, also indicated. Global symmetry and parity are shown as in Fig.~\ref{fg:N=3}. Dotted lines like `\,\dottedline\,' represent the action of extended-symmetry generators $\widehat{\F}_1^\pm$, joining together representations of the global symmetry algebra. \textit{Right.} The corresponding spectrum.}
	\label{fg:N=5}
\end{figure}

\textit{$N=7$.} The two-particle sector has dimension~$21$. Its states have the following fermionic description.
\begin{itemize}[topsep=1ex,itemsep=-.5ex]
	\item The $6$ global descendants $\ket{0,n} \propto \F_1^+ \, \ket{n}$, $1\leqslant n \leqslant 6$, come from the motifs $\{n\}$.
	\item The $6$ Fock states $\ket{1,3}$, $\ket{1,5}$, $\ket{2,4}$, $\ket{2,6}$, $\ket{3,5}$, $\ket{4,6}$ are protected by the selection rules, corresponding to the motifs $\{1,3\}$ etc.
	\item The other two-particle eigenstates correspond to blocks:
	\begin{itemize}[topsep=-.5ex]
		\item The Fock state $\ket{1,2} \propto \widehat{\F}_1 \, \ket{3}$ is an extended-symmetry descendant for the motif $\{3\}$.
		\item The $2$ Fock states $\ket{1,4}$ and $\ket{2,3}$ are mixed. Its diagonalisation produces eigenstates $\ket{1,4}'$, corresponding to the motif $\{1,4\}$, along with $\ket{2,3}' \propto \widehat{\F}_1 \, \ket{5}$ belonging to $\{5\}$.
		\item The $3$ Fock states $\ket{1,6},\ket{2,5},\ket{3,4}$ correspond to a $3\times3$ block of $\HH$. Diagonalisation yields eigenstates with motifs $\{1,6\}$ and $\{2,5\}$, plus a global descendant $\ket{3,4}' \propto \F_2^+ \, \ket{\varnothing}$ belonging to the empty motif.
		\item The $2$ Fock states $\ket{3,6},\ket{4,5}$ give, by diagonalisation, an eigenstate with motif $\{3,6\}$, and $\ket{4,5}' \propto \widehat{\F}_1 \, \ket{2}$.
		\item The Fock state $\ket{5,6} \propto \widehat{\F}_1 \, \ket{4}$ is again an extended-symmetry descendant for the motif $\{4\}$.
	\end{itemize}
\end{itemize}
Observe that the blocks of the same size are related by parity, and the Fock- and eigenstates in the middle block are parity-selfconjugate. The one-fermion Fock states with motifs $\{2\},\{3\},\{4\},\{5\}$ have a global as well as an extended-symmetry descendant. From this length onwards, the sectors with $\leqslant 2$ particles do no longer determine the full Hilbert space. In addition, we start to get `accidental degeneracies' between different motifs: from Fig.~\ref{fg:N=7} we see that $\{1\}$ and $\{3,5\}$ have the same values for the conserved charges, and by parity the same is true for $\{6\}$ and $\{2,4\}$.
\medskip

\begin{figure}[h]
	\begin{tabular}[c]{lccccccccccccccccc} 
		\toprule
		motif & $\{\,\}$ & $\{1\}$ & $\{2\}$ & $\{3\}$ & $\{4\}$ & $\{5\}$ & $\{6\}$ & $\{1,3\}$ & $\{1,4\}$ & $\{1,5\}$ & $\{1,6\}$ & $\{2,4\}$ & $\{2,5\}$ & $\{2,6\}$ & $\{3,5\}$ & $\{3,6\}$ & $\{4,6\}$ \\
		\midrule
		$p$ & $0$ & $\frac{2\pi}{7}$ & $\frac{4\pi}{7}$ & $\frac{6\pi}{7}$ & $\frac{8\pi}{7}$ & $\frac{10\pi}{7}$ & $\frac{12\pi}{7}$ & $\frac{8\pi}{7}$ & $\frac{10\pi}{7}$ & $\frac{12\pi}{7}$ & $0$ & $\frac{12\pi}{7}$ & $0$ & $\frac{2\pi}{7}$ & $\frac{2\pi}{7}$ & $\frac{4\pi}{7}$ & $\frac{6\pi}{7}$  \\
		$E^\LL$ & $0$ & $\mathllap{-}6$ & $2$ & $\mathllap{-}4$ & $4$ & $\mathllap{-}2$ & $6$ & $\mathllap{-}10$ & $\mathllap{-}2$ & $\mathllap{-}8$ & $0$ & $6$ & $0$ & $8$ & $\mathllap{-}6$ & $2$ & $10$ \\
		$E$ & $0$ & $6$ & $2$ & $4$ & $4$ & $2$ & $6$ & $10$ & $10$ & $8$ & $12$ & $6$ & $4$ & $8$ & $6$ & $10$ & $10$ \\
		degen. & $8$ & $6$ & $10$ & $12$ & $12$ & $10$ & $6$ & $4$ & $6$ & $6$ & $4$ & $6$ & $8$ & $6$ & $6$ & $6$ & $4$ \\
		\bottomrule\hline
	\end{tabular}
	\caption{The spectrum for $N=7$ in the sectors of the Fock space with $\leqslant 2$ particles.} \label{fg:N=7}	
\end{figure}

Continuing in this way one arrives at the general description of the two-particle sector as in the main text.

\subsection{Summary of open conjectures}

\noindent
While some key properties of our fermionic model are related to the integrability of the parent model, most of the fermionic features presented in the main text are special to the free-fermion point $\q=\I$ nand therefore completely new. We are able to establish most of these features rigorously. Here we summarise which properties that were mentioned above have so far eluded a proper proof and remain conjectures.

Recall that the reason why all the results we established in \textsection\ref{sec: model at i} for our parent model specialised at $\q = \I$ carry over to the new fermionic setting of \textsection\ref{sec_sm:fermionic_rep} hinges on the formal statement that these two representations of the TL algebra are isomorphic. This will be established in \textsection\ref{sec: XXZ vs ferm}. One of the major new properties of the fermionic model compared to the parent model is the symmetry \eqref{Hfull=0 at q=i} of $\HH^{\LL}$ under parity. We will establish this using complex analysis in \textsection\ref{sec: proof vanishing H}.

In \textsection\ref{sec:techn_G_fermions} we will give a thorough analysis of the action of the quasitranslation $\G$ on the basis $f_k^+$. This provides a solid basis for our various statements about the Fourier modes $\tilde{\Phi}_k$. The only exception is the simple formula \eqref{explicit alpha} for anticommutation relations between the $\q$-translated fermions \eqref{defPsi}, which we have not managed to prove completely. The anticommutation relations \eqref{explicit alpha} are the \emph{main conjecture} of this paper. Although it can be straightforwardly derived for the first few $n$ using the explicit action \eqref{linearity of G on the g} of $\G$ on the fermions $g_{k}$, we have not been able to establish it properly for all $n$ and $N$. As we noted, upon Fourier transforming, this conjecture is equivalent to formula \eqref{tilde alpha}. Tantalisingly, we managed to establish \eqref{tilde alpha} \emph{up to a possible sign}; this is the content of the discussion in \textsection\ref{sec: proof anticomm}.

Some of our results rely on this main conjecture:
\begin{itemize}
	\item the explicit formula \eqref{eq: conj Mn1} for $M_{n1}$ and $\bar{M}_{n1}$ which would provide an exact expression for the phase appearing in the $\PP$-transformation of $\q$-translated fermions as explained in \textsection\ref{sec: C transf};
	\item the coefficients in the diagonal expressions \eqref{psiquad} of $\HH^{\LL}$ and \eqref{psi2quad} of the quadratic part of $\HH$ in the Fourier-transformed fermions.
\end{itemize}
More precisely, these formulas depend on \eqref{tilde alpha} and are thus firmly established up to a possible sign. 

Our \emph{second conjecture} is the formula \eqref{pot} for the quartic potential, including the stronger selection rule, which is entirely based on numerical observations. Note that \textsection\ref{sec: examples} provides examples for low $N$ that illustrate with the discussion in the main text regarding the consequence of \eqref{pot} on the two-particle spectrum.

\subsection{Technical proofs} \label{app:techn_prfs}

\noindent
This section contains various technical proofs of claims from the preceding sections.

\subsubsection{Action of $\G$ on the fermions} \label{sec:techn_G_fermions}

\noindent
The aim of this part is to give further details about the results mentioned in \textsection\ref{sec:fermions} that give intel about the action of the conjugation operator $x \mapsto \G^{-1} \, x \, \G$ on the one-particle fermionic operators, i.e.\ the $N$-dimensional space spanned by $f_1^+, \dots, f_N^+$ in the case of creation operators and the one spanned by $f_1, \dots, f_N$ in the case of annihilation operators.
In order to fully appreciate the symmetry in formulas between creation and annihilation operators and to render the notation more compact we will use the notation $f_i^- \equiv f_i$ for annihilation operators, and similarly for $g_i^-$ et cetera, in the following.
\medskip

\paragraph{Ingredients.} 
In the main text and in section \textsection\ref{sec_sm:fermionic_rep} we have separately treated the two-site fermions $g_i^\pm$ with $1\leqslant i<N$ and
the generators of the global symmetry $\Phi_0^\pm=\sum_{i=1}^Nf_i^\pm$, because they play different roles for the hamiltonians: the former appear in the expression of the hamiltonians, while the latter are symmetries. In the following a more uniform treatment will be useful, 
so we complete the family $\bigl(g_{1}^{\pm},\dots ,g_{N-1}^{\pm}\bigr)$ defined in \eqref{gfer} by 
\begin{equation} \label{g_N}
	g_{N}^{\pm} \equiv f_{N}^{\pm} + f_{1}^{\pm} \, .
\end{equation}
Since $N$ is odd, we get an alternative basis $\bigl(g_{1}^{\pm},\dots ,g_{N-1}^{\pm},g_N^{\pm}\bigr)$ for the span of $(f_{1}^{\pm},\dots ,f_{N}^{\pm})$: for example, we have
\begin{equation} \label{f_1 in the g}
    f_1^\pm = -\frac{1}{2} \sum_{i=1}^{N} (-1)^i \, g_i^\pm \, .
\end{equation}

Now consider the quasi-translation. Since $e_i^2 = 0$ we have $(1+t_i \,e_i)^{-1}=(1-t_i \,e_i) $, so
\begin{equation} \label{qshiftinv at i}
	\G^{-1} = (1-t_{1}\, e_{1}) \cdots (1-t_{N-1} \, e_{N-1}) \, .
\end{equation}
To work out $\G^{-1} \, f_{i}^{\pm} \, \G$, it is natural to compute the result of successive conjugations by $(1-t_{j} \, e_{j})$ in increasing order. These computations can be performed with the commutation relations 
\begin{equation}  \label{fpmwithTL}
    \bigl[e_j,f_j^\pm\bigr] = \pm(-1)^jg_j^\pm \,, \quad
    \bigl[e_j,f_{j+1}^\pm\bigr] = \mp(-1)^{j}g_j^\pm\,, \qquad
    1 \leqslant j \leqslant N-1 \,  .
\end{equation}
Note that we retrieve in particular the analogous formulas for the $\bigl[e_j, g_k^\pm\bigr]$ that were given in \eqref{gwithTL}. In the process, the following quantities naturally arise:
\begin{equation} \label{def yk}
    y_i^{\pm} \equiv (1-t_{1} \, e_1) \cdots (1-t_{i-1}\, e_{i-1}) \, g_i^\pm \, (1 + t_{i-1} \, e_{i-1}) \cdots (1 + t_{1} \, e_1) \ , \qquad 1 \leqslant i \leqslant N \, ,
\end{equation}
which can be characterised by the recurrence relation 
\begin{equation} \label{rec yk}
    y_{i+1}^{\pm} = g_{i+1}^{\pm} \pm (-1)^i \, t_{i} \, y_{i}^{\pm} \, , \qquad y_1^\pm = g_1^\pm \, .
\end{equation}
Since 
  $t_N=0$,  one can formally extend the domain of $i$ in the above recurrence relation (but not in the definition \eqref{def yk})  by periodicity  to any integer value
  so that 
\begin{equation} \label{per yk}
y_{i+N}^\pm = y_i^\pm\, .
\end{equation}
An explicit formula for the $y_j^\pm$ follows by induction
\begin{equation} \label{explicit yk}
    y^{\pm}_{j} = \sum_{i=1}^{j} \, (\pm1)^{i-j} \, s_{i,j} \, t_{i,j} \, g_{i}^{\pm} \, , \qquad 
    1 \leqslant j \leqslant N \, ,
\end{equation}
where  $t_{i,j}$ was defined in \eqref{def t_kl SM} and $s_{i,j}$ in \eqref{sij sign}. Observe that $ y_{N}^\pm \neq 0$. It is now clear that the $y_{i}^{\pm}$ form another basis that is triangular in the $g_{i}^{\pm}$.
In particular, we have 
\begin{equation} \label{yk leading term}
    y_{i}^{\pm} = f_{i+1}^{\pm} + \text{lower} \ , \qquad 1 \leqslant i < N ,
\end{equation}
where `$\text{lower}$' stands for terms that are supported by $f_{1}^{+},\dots ,f_{i}^{+}$.

\medskip
\paragraph{Action of quasi-translation.}
Using the ingredients introduced above, a direct calculation leads to the main formula
\begin{equation} \label{Ginv on the fk}
    \G^{-1} \, f_{i}^\pm \, \G = f_{i}^{\pm} \pm (-1)^{i-1} \, \bigl(t_{i-1} \, y_{i-1}^{\pm} + t_{i} \, y_{i}^\pm \bigr) = \mp (-1)^{i} \, t_{i} \, f_{i+1}^{\pm} + \text{lower} \,  
\end{equation}
with
\begin{align} \label{explicit Phi1}
    \Phi_2^\pm \equiv \G^{-1} \, f_{1}^\pm \, \G &= f^{+}_{1} \pm t_1 \, g^+_1 \, .
   \end{align}
By induction, the quasi-translated fermions $\Phi_{i}^{\pm}$ defined in \eqref{defPsi} are therefore triangular in the $f_{j}^{\pm}$:
\begin{equation} \label{triangularity of the phi}
	\Phi_{i} = s_{1 i} \, t_{1i} \, f_{i} + \text{lower} \ , \qquad \Phi_{i}^{+} = (-1)^{i-1} \, s_{1i} \, t_{1,i} \, f_{i}^{+} + \text{lower} \ , \qquad 1 \leqslant i \leqslant N \, .
\end{equation}
In particular, we get
\begin{equation}
    \Phi_N^\pm =  s_{1,N} \, t_{1,N} \, f_N^+  + \text{lower} = N f_N^+  + \text{lower}\, ,
\end{equation}
where in the last equality we used  the identity $s_{1,N}=(-1)^{(N-1)/2}$ and
\begin{equation}
    t_{1,N} \equiv t_1\dots t_{N-1}=(-1)^{(N-1)/2} \,\lim_{z \rightarrow 1} \frac{f(z^N)}{f(z)} =s_{1,N}\,N\, .
\end{equation}
Given that $f^\pm_N$ only appears in $\Psi_N^\pm$ we conclude that in 
the Fourier-transformed fermions we have
\begin{equation} \label{fourier fermions leading term}
    \tilde{\Phi}_n^\pm = f_N^\pm + \text{lower} \, , \qquad 1 \leqslant n \leqslant N \, .
\end{equation}
Similarly we can compute the action of the quasi-translations on the two-site fermions. It follows directly from \eqref{Ginv on the fk} and  \eqref{rec yk} that
\begin{align} \label{Ginv on gk - telescopic}
    \G^{-1} \, g_{i}^\pm \, \G &= g_i^\pm \pm (-1)^{i} \, \bigl(t_{i+1} \, y_{i+1}^\pm - t_{i-1} \, y^\pm_{i-1} \bigr) \\ \label{Ginv on gk - yk only}
    &= y_{i}^{\pm} \pm (-1)^{i} \, t_{i+1} \, y_{i+1}^{\pm}\\  \label{Ginv on gk - triangular}
    &= \bigl(1 + t_{i} \, t_{i+1}\bigr) y_{i}^{+} \pm (-1)^{i} \, t_{i+1} \, g_{i+1}^{+} \, .
\end{align}
Using \eqref{explicit yk} in the formula \eqref{Ginv on gk - triangular}  we obtain the expressions for  $\gamma,\bar{\gamma}$  in \eqref{linearity of G on the g}.
\medskip
\paragraph{Recursive equation for the Fourier-transformed fermions.}
Recall the definition of $ g_N^\pm$ from \eqref{g_N}.
We will establish a necessary condition for a one-particle fermionic operator that is a linear combination of the $g_i^+$, or of the $g_i^-$, to be an eigenmode for the conjugation by $\G$. We will treat both cases simultaneously and write $\psi^\pm \equiv \sum_{i=1}^{N} b_i^\pm \, g_i^{\pm}$. Then we consider
\begin{equation}
    \G \, \psi^\pm \, \G^{-1} = \rho_\pm \, \psi^\pm \, .
\end{equation}
To be able to apply our previous findings, it is convenient to rewrite it as $\G^{-1} \, \psi ^\pm\, \G = \rho_\pm^{\,-1} \psi^\pm$. Using formula \eqref{Ginv on gk - yk only} on the left-hand side and the recursion relation \eqref{rec yk} on the right-hand side, we then get the following equation in the basis of  $y_{i}^{\pm}$:
\begin{equation} 
    \sum_{i=1}^N b^\pm_{i} \, \bigl(y_{i}^{\pm} \pm (-1)^{i} \, t_{i+1} \, y_{i+1}^{\pm} \bigr) = \rho^{-1}_{\pm} \sum_{i=1}^N b^\pm_{i} \, \bigl( y_{i}^{\pm} \pm (-1)^{i} \, t_{i-1} \, y_{i-1}^{\pm} \bigr) \, .
\end{equation}
Recall that the range of $y_i^\pm$ was extended by periodicity. It is natural to do the same for the $b^\pm_i$ so that we can identify coefficients in the last equation for all $i$ to obtain
\begin{equation} \label{G ev equation in the g}
    \pm \bigl(1-\rho_\pm \bigr) \, b_{i}^\pm = (-1)^{i} \, t_{i} \, \bigl(b_{i+1}^\pm - \rho_\pm \, b_{i-1}^\pm \bigr) \, , \quad b_{i+N}^\pm \equiv b_i^\pm
    \, , \qquad i \in \mathbb{Z} \, .
\end{equation} 
Depending on the value of $\rho_\pm$, there is much to tell about $\psi^\pm$ from this equation:
\begin{enumerate}
    \item When $\rho_\pm \neq 1$, the fact that $t_{N}=0$ clearly leads to
        \begin{equation}  \label{bN vanishing}
            b_{N}^\pm=0 \ .
        \end{equation}
    \item When $\rho_\pm = 1$, \eqref{G ev equation in the g} simply yields $b_{i+1}^\pm = b_{i-1}^\pm$ for any $1 \leqslant i \leqslant N-1$. Due to periodicity and $N$ being odd, this implies that $b_{i}^\pm$ is actually a constant sequence, hence 
        \begin{equation} \label{coeffs of stabilised elmt}
            \psi^\pm \propto \sum_{i=1}^N g_i^\pm \propto \sum_{i=1}^N f_i^\pm \ .
        \end{equation}
\end{enumerate}

Let us now apply these results to the Fourier modes $\psi^\pm = \tilde{\Phi}_{n}^{\pm}$ defined in \eqref{Phi_modes}, which have eigenvalues $\rho_\pm = \omega^{\pm n}$ for the quasi-translation operator and whose coefficients we shall note $b_j^+ = \bar{M}_{nj}$ and $b_j^- = M_{nj}$ so that
\begin{equation} \label{NxN M}
	\tilde{\Phi}_n = \sum_{j=1}^N M_{nj} \; g_j \, , \quad
	\tilde{\Phi}_n^+ = \sum_{j=1}^N \bar{M}_{nj} \; g_j^+ \, , \qquad 0 \leqslant n < N \, .
\end{equation}
Equation \eqref{G ev equation in the g}  then becomes
\begin{equation} \label{equation M Mbar}
	\begin{aligned} 
    (1 - \omega^n) \, \bar{M}_{nj} & = (-1)^{j} \, t_j \, \bigl( \bar{M}_{n, j+1} - \omega^n \, \bar{M}_{n,j-1}\bigr) \, , \\ 
    (\omega^{-n} - 1) \, M_{nj} &= (-1)^{j} \, t_j \, \bigl(M_{n, j+1} - \omega^{-n} M_{n,j-1} \bigr) \, .
    \end{aligned}
\end{equation}
Once again, note that these equations are also valid for $n=0$ and for $j=N$, and are $N$-periodic in both indices.

To ensure uniformity in the treatment of all Fourier-transformed fermions by \eqref{equation M Mbar}, we extended our notations $M$ and $\bar{M}$ compared to \textsection\ref{sec_sm:fermionic_rep} from $(N-1) \times (N-1)$ to $N\times N$ matrices. We now have the index $n$ running from $0$ to $N-1$ and the index $j$ from $1$ to $N$ in a way that we shall clarify without waiting. First, notice that formula \eqref{eq:zero_modes} announced for $\tilde{\Phi}_0^\pm$ immediately follows from relation \eqref{coeffs of stabilised elmt} since the multiplicative constant that remained to be found is actually determined by \eqref{fourier fermions leading term}. Moreover, in agreement with our original definitions of $M$ and $\bar{M}$ in \textsection\ref{sec_sm:fermionic_rep}, \eqref{bN vanishing} reveals that \eqref{NxN M} gives the exact same decomposition of the $\tilde{\Phi}_n$ when $1 \leqslant n \leqslant N-1$ as in \eqref{Psig} since the extra $g_N^\pm$ term vanishes. Put together, these results allow us to spell out the matrix extension we have performed as 
\begin{equation} \label{matrix extension}
    \begin{aligned}
        \bar{M}_{nN} &= M_{nN} = 0 \, , \qquad 1 \leqslant n \leqslant N-1 \, ,\\
        \bar{M}_{0j} &= M_{0j} = \frac{1}{2} \, , \qquad\: 1 \leqslant j \leqslant N \, ,  
    \end{aligned}
\end{equation}
i.e.\ we first added an $N$th column with $0$s and then a $0$th row of $1/2$s to both our matrices $M$ and $\bar{M}$.

Equation \eqref{equation M Mbar} can be seen as a pair of disjoint second-order difference equations in the index $j$ and as such, they open the way to efficient computations of the Fourier-transformed fermions $\tilde{\Phi}_n$ by recursion as long as we have two initial conditions at hand. Let us focus on the nontrivial case of $1 \leqslant n \leqslant N-1$. 
While \eqref{matrix extension} provides a first condition at index $j=N$, it turns out that a second constraint at $j=N-1$ can be read off from \eqref{fourier fermions leading term} and \eqref{matrix extension} by noticing that in the change of basis from the $g^\pm_k$ to the $f^\pm_k$, $f^\pm_N$ only appears in $g^\pm_N$ and $g^\pm_{N-1}$, which leads to
\begin{equation}
\label{second condition}
    \bar{M}_{n, N-1}  = M_{n, N-1}  = 1 \, .
\end{equation}

This point of view provides an alternative way to retrieve the symmetry properties \eqref{M Mbar complex} between $M$ and $\bar{M}$. Indeed, comparing the two equations in \eqref{equation M Mbar}, we see that the second is obtained from the first one by complex conjugation and
multiplying the left-hand side by $-1$. Also, the two equations are invariant under simultaneous complex conjugation and sending $n\to N-n$. As both initial conditions \eqref{matrix extension} and \eqref{second condition} are preserved by all these operations, we can readily deduce \eqref{M Mbar complex}, namely
\begin{equation} \label{particule hole phi}
	\bar{M}_{nj} =(-1)^j \, M_{nj}^*\, ,\qquad M_{N-n,j}=M_{nj}^*\, , \qquad \bar{M}_{N-n,j} = \bar{M}_{nj}^*\,.
\end{equation}
        
\subsubsection{Discrete symmetries} \label{sec:descr_symm}

\noindent 
We are now in a position to establish the action of the $\PP$, $\TT$ and $\CC$ transformations on our Fourier modes.

\paragraph{$\PP \TT$ transformation.} \label{sec: technical PT}
It is clear that definitions \eqref{eq: def P} and \eqref{eq: def T} of $\PP$ and $\TT$ in the basis $f_1^\pm, \dots, f_N^\pm$ of one-particle fermionic operators preserve commutations relations \eqref{fermtranssm}. As such, $\PP$ and $\TT$ are well-defined on the whole space of fermionic operators as algebra endomorphisms. Note that $\TT$ is $\mathbb{C}$-antilinear and that both $\PP$ and $\TT$ square to $1$ as expected.
From these definitions, it follows that 
\begin{equation} \label{eq: PT on the gk}
    \PP\bigl(g_i^\pm\bigr) \equiv g_{N-i}^{\pm} \, , \quad \TT\bigl(g_i^\pm\bigr) \equiv g_{i}^{\pm} \, , \qquad 1 \leqslant i \leqslant N-1 \, ,
\end{equation}
and we readily obtain
\begin{equation}
    \PP(e_i) = e_{N-i} \, , \quad \TT(e_i) = e_i \, , \qquad 1 \leqslant k \leqslant N-1 \, .
\end{equation}
This leads to the results \eqref{parity} and \eqref{charges under T} of the main draft about how conserved charges behave under $\PP$ and $\TT$.

Let us now establish how our Fourier-transformed fermions $\tilde\Phi_k^\pm$ behave under these two discrete symmetries.
First, it is clear by formula \eqref{eq:zero_modes} that 
\begin{equation}
    \PP\bigl(\tilde\Phi_0^\pm\bigr) = \tilde\Phi_0^\pm \, , \qquad \TT\bigl(\tilde\Phi_0^\pm\bigr) = \tilde\Phi_0^\pm \ .
\end{equation}
For $n \neq 0$, since $\PP(\G)=\G^{-1}$ and $\TT(\G)=\G$, the relation $\G \, \tilde{\Phi}_n^\pm \G^{-1} =\omega^{\pm n}\, \tilde{\Phi}_n^\pm$ transforms as 
\begin{equation} \label{PT on phi ev equation}
    \G^{-1} \, \PP\bigl( \tilde{\Phi}_n^\pm\bigr)\, \G =\omega^{\pm n}\, \PP\bigl( \tilde{\Phi}_n^\pm\bigr) \, , \qquad \G \, \TT\bigl( \tilde{\Phi}_n^\pm\bigr)\, \G^{-1} =\omega^{\mp n}\, \TT\bigl( \tilde{\Phi}_n^\pm\bigr) \, .
\end{equation}
In other words, both $\PP\bigl( \tilde{\Phi}_n^\pm\bigr)$ and $\TT\bigl( \tilde{\Phi}_n^\pm\bigr)$ have the same quasi-momentum as $\tilde{\Phi}_{N-n}^\pm$, which means that they are all proportional to one another. To determine the corresponding coefficient of proportionality, note that we can alternatively compute these quantities from the action \eqref{eq: PT on the gk} of $\PP$ and $\TT$ on the bases $g_j^\pm$ as
\begin{equation} \label{P phi in gk}
    \begin{aligned}
        \PP\bigl(\tilde{\Phi}_n\bigr) &= \sum_{j=1}^{N-1} \! M_{nj} \; g_{N-j} \, , \qquad 
        &&\PP\bigl(\tilde{\Phi}_n^+\bigr) = \sum_{j=1}^{N-1} \! \bar{M}_{nj} \; g_{N-j}^+ \, \\
        \TT\bigl(\tilde{\Phi}_n\bigr) &= \sum_{j=1}^{N-1} \! M_{nj}^* \; g_{j} \, , \qquad 
        &&\TT\bigl(\tilde{\Phi}_n^+\bigr) = \sum_{j=1}^{N-1} \! \bar{M}_{nj}^* \; g_{j}^+ \, .
    \end{aligned}
\end{equation}
Focussing on the coefficient of $g_{N-1}^+$ we find $\PP\bigl(\tilde{\Phi}_n^+\bigr) = (\bar{M}_{n1} / \bar{M}_{N-n,N-1}) \, \tilde{\Phi}_{N-n}^+$ and $\TT\bigl(\tilde{\Phi}_n^+\bigr) = (\bar{M}_{n,N-1}^* / \bar{M}_{N-n,N-1}) \, \tilde{\Phi}_{N-n}^+$, and similarly for the annihilation operators. Due to the result \eqref{second condition} we conclude that 
\begin{equation} \label{PT phi explicit}
    \begin{aligned}
        &\PP\bigl(\tilde{\Phi}_n\bigr) = M_{n1} \, \tilde{\Phi}_{N-n} \, , \qquad 
        &&\PP\bigl(\tilde{\Phi}_n^+\bigr) = \bar{M}_{n1} \, \tilde{\Phi}_{N-n}^+ \, , \\    
        &\TT \bigl(\tilde{\Phi}_n\bigr) = \tilde{\Phi}_{N-n} \, , \qquad 
        &&\TT\bigl(\tilde{\Phi}_n^+\bigr) = \tilde{\Phi}_{N-n}^+ \, .
    \end{aligned}
\end{equation}
As a byproduct, since $\PP^2=1$ because of
\begin{equation} \label{P phi ter}
    \begin{aligned}
    \tilde{\Phi}_n & = \PP^2\bigl(\tilde{\Phi}_n\bigr) =M_{n1}\, M_{N-n,1}\,\tilde{\Phi}_{n} \, , \\
    \tilde{\Phi}_n^+ & = \PP^2\bigl(\tilde{\Phi}_n^+\bigr) = \bar M_{n1}\, \bar M_{N-n,1}\,\tilde{\Phi}_{n}^+ \, .
    \end{aligned}
\end{equation}
our previous formulas \eqref{particule hole phi} allow us to conclude that  
\begin{equation} \label{unimodular Mn1}
    M_{n1}\,M_{n1}^*= \bar M_{n1}\,\bar M_{n1}^*=1 \ .
\end{equation}
As explained in \textsection\ref{sec: proof anticomm}, this is compatible with conjecture \eqref{tilde alpha} that can itself be written using \eqref{anticomm and first coeff} and \eqref{particule hole phi} as
\begin{equation} \label{eq: conj Mn1}
    M_{n1} = (-1)^{n+1} \, \I \, \omega^{-n} \, , \qquad
    \bar{M}_{n1} = (-1)^{n+1} \, \I \, \omega^{n} \ .
\end{equation}
Finally, \eqref{PT phi explicit} can be expressed in the basis of rescaled fermions $\tilde\Psi_n^\pm$ \eqref{sm:Psi_tilde} as 
\begin{equation} 
    \begin{aligned}
        \PP\bigl(\tilde{\Psi}_0\bigr)&  = \tilde{\Psi}_{0} \, , \qquad && \mspace{2mu}
        \PP\bigl(\tilde{\Psi}_n\bigr) =  \mathrm{i}^N \,(-1)^{N-n} \,M_{n1} \, \tilde{\Psi}_{N-n} \, , \ \ \\
        \PP\bigl(\tilde{\Psi}_0^+\bigr) & = \tilde{\Psi}_{0}^+ \, , && 
        \PP\bigl(\tilde{\Psi}_n^+\bigr) =  \mathrm{i}^N \,(-1)^{N-n} \,\bar{M}_{n1} \, \tilde{\Psi}_{N-n}^+ \, , 
    \end{aligned}
    \quad 1 \leqslant n \leqslant N-1 \, ,    
\end{equation}
which, using the conjecture \eqref{eq: conj Mn1},  simplifies to
\begin{equation}  \label{P on psi after conj}
    \PP\bigl(\tilde{\Psi}_0^\pm\bigr) = \tilde{\Psi}_{0}^\pm \, , \qquad \PP\bigl(\tilde{\Psi}_n^\pm\bigr) =  (-1)^{(N+1)/2} \,\omega^{\pm n} \, \tilde{\Psi}_{N-n}^\pm \, , 
     \quad 1 \leqslant n \leqslant N-1 \, .
\end{equation}
Likewise we obtain
\begin{equation}  \label{T on psi}
    \TT\bigl(\tilde{\Psi}_0^\pm\bigr) = -\tilde{\Psi}_{0}^\pm \, , \qquad
    \TT\bigl(\tilde{\Psi}_n^\pm\bigr) = (-1)^{(N+1)/2} \, \tilde{\Psi}_{N-n}^\pm \, , 
    \quad 1 \leqslant n \leqslant N-1 \, .    
\end{equation}
Combining \eqref{P on psi after conj} and \eqref{T on psi} gives 
\begin{equation} \label{PT on psi}
    \PP\, \TT\bigl(\tilde{\Psi}_0^\pm\bigr) = -\tilde{\Psi}_{0}^\pm , \qquad \PP\, \TT\bigl(\tilde{\Psi}_n^\pm\bigr) = \omega^{\pm n}\, \tilde{\Psi}_{n} \, , 
    \quad 1 \leqslant n \leqslant N-1 \, .    
\end{equation}

\paragraph{Charge conjugations.} \label{sec: C transf}
As before, definition \eqref{sm: def parthole} for the particle-hole transformation $\CC'$ clearly preserves anticommutation relations \eqref{fermtranssm} and as such defines an involutive algebra endomorphism of the space of fermionic operators. It is immediately checked that \begin{equation}
    \CC'(g_i^\pm) \equiv g_{i}^{\mp} \, , \qquad 1 \leqslant i \leqslant N-1 \, ,
\end{equation}
so that 
\begin{equation}
    \CC'(e_i) = -e_i \, , \qquad 1 \leqslant i \leqslant N-1 \, .
\end{equation}

At the moment, we have no easy way to express the action of $\CC'$ on the Fourier-transformed fermions. However, it is possible to relate it to an alternative version $\CC$ of the particle-hole transformation that acts more naturally in the Fourier-transformed fermions as the following $\mathbb{C}$-\emph{anti}linear charge conjugation:
\begin{equation} \label{C def}
    \CC\bigl(\tilde\Phi_0^\pm\bigr) \equiv \tilde\Phi_0^\mp \, , \qquad 
    \CC\bigl(\tilde\Phi_n^\pm\bigr) \equiv \I \, \tilde\Phi_n^\mp \, , \quad 
    1 \leqslant n \leqslant N-1 \, .
\end{equation}
There is no need to invoke conjecture \eqref{tilde alpha} to establish that $\CC$ preserves the anticommutation relations between the $\tilde{\Phi}_n^\pm$ as this follows from our discussion in \textsection \ref{sec: proof anticomm}, and we obtain an algebra endomorphism on the whole space of fermionic operators that clearly satisfies $\CC^2=1$ by antilinearity. Note that definition \eqref{C def} can be expressed in the basis of rescaled Fourier-transformed fermions \eqref{sm:Psi_tilde} as 
\begin{equation}
    \CC\bigl(\tilde\Psi_{0}^{\pm}\bigr) = -\tilde{\Psi}_{0}^{\mp} \, , \qquad 
    \CC\bigl(\tilde{\Psi}_{n}^{\pm}\bigr) = (-1)^{n} \, \tilde{\Psi}_{n}^{\mp} \, , \quad 
    1 \leqslant n \leqslant N-1 \, ,
\end{equation} 
and it is now clear that the charge conjugation $\CC$ corresponds to the usual particle-hole transformation in our Fock basis \eqref{freespectrum} up to a sign. It is possible to express $\CC$ in terms of $\CC'$ using the $\mathbb{C}$-antilinear operator $\UU$ that we already defined in \eqref{def U} on the space of one-particle operators as
\begin{equation} \label{def U pm}
    \UU\bigl(\tilde\Phi_{0}^{\pm}\bigr) = \tilde\Phi_{0}^{\pm} \, , \qquad 
    \UU\bigl(g_{j}^{\pm}\bigr)= (-1)^{j} \,\I\, g_{j}^{\pm} \, , \quad 
    1 \leqslant j \leqslant N-1 \, .
\end{equation}
It is easily checked that this operator preserves anticommutation relations and is thus once again well-defined on the whole algebra of fermionic operators. In particular, we get
\begin{equation}
    \UU(e_i) = -e_i \, .
\end{equation}
It furthermore satisfies $\UU^2=1$. The purpose of the phases we introduced in the previous definitions was to reproduce the relations \eqref{particule hole phi} that we have established coefficient-wise between the $\tilde{\Phi}_n^+$ and the $\tilde{\Phi}_n^-$ in order to obtain
\begin{equation} \label{C from C'}
    \CC = \UU \, \CC' = \CC' \, \UU \, ,
\end{equation}
which is \eqref{eq:C=C'U}.
This relation makes it easy to compute the action of $\CC$ on the conserved charges as we first obtain 
\begin{equation}
    \CC(e_i) = e_i \, , \qquad 1 \leqslant i \leqslant N-1 \, ,
\end{equation}
and then get the announced action \eqref{charges under charge transf} of $\CC$ on our model.

As a side note, it is also possible to compute the action of $\UU$, thus of $\CC$, on the $f_j^\pm$. Computations are straightforward though a bit tedious: since $\UU \CC' = \CC' \UU$, it is only needed to compute it in the case of, say, annihilation operators, and if one first establishes that $\UU (g_N) = \I \, (f_{1} - f_{N}) + 2\, \tilde{\Phi}_{0}$ then a change of basis as in \eqref{f_1 in the g} leads to
\begin{equation}
    (-1)^{j+1} \, \UU(f_{j}) = (1+\I) \bigl(f_{1}+\dots+ f_{j-1}\bigr) + f_{j} + (1-\I) \bigl(f_{j+1} + \dots + f_{N}\bigr) \, , \qquad 1 \leqslant j \leqslant N \, .
\end{equation}

\subsubsection{Anticommutation relations of Fourier-transformed fermions} \label{sec: proof anticomm}

\noindent
The purpose of this part is to establish what we know about the fermionic relations between the $\tilde{\Phi}_n^\pm$. First, as we have seen in \textsection\ref{sec_sm:fermionic_rep},
the anticommutation relations \eqref{fermtranssm} for $f_i^\pm$ and quasi-translation invariance guarantee that 
\begin{equation} \label{anticomm scalar}
    \begin{gathered}
        \{ \tilde{\Phi}_n , \tilde{\Phi}_m \} = \{ \tilde{\Phi}_n^+ , \tilde{\Phi}_m^+ \} = 0 \, , \\
         \{ \tilde{\Phi}_n , \tilde{\Phi}_m^+ \} = \tilde \alpha(n)\, \delta_{nm} \, ,
    \end{gathered}
    \qquad\quad \text{for all} \ n, m \, .
\end{equation}
It only remains to determine the exact expression of the coefficients $\tilde \alpha(n)$. Compared to the fully conjectural approach of \textsection\ref{sec_sm:fermionic_rep} based on \eqref{explicit alpha}, the previous sections \textsection\ref{sec:techn_G_fermions} and \textsection\ref{sec: technical PT} allow us to get direct information about the $\tilde \alpha(n)$ that is compatible with numerical experiments.

For $n=0$, using the explicit formula \eqref{eq:zero_modes} we obtain
\begin{equation} \label{anticomm at 0}
    \{ \tilde{\Phi}_0 , \tilde{\Phi}_0^+ \} = -1 \, .
\end{equation}
In the other cases, applying time reversal and parity using formula \eqref{PT phi explicit}, one gets
\begin{equation}
    \begin{aligned}
        \PP\bigl( \bigl\{ \tilde{\Phi}_n, \tilde{\Phi}_n^+ \bigr\} \bigr) &= M_{n1} \, \bar{M}_{n1} \, \bigl\{ \tilde{\Phi}_{N-n}, \tilde{\Phi}_{N-n}^+ \bigr\}\, \\
        \TT\bigl( \bigl\{ \tilde{\Phi}_n, \tilde{\Phi}_n^+ \bigr\} \bigr) &=  \bigl\{ \tilde{\Phi}_{N-n}, \tilde{\Phi}_{N-n}^+ \bigr\}\, .  
    \end{aligned}
\end{equation}
By formula \eqref{particule hole phi}, we have $M_{n1} \, \bar{M}_{n1} = - M_{n1} \, {M}_{n1}^*$, and we know by relation \eqref{unimodular Mn1} that $M_{n1}$ is unimodular, so that
\begin{equation} \label{PT on anticomm }
    \tilde{\alpha}(n) = -\tilde{\alpha}(N-n) \, , \qquad \tilde{\alpha}(n)^* = \tilde{\alpha}(N-n)\, .
\end{equation}
In particular, $\tilde \alpha(n)$ is purely imaginary.
Moreover, it follows from \eqref{anticomm scalar} and the inverse Fourier transform 
\begin{equation} \label{inverse Fourier}
    f_1 = \sum_{n=1}^N \omega^{n} \, \tilde{\Phi}_n 
\end{equation}
due to definition \eqref{Phi_modes} and $\Phi_1=f_1$ that $\{ f_1, \tilde{\Phi}_n^+ \} = \omega^{n} \, \{ \tilde{\Phi}_n, \tilde{\Phi}_n^+ \}$. In decomposition \eqref{Psig}, since $f^+_1$ appears only in $g_1^+$,  it is clear using the anticommutation relations \eqref{fermtranssm} that $\{ f_1, \tilde{\Phi}_n^+ \} = -\bar{M}_{n1}$ and it follows that  
\begin{equation} \label{anticomm and first coeff}
     \bigl\{ \tilde{\Phi}_n, \tilde{\Phi}_n^+ \bigr\} = -\omega^{-n} \, \bar{M}_{n1} \, .
\end{equation}
In particular, $\tilde \alpha(n)$ is unimodular, and since we have seen that it is also purely imaginary, we get
\begin{equation} \label{phitilde anticomm unsigned}
   \tilde \alpha(n)= \bigl\{ \tilde{\Phi}_n, \tilde{\Phi}_n^+ \bigr\} =\pm\I \, , \qquad 1\leqslant n<N\, .
\end{equation}
This sign is partially fixed by \eqref{PT on anticomm }, in a way which is compatible with conjecture \eqref{tilde alpha}. However, we were able to fix the sign in \eqref{phitilde anticomm unsigned} exactly as in the conjecture only by resorting to numerical experiments.

\subsubsection{Link between \textsc{xxz} and fermionic TL representations} \label{sec: XXZ vs ferm}

\noindent
In this Supplemental Material, we have somewhat carelessly used $e_j$ to denote the Temperley--Lieb generator either  
\begin{enumerate}
    \item for the specialisation \eqref{eq:TL_q=i} to $\q=\I$ of the standard spin-chain representation \eqref{eq:TL_spin} of the parent model,
    \item in the fermionic representation \eqref{gTL} that is specific for $\q = \I$, defined in terms of two-site Jordan--Wigner fermions \eqref{gfer_supp}.
\end{enumerate}
The explicit matrices, however, do not coincide.
Nevertheless, our switch from the spin-chain to the fermionic representation is allowed by the fact that the two representations are isomorphic. It is the purpose of this section to establish this statement. To this end let us here distinguish the two representations by writing $e_j^{\mathrm{sp}}$ for the spin-chain representation~\eqref{eq:TL_q=i} and $e_j^{\mathrm{ff}}$ for the (free) fermionic representation~\eqref{gTL}.

Let us first recall the standard result
\begin{equation} \label{eq: quad TL_spin}
    e_j^{\mathrm{sp}} = \I \, (-1)^{j+1} \, e_j^{\mathrm{ff}} \, , \qquad 1 \leqslant j \leqslant N-1 \, .
\end{equation}
To see this, rewrite both sides in terms of Pauli matrices, so that the matrix expression \eqref{eq:TL_spin} at $\q = \I$ becomes
\begin{equation}
    -e_{j}^\text{sp} = \sigma_j^+ \sigma_{j+1}^- + \sigma_j^- \sigma_{j+1}^+ + \frac{\I}{2} \,(\sigma_j^z-\sigma_{j+1}^z) \, ,
\end{equation}
while for $e_j^{\mathrm{ff}}$ from \eqref{gTL} one uses the Jordan--Wigner fermions \eqref{JW transf}.

To establish the isomorphism we will be using \eqref{eq: quad TL_spin} to exhibit an explicit transformation of the space of states that intertwines the $e_j^{\mathrm{sp}}$ and $e_j^{\mathrm{ff}}$. First, let us work at the level of the algebra $\mathcal{F}$ of fermionic operators, and denote by $\mathcal{F}^+$ and $\mathcal{F}^-$ its subalgebras of creation and annihilation operators, respectively. In the subspace of one-particle fermionic operators, we define a transformation $\varphi$ by 
\begin{equation}
    \begin{aligned}
        \varphi(\tilde\Phi_0) & = \tilde\Phi_0 \, , \quad \varphi(\tilde\Phi_0^+) = \tilde\Phi_0^+ \, , \\
        \varphi(g_{j}) & = g'_j \, , \quad \ \, \varphi(g_{j}^{+}) = g_j^{\prime\,+} \, , \qquad 1 \leqslant j \leqslant N-1 \, ,
    \end{aligned}
\end{equation}
where we denoted
\begin{equation} \label{eq: def g prime}
    \begin{aligned}
        g'_{2j} & \equiv -\mathrm{i} \, g_{2j} \, , \qquad &&g'_{2j+1} \equiv \hphantom{\mathrm{i} } \, g_{2j+1} \, , \\
        g_{2j}^{\prime\,+} &\equiv \hphantom{-\mathrm{i}} \, g_{2j}^{+} \, , \qquad &&g_{2j+1}^{\prime\,+} \equiv  \mathrm{i} \, g_{2j+1}^{+} \, .
    \end{aligned}    
\end{equation}
It is straightforward to check that the $g'_j$ and $g_j^{\prime\,+}$ satisfy the same anticommutation relations \eqref{gfer_supp} as the $g_j$ and $g_j^{+}$. Moreover, $\tilde{\Phi}_0$ and $\tilde{\Phi}_0^+$ anticommute with all the $g_j$ and $g_j^+$, which is enough to conclude that $\varphi$ preserves all fermionic anticommutation relations and induces a transformation of the whole algebra $\mathcal{F}$. Phases in \eqref{eq: def g prime} were so defined that by \eqref{eq: quad TL_spin}, we now have
\begin{equation}
    \varphi( e_j^{\mathrm{sp}} ) = e_j^{\mathrm{ff}} \, , \qquad 
    \varphi( e_j^{\mathrm{ff}} ) = -e_j^{\mathrm{sp}} \, .
\end{equation}
It is clear that $\varphi$ is bijective, with inverse $\varphi^{-1} = \TT \, \varphi$, and preserves the subalgebras $\mathcal{F}^+$ and $\mathcal{F}^-$. Note that the square of this transformation corresponds, at the level of the free-fermion Temperley--Lieb algebra, to the mapping
\begin{equation}
    e_j \longmapsto -e_j \, ,
\end{equation}
which can for instance also be realised by applying the usual particle-hole transformation $\CC'$ from \eqref{particle hole}.

To conclude, recall that our states live in the Fock space associated to $\mathcal{F}$ with the same vacuum $\ket{\varnothing}$. Although we did introduce a new basis for this space in \eqref{freespectrum} and labelled it canonically, note that this is still the usual Fock space associated to the Jordan--Wigner fermions \eqref{JW transf}.
Despite the phases introduced in definition \eqref{eq: def fk} compared to the standard fermions \eqref{JW transf}, a basis of our space is given as usual by the $2^{N}$ states $f_{k_{1}}^{+} \cdots f_{k_{M}}^{+} \ket{\varnothing}$ indexed by $0 \leqslant M \leqslant N$ and $1 \leqslant k_{1} < \dots < k_{M} \leqslant N$. Working in this basis, it is easy to check that the transformation
\begin{equation}
    \hat{\varphi} \bigl( F \, \ket{\varnothing} \big) \equiv \varphi(F) \, \ket{\varnothing}
\end{equation}
is well defined, bijective and intertwines $e_k^{\mathrm{sp}}$ with $e_k^{\mathrm{ff}}$, as announced.

\subsubsection{Proof that $H^\mathrm{full}$ vanishes at $\q=\I$} \label{sec: proof vanishing H}

\paragraph{Main discussion.} \label{proof of vanishing hij}
The purpose of this part is to show that for any $1 \leqslant i \leqslant j < N$, the coefficients $h^{\LL,\RR}_{i j}$ defined in \eqref{notation hij} satisfy \eqref{hL=-hR}, i.e.\
\begin{equation} \label{hL+hR=0}
	h^{\LL}_{i j} + h^{\RR}_{i j} = 0 \, .
\end{equation}
This will imply that $\HH^\LL = -\HH^\RR$ and thus $\HH^\text{full} = 0$ as in \eqref{Hfull=0 at q=i}.

Introduce the notation 
\begin{equation} \label{def eta}
	\eta_{i j} \equiv \sum_{k=1}^{N} \varsigma_{i, j, k}^{\LL} \, ,
\end{equation}
where the summands are the coefficients of the nested TL commutators of $\Ss^{\LL, \RR}_{[i,i+k]}$, see \eqref{def slmk}. Note that the definition of $\eta_{ij}$ makes sense for any nonnegative $i,j$. Then it will be shown in \textsection\ref{proof closure of loop} (p.\,\pageref{proof closure of loop}) that the sums in the expression \eqref{telescoped h^L} for $h^{\LL}_{ij}$ can be `closed' in terms of $\eta$ as follows:
\begin{equation}
	\label{loop closure for hfull}
	h^{\LL}_{i j} + h^{\RR}_{i j} = \eta_{\mspace{1mu}0,\mspace{1mu}j-i} - \eta_{i j} - (-1)^{j-i} \, \eta_{N-j,N-i} \, .
\end{equation}
To prove that this quantity vanishes, we will use an analytic argument based on an extension of our notations to the following meromorphic functions for $k \leqslant l$ and arbitrary nonnegative $i,j$: 
\begin{equation} \label{useful mero functions}
	\begin{aligned}
		t_{k, l}(z) &\equiv \prod_{m=k}^{l-1} \! f(z \, \omega^{m}) \, , \\
		\varsigma_{i j}(z) &\equiv (-1)^{l} \, t_{-i, 0}(z) \, t_{-j, 0}(z) \, , \\
		\eta_{i j}(z) &\equiv \sum_{k=1}^{N} \varsigma_{ij}(z \, \omega^{k}) \, .
	\end{aligned}
\end{equation}
It is clear from definition \eqref{def tk} that $t_{k, l}(\omega^{r})=t_{k+r, l+r}$ for any integer $r$, hence $\varsigma_{i j}(\omega^{r})=\varsigma^{\LL}_{i, j, r}$ by definition \eqref{def slmk} and $\eta_{i j}(1)=\eta_{i j}$.

The following two major results proved in \textsection\ref{proof LR symm} and \textsection\ref{proof eta(z)} respectively will now allow us to compute $\eta_{i j}$  as the limit of the meromorphic function $\eta_{i j}(z)$ at $z \rightarrow 1$:
\begin{itemize}
	\item As rational fractions, we have the following analogue of $\LL$-$\RR$ symmetry:
	\begin{equation}
		\label{LR symmetry on eta}
		\eta_{N-i,N-j}(z)= (-1)^{j-i} f\bigl(z^{N}\bigr)^{\!2} \, \eta_{i j}(-z^{-1}) \, .
	\end{equation}
	\item  There exists constants $C_{ij}$ in terms of which the following equality holds true for $z \in \mathbb{C}$ generic:
	\begin{equation}
		\label{expression of eta(z)}
		\eta_{i j}(z) = \frac{N}{\I^{\mspace{1mu}j-i}} \biggl( 1 - \frac{1}{1+z^{N}} \, \bigl(1-(-1)^{j-i}\bigr) \biggr) + \frac{z^{N-1}}{(1+z^{N})^{2}} \, C_{i j} \, .
	\end{equation}
	Moreover, $C_{ij} = 0$ when $i=j$.
\end{itemize}
Before proving these results, let us use them to derive an expression for $h_{i j}^{\LL} + h_{ij}^{\RR}$ that will let us establish \eqref{hL=-hR}. To begin with, it directly follows from \eqref{expression of eta(z)} that 
\begin{equation}
	\eta_{i j}(1) - \eta_{\, 0, \, j-i \,}(1) = \frac{C_{ij}}{4} \, .
\end{equation}
Besides, straightforward asymptotic expansions of \eqref{expression of eta(z)} around $1$ lead to
\begin{equation}
	\eta_{i j}(-z^{-1}) = \frac{C_{ij}}{N^{2}(z-1)^{2}} + \mathcal{O}\biggl(\frac{1}{z-1}\biggr) \, .
\end{equation}
Now notice that $\frac{f(z)}{z-1} \rightarrow_{z \to 1}-\I N/2$ so by left/right symmetry \eqref{LR symmetry on eta}, we get
\begin{equation}
	\eta_{N-j,N-i}(1)=(-1)^{j-i+1} \frac{C_{ij}}{4} \, .
\end{equation}
This leads us to \eqref{hL=-hR} as expected, since by \eqref{loop closure for hfull} we can write $h_{i j}^{\LL} + h_{i,j}^{\RR}=\eta_{\, 0,j-i} - \eta_{i j} - (-1)^{j-i}\eta_{N-j, N-i}=0$. 
\bigskip

\paragraph{Proof of the closure of the summation loop \eqref{loop closure for hfull}.} \label{proof closure of loop} 
We have seen in \eqref{telescoped h^L} that $h^{\LL}_{i j}$ can be written as $A_{ij} - B_{ij}$ with 
$A_{ij} = \sum_{m=j}^{N-1}\varsigma^{\LL}_{0,j-i,m-i}$ and $B_{ij}=\sum_{m=j}^{N-1}\varsigma^{\LL}_{i, j, m}$.
Moreover, by \LL-\RR\ symmetry \eqref{hij LR symm}, we see that 
\begin{equation}
	h^{\LL}_{ij} + h^{\RR}_{ij} = \bigl( A_{ij} + (-1)^{j-i}A_{N-j, N-i} \bigr) - \bigl( B_{ij} + (-1)^{j-i}B_{N-j, N-i} \bigr) \, .
\end{equation}
As soon as $0 \leqslant m \leqslant j-1$, the value $0$ is within the bounds of the products in the definition \eqref{def slmk} of $\varsigma_{ijm}^{\LL}$. Since $t_{0}=0$, it follows that $\varsigma_{i, j, m}^{\LL}=0$ for these values of $m$, and the sum in $B_{ij}$ can be formally completed to obtain $B_{ij}=\eta_{ij}$ and $B_{N-j,N-i}=\eta_{N-j,N-i}$.

Moreover, recall that by definition \eqref{def slmk},  $\varsigma_{0,j-i,m-i}^{\LL}=t_{m-j+1,m-i+1}$, which implies that $A_{ij}$ can be reindexed as $A_{ij} = \sum_{m=1}^{N-j}t_{m,m+j-i}$.
At the same time, we get $(-1)^{j-i}A_{N-j, N-i} =(-1)^{j-i}\sum_{m=1}^{i} t_{m,m+j-i}$ and, using the property $t_{k} = - t_{N-k}$, successive changes of indices lead to $(-1)^{j-i}A_{N-j, N-i} = \sum_{m=N-j+1}^{N-(j-i)}t_{m,m+j-i}$.
Adding up the expressions, we have established that $A_{ij} + (-1)^{j-i}A_{N-j ,N-i} = \sum_{m=1}^{N-(j-i)}t_{m,m+j-i}$.
It is moreover easily seen that for the remaining values $N-(j-i)+1 \leqslant m \leqslant N$ of the summation index, the summand $t_{m,m+j-i}$ vanishes since the product \eqref{def t_kl SM} defining it contains $t_{N}=0$.
We then obtain $A_{ij} + (-1)^{j-i}A_{N-j ,N-i} = \sum_{m=1}^{N}t_{m,m+j-i}$, which is precisely $\eta_{\mspace{1mu}0,j-i}$.
Finally, we get the equation \eqref{loop closure for hfull}.
\bigskip

\paragraph{Proof of \LL-\RR\ symmetry \eqref{LR symmetry on eta}.} \label{proof LR symm} 
Recall that we take $N$ to be odd.
By definitions \eqref{f at i} and \eqref{useful mero functions}, $t_{1,N+1}(z) = \prod_{k=1}^{N} \bigl(-\I \, \frac{z \omega^{k}-1}{z \omega^{k}+1}\bigr)$, which can be reindexed by $N$-periodicity as $\prod_{k=1}^{N} \bigl(-\I \frac{z-\omega^{k}}{z+\omega^{k}}\bigr)$. Recognising $\prod_{k=1}^{N}(z-\omega^{k})$ as the well-known factorisation of the polynomial $z^{N}-1$ and taking note that $(-z)^{N}=-z^{N}$ since $N$ is odd, we get $t_{1,N+1}(z) =(-\I)^{N}\prod_{k=1}^{N} \frac{z-\omega^{k}}{z+\omega^{k}} = (-\I)^{N} \frac{z^{N}-1}{z^{N}+1}$, which leads us by definition \eqref{f at i} to
\begin{equation} \label{closed product}
	t_{1,N+1}(z)=(-1)^{(N-1)/2} \, f\bigl(z^{N}\bigr) \, .
\end{equation}
Let us now fix $j \geqslant 0$.
A straightforward calculation shows that $f(-z^{-1})=f(z)^{-1}$ and a suitable reindexing of the product \eqref{useful mero functions} leads to $t_{-j,0}(-z^{-1}) = t_{1,j+1}^{-1}(z)$. It follows that $t_{j+1,N+1}(z) = t_{1,N+1}(z) \, t_{-j,0}(-z^{-1})$, and the previous property \eqref{closed product} leads to
\begin{equation} \label{reversed product}
	t_{j,N}(z \, \omega)=(-1)^{(N-1)/2} \, f(z^{N}) \, t_{-j,0}(-z^{-1})
\end{equation}
after shifting indices in \eqref{useful mero functions}.

We are now in a position to establish \eqref{LR symmetry on eta}.
Fix $i$ and $j$. The last result \eqref{reversed product} precisely gives $t_{i,N}(z \, \omega) \, t_{j,N}(z \, \omega) = f(z^{N})^{2} \, t_{-i,0}(-z^{-1}) \, t_{-j,0}(-z^{-1})$.
By definition \eqref{useful mero functions}, this can be rewritten as 
\begin{equation}
	(-1)^{N-j} \, \varsigma_{N-j, N-i}(z \, \omega) = f(z^{N})^{2} \, (-1)^{i} \, \varsigma_{ij}(-z^{-1}) \, .
\end{equation}
Replacing $z$ by $z \, \omega^{k}$ and summing over all $k$, we obtain
\begin{equation}
	\sum_{k=1}^{N} \varsigma_{N-j,N-i} (z\, \omega^{k+1}) = -(-1)^{j-i} \sum_{k=1}^{N} f\bigl((z \, \omega^{k})^{N}\bigr)^{2} \, \varsigma_{ij}(-z^{-1}\,\omega^{-k}) \, .
\end{equation}
Finally, $f((z\,\omega^{k})^{N})=f(z^{N})$ can be factored out and sums can be reindexed to obtain
\begin{equation*}
	\eta_{N-i,N-j}(z)= (-1)^{j-i} f\bigl(z^{N}\bigr)^{\!2} \, \eta_{i j}(-z^{-1}) \, ,
\end{equation*}
which is \eqref{LR symmetry on eta}, as we wished to show.
\bigskip

\paragraph{Proof of $\eta_{ij}(z)$ from expression  \eqref{expression of eta(z)}.} \label{proof eta(z)} 
In what follows, we will fix indices $1 \leqslant i \leqslant j \leqslant N-1$ and establish that when $\alpha$ is taken in what we will call $A$, we have
\begin{equation*}
	\eta_{i j}(\alpha) = \frac{N}{\I^{\mspace{1mu}j-i}} \biggl( 1 - \frac{1}{1+\alpha^{N}} \bigl(1-(-1)^{j-i} \bigr) \biggr) + \frac{\alpha^{N-1}}{(1+\alpha^{N})^{2}} \, C_{i j} \, , 
\end{equation*}
where $C_{i j}$ will be properly defined below in such a way that $C_{i j}=0$ when $i=j$.
This equality between rational fractions was precisely \eqref{expression of eta(z)}, and since it will hold for an infinite number of $\alpha$, namely on $A$, this will be enough to conclude that it is also true generically on $\mathbb{C}$ as a whole.
The proof consists of two steps. 
\medskip

\textit{Step 1: $\eta_{ij}(\alpha)$ in terms of residues.}
First, let us list out the singularities of the function $\varsigma_{ij}(z)$. Recall that by \eqref{f at i} $f(z)$ has exactly one pole, namely $-1$, which is simple, and has exactly one zero at $z=1$.
By \eqref{useful mero functions}, for any $k \leqslant l$, all the poles of $t_{k,l}(z)$ are simple and exactly given by $z = -\omega^{-p}$ for $k \leqslant p \leqslant l-1$.
It follows from \eqref{useful mero functions} that the zeroes (resp.\ poles) of $\varsigma_{ij}(z)$ lie in the set $\mathcal{Z}$ consisting of $\omega^{k}$ for $1 \leqslant k \leqslant N$ (resp.\ in the set $\mathcal{P}\equiv -\mathcal{Z}$ consisting of the $-\omega^{k}$). Notice that $\mathcal{P}$ and $\mathcal{Z}$ are disjoint since $N$ is odd.
More precisely, the poles of $\varsigma_{ij}(z)$ are exactly as follows:
\begin{itemize}
	\item double poles $\mathcal{P}_{2}=\{-\omega^{k} \mid 1 \leqslant k \leqslant i\}$,
	\item simple poles $\mathcal{P}_{1}=\{-\omega^{k} \mid i+1 \leqslant k \leqslant j\}$. 
\end{itemize}

For any $\alpha$, let us now define $\varphi_{i j; \alpha}(z)$ as the meromorphic function 
\begin{equation}
	\varphi_{ij; \alpha}(z) \equiv \varsigma_{i j}(\alpha\, z) \sum_{\omega^{k} \in \mathcal{Z}} \frac{1}{z-\omega^k} \, .
\end{equation}
If $\alpha$ is taken in a certain neighbourhood $A$ of $1$ inside the unit circle (excluding $1$ itself), it is clear that $\alpha^{-1} \, \mathcal{P}$, $\alpha^{-1} \mathcal{Z}$ and $\mathcal{Z}$ are pairwise disjoint sets.

In what follows, we will always suppose that $\alpha \in A$.
Under this assumption, $\varsigma_{ij}(\alpha\,z)$ is regular at $\omega^{i} \in \mathcal{Z}$, so that Cauchy's integral formula and definition \eqref{def eta} lead to
\begin{equation}
	\eta_{i j}(\alpha) = \sum_{z \in \mathcal{Z}} \mathrm{Res}_{z}(\varphi_{i j; \alpha})  \, . 
\end{equation}
To compute this quantity, note that the poles of $\varphi_{i j; \alpha}$ can be exactly identified as follows:
\begin{itemize}
	\item simple poles: $\alpha^{-1} \, \mathcal{P}_{1} \cap \mathcal{Z}$,
	\item double poles: $\alpha^{-1} \, \mathcal{P}_{2}$.
\end{itemize}
Now, recall that the residue at infinity $\mathrm{Res}_{\infty}(f)$ of any meromorphic function $f(z)$ is defined as $\mathrm{Res}_{0}\bigl(-\frac{1}{z^2} f(z^{-1})\bigr)$.
A version of the residue theorem then yields that the sum of all residues of $\varphi_{i j; \alpha}$ vanishes, i.e.\
\begin{equation}
	\mathrm{Res}_{\infty}(\varphi_{i j; \alpha}) + \!\! \sum_{p \in \mathcal{P}_{1} \cup \mathcal{P}_{2}} \!\!\!\!\! \mathrm{Res}_{\alpha^{-1}p}(\varphi_{i j; \alpha}) + \sum_{z \in \mathcal{Z}} \mathrm{Res}_{z}(\varphi_{i j; \alpha}) = 0  \, .
\end{equation}
We finally get the formula
\begin{equation}
	\eta_{i j}(\alpha) = -\mathrm{Res}_{\infty}(\varphi_{i j; \alpha}) - \!\! \sum_{p \in \mathcal{P}_{1} \cup \mathcal{P}_{2}} \!\!\!\!\! \mathrm{Res}_{p}(\varphi_{i j; \alpha}) \, .
\end{equation}

In the following, we will establish that this identity can be rewritten exactly as \eqref{expression of eta(z)}, which was
\begin{equation*}
	\eta_{i j}(\alpha) = \frac{N}{\I^{\mspace{1mu}j-i}} \biggl( 1 - \frac{1}{1+\alpha^{N}} \bigl(1-(-1)^{j-i} \bigr) \biggr) + \frac{\alpha^{N-1}}{(1+\alpha^{N})^{2}} \, C_{i j} \, , 
\end{equation*}
where $C_{i j}$ will be properly defined below.
More precisely, we will prove that this equality between rational fractions holds for an infinite number of points, namely on $A$, which is enough to conclude that it is also true generically on $\mathbb{C}$ as a whole.
\medskip

\textit{Step 2: Computing residues.} 
For the commodity of the reader, let us recall here that for any meromorphic function $f$ and point $a \in \mathbb{C}$, if $\lim_{z\to a}(z-a)f(z)$ is finite and nonzero, then it is equal to $\mathrm{Res}_{a}(f)$. This result is readily adapted to $a= \infty$ following the definition $\mathrm{Res}_{\infty}(f) \equiv \mathrm{Res}_{0}(-\frac{1}{z^{2}}f(z^{-1}))$: if $\lim_{z\to \infty} \bigl( -zf(z) \bigr)$ is finite and nonzero, then it is equal to $\mathrm{Res}_{\infty}(f)$.

In addition, it will be worth noticing for what follows that $\sum_{j=1}^{N} \frac{1}{z- \omega^{j}} =  \frac{N z^{N-1}}{z^{N} -1}$, which can be checked by comparing the logarithmic derivatives of both sides of the identity $\prod_{j=1}^{N}(z-\omega^{j})=z^{N}-1$.
With this result, it becomes easy to obtain $\mathrm{Res}_{\infty}(\varphi_{i j; \alpha})$ since on one hand, $z \, \frac{N z^{N-1}}{z^{N}-1} \to N$ as $z \to \infty$ and on the other, \eqref{f at i} and \eqref{useful mero functions} respectively lead to $f(z) \to -\I$ and $\varsigma_{ij}(z) \to (-1)^{i} \, (-\I)^{i+j} = (-\I)^{j-i}$. Explicitly,
\begin{equation}
	\mathrm{Res}_{\infty}(\varphi_{i j; \alpha}) = -\frac{N}{\I^{\mspace{1mu}j-i}} \, .
\end{equation}

As for the other terms of the sum, let us show that
\begin{equation} \label{technical residue statement}
	\sum_{p \in \mathcal{P}_{1} \cup \mathcal{P}_{2}} \!\!\!\!\! \mathrm{Res}_{p}(\varphi_{i j; \alpha})  = \frac{1}{\I^{\mspace{1mu}j-i}}\frac{N}{1+\alpha^{N}}(1-(-1)^{j-i}) - \frac{\alpha^{N-1}}{(1+\alpha^{N})^{2}} C_{i j}    
\end{equation}
Provided the constants $C_{i j}$ are properly defined and respect $C_{i i}=0$, this will conclude the proof of \eqref{expression of eta(z)}. 

Surprisingly enough, the key to get to the result is to write $\varphi_{i j; \alpha}(z) = \varsigma_{i j}(\alpha\, z) \sum_{\omega^{k} \in \mathcal{Z}} \frac{1}{z-\omega^k}$ as the product 
\begin{equation}
	\varphi_{i j; \alpha}(z) = \alpha \, \sigma_{i j}(\alpha\, z)\kappa(z)  
\end{equation}
where $\sigma_{ij}(z) \equiv \varsigma_{ij}(z)/z$ and $\kappa(z) \equiv z \sum_{j=1}^{N} \frac{1}{z- \omega^{j}} = \frac{N}{1-z^{-N}}$. 
In this setting, since $\alpha \in A$, $\sigma_{i j}(\alpha z)$ exhibits a simple (resp. double) pole at any $\alpha^{-1}p$ for $p \in \mathcal{P}_{1}$ (resp. $\mathcal{P}_{2}$) while $\kappa$ stays regular at these values. By usual formulas in the case of lower-order poles, we can calculate residues of $\varphi_{i j; \alpha}$ as the sum of two terms
\begin{equation}
	\sum_{p \in \mathcal{P}_{1} \cup \mathcal{P}_{2}} \!\!\!\!\! \mathrm{Res}_{\alpha^{-1}p}(\varphi_{i j; \alpha})  = R_{1}+ R_{2}  
\end{equation}
where
\begin{align}
	R_{1} &\equiv \!\!\!\! \sum_{p \in \mathcal{P}_{1}\cup \mathcal{P}_{2}} \!\!\!\!\! \kappa(\alpha^{-1}p) \, \alpha \, \mathrm{Res}_{\alpha^{-1}p}(\sigma_{i j}(\alpha \, z)) \, , \\
	\label{technical def of R2}
	R_{2} &\equiv \sum_{p \in \mathcal{P}_{2}} \! \kappa'(\alpha^{-1}p) \, \alpha \,\frac{d}{dz} \, (z-\alpha^{-1}p)^{2} \, \sigma_{i j}(\alpha \, z) \Big|_{z=\alpha^{-1}p}  \, .
\end{align}
Both terms can be substantially simplified and will lead to \eqref{technical residue statement}. 

Let us start with $R_1$.
It is straightforward to check that $\kappa$ takes the same value at each $\alpha^{-1} \, p = -\alpha^{-1} \, \omega^{k} \in \mathcal{P}_{1} \cup \mathcal{P}_{2}$, namely $\kappa(p)= \frac{N}{1+\alpha^{N}}$, which leads to the factorised expression
\begin{equation}
	R_{1} = \frac{N}{1+\alpha^{N}} \!\! \sum_{p \in \mathcal{P}_{1}\cup \mathcal{P}_{2}}  \!\!\!\!\! \alpha \, \mathrm{Res}_{\alpha^{-1}p}(\sigma_{i j}(\alpha \, z)) \, .
\end{equation}
Moreover, we have $\alpha \, \mathrm{Res}_{\alpha^{-1}p}(\sigma_{i j}(\alpha \, z)) = \mathrm{Res}_{p}(\sigma_{i j}(z))$ by the usual change of variable formula.
Dressing the list of its poles, an application of the residue theorem on the function $\sigma_{i j}$ readily yields
\begin{equation}
	\sum_{p \in \mathcal{P}_{1} \cup \mathcal{P}_{2}} \!\!\!\!\! \mathrm{Res}_{p}(\sigma_{i j}) = -\mathrm{Res}_{\infty}(\sigma_{i j}) - \mathrm{Res}_{\, 0}(\sigma_{i j})  \, .
\end{equation}
The right-hand side of the last expression is then easily computed:
\begin{itemize}
	\item $\mathrm{Res}_{\infty}(\sigma_{i j})=\mathrm{Res}_{0}(\frac{-1}{z^2} \, \sigma_{i j}(z^{-1}))$ by definition and $\mathrm{Res}_{\infty}(\sigma_{i j})=-1/{{\I^{\mspace{1mu}j-i}}}$ since $\varsigma_{ij} \to (-\I)^{\mspace{1mu}j-i}$ as $z \to \infty$,
	\item $\mathrm{Res}_{\, 0}(\sigma_{i j})= \varsigma_{i j}(0)$ by Cauchy's integral formula, and we get $\varsigma_{ij}(0) = (-1)^{i} \, \I^{\mspace{1mu}j+i} = {\I^{\mspace{1mu}j-i}}$ by \eqref{useful mero functions} using $f(0)=\I$.
\end{itemize}
Finally
\begin{equation}
	R_{1} = \frac{1}{\I^{\mspace{1mu}j-i}} \, \frac{N}{1+\alpha^{N}} \, \bigl(1-(-1)^{j-i} \bigr) \, .    
\end{equation}

It only remains to compute $R_2$. First, working out $\kappa'(z)=-N^{2} \frac{z^{1-N}}{(1-z^{-N})^{2}}$ explicitly at any $\alpha^{-1}p$ for $p \in \mathcal{P}_{2}$ leads to $\kappa'(\alpha^{-1}p)=N^{2} \frac{\alpha^{N-1}}{(1+\alpha^{N})^{2}} \, p$ since $p^{N} = -1$.
Based on these calculations, \eqref{technical def of R2} can be rewritten as
\begin{equation}
	R_{2} = -\frac{\alpha^{N-1}}{(1+\alpha^{N})^{2}} \, C_{i j} \, , \qquad 
	C_{ij} \equiv \sum_{p \in \mathcal{P}_{2}} \!  N^{2} \, p \, \alpha \, \frac{d}{dz} \, \bigl(z-\alpha^{-1} \, p \bigr)^{2} \, \sigma_{i j}(\alpha\, z)\big|_{z=\alpha^{-1}p} \; .
\end{equation}
Moreover, a change of variables in the derivative yields 
\begin{equation}
	\alpha \, \frac{d}{dz} \, (z-\alpha^{-1}p)^{2} \, \sigma_{i j}(\alpha \, z) \big|_{z=\alpha^{-1}p} = \frac{d}{dz} \, (z-p)^{2} \, \sigma_{i j}(z) \big|_{z=p} \, ,    
\end{equation}
revealing that $C_{ij}$ is independent of $\alpha$.
Finally, $C_{i j}$ clearly vanishes when $i =j$ since $\mathcal{P}_{2}$ is then empty, which concludes the proof of \eqref{technical residue statement} and of \eqref{expression of eta(z)}.

\end{document}